\documentclass[useAMS,usenatbib,twocolumn]{aastex61}

\usepackage{mathtools}

\expandafter\def\csname ver@subfig.sty\endcsname{} %//// ---------- watch out!!!!

\usepackage{float}
\usepackage{hyperref}
\hypersetup{
     colorlinks = true,
     linkcolor = blue,
     urlcolor = red,
     citecolor = blue
}

\received{}
\revised{}
\accepted{}
\submitjournal{ApJ}

\usepackage{footnote}
\usepackage{tablefootnote}

\usepackage{enumerate}
\usepackage{enumitem}

\def\apj{{ ApJ}}
\def\apjl{{ ApJL}}
\def\apjs{{ ApJS}}

\def\aap{{ A\&A}}
\def\aj{{ AJ}}
\def\mnras{{ MNRAS}}
\def\araa{{ ARA\&A}}

\def\sci{{ Science}}
\def\nat {{ Nature}}
\def\pasj{{ PASJ}}

\def\prd{{ Phys. Rev. D}}

\def\pasa{{ PASA}}

\newcommand{\SmallEntryGap}{\vspace{0.1cm}}

\shorttitle{Fast radio bursts population studies}
\shortauthors{Bhattacharya \& Kumar}

\begin{document}

%%%%%%%%%%%%%%%%%%%%%%%%%%%%%%%%%%%%%%%%%%%%%%%%

\title{Population modelling of FRBs from source properties} 
%[FRB population studies]

\correspondingauthor{Mukul Bhattacharya}
\email{mukul.b@utexas.edu (MB)} 

\author{Mukul Bhattacharya}
%\affiliation{Department of Astronomy, University of Texas at Austin, Austin, TX 78712, USA}
\affiliation{Department of Physics, University of Texas at Austin, Austin, TX 78712, USA}

\author{Pawan Kumar}
\affiliation{Department of Astronomy, University of Texas at Austin, Austin, TX 78712, USA}

%\author[Bhattacharya \& Kumar]
%{Mukul Bhattacharya,$^1$\thanks{E-mail: mukul.b@utexas.edu (MB)} 
%and Pawan Kumar,$^2$\thanks{pk@astro.as.utexas.edu}\\ \\
%and Duncan Lorimer$^3$\thanks{Duncan.Lorimer@mail.wvu.edu}\\ \\ 
%$^{1}$ Department of Physics, University of Texas at Austin, Austin, TX 78712, USA\\
%$^{2}$ Department of Astronomy, University of Texas at Austin, Austin, TX 78712, USA}%\\
%$^{3}$ Department of Physics and Astronomy, West Virginia University, Morgantown, WV 26506, USA}

%\date{Accepted . Received ; in original form }

%\pagerange{\pageref{firstpage}--\pageref{lastpage}} \pubyear{2020}

%\maketitle

%\label{firstpage}

\begin{abstract}
We present a method to estimate the source properties of FRBs from observations by assuming a fixed DM contribution from a MW-like host galaxy, pulse temporal broadening models for turbulent plasma and a flat FRB energy spectrum. We then perform Monte Carlo simulations to constrain the properties of the FRB source, its host galaxy and scattering in the intervening plasma from the observational data of FRBs detected with Parkes. % published until February 2019.
%current observations. 
The typical scatter broadening of the intrinsic pulse is found to be considerably small $\lesssim 10^{-2}-1\ {\rm ms}$ from physical models,
with the ISM contribution suppressed significantly relative to IGM. The intrinsic width for non-repeating FRBs is broadened by a factor $\sim 2-3$ on average primarily due to dispersive smearing.
%, as opposed to the FRB 121102 bursts where most of the observed width is intrinsic. 
From the simulations, we find that the host galaxy DM contribution is likely to be comparable to the Galactic contribution and the FRB energy decreases significantly at high frequencies with a negative spectral index. %smaller than
%over the emitted frequency range with a spectral index within -1.5 to -3.0. 
%We find that the FRB spatial density increases with distance at small redshifts and then drops significantly at larger distances with a distribution peaked around redshift $\sim 1.0-1.5$.
The FRB spatial density is found to increase up to redshift $\sim 2.0$ and then drops significantly at larger distances. %The current observations suggest that a shallow energy spectrum is very unlikely for the inferred host galaxy DM contribution of FRB 121102. 
We obtain the energy distribution for FRB 121102 with repetition rate $\sim 0.1-0.3\ {\rm hr^{-1}}$ and exponential energy cutoff that is significantly smaller compared to typical FRB energies. We find that the probability of observing none of the other FRBs to be repeating at Parkes is $\sim 0.8-0.9$ with the current follow-up data insufficient to suggest more than one class of FRB progenitors.  
\end{abstract}

%\end{keywords}

\section{Introduction}
\label{Intro}
%\vspace{0.05in}
Fast radio bursts (FRBs) are radio transients with millisecond duration and Jy-level brightness, mostly detected from high Galactic latitudes \citep{Lorimer07,Thornton13}. The physical origin of these bursts is still unknown, primarily due to their short durations and the low angular resolutions of the current radio surveys. The frequency dependence of the arrival time delay ($\propto \nu^{-2}$) and the pulse width evolution ($\propto \nu^{-4}$) of FRBs are both consistent with propagation through cold, turbulent plasma suggesting their astrophysical origin. 
Until date, more than 110 apparently non-repeating FRBs and $\sim$20 repeating FRBs have been published \footnote{\href{http://www.astronomy.swin.edu.au/pulsar/frbcat/}{FRB catalogue} lists the properties of all discovered FRBs \citep{Petroff16}}, which were discovered in frequency bands between 300 MHz and 8 GHz \citep{Shannon18,CHIME19}. 
Many more of these energetic bursts are expected to be detected in the near future with several broad-band and wide-field surveys becoming operational in full capacity. 
%improving sensitivities of the upcoming radio transient surveys.

The host galaxy identification has now become possible from the real-time arcsecond localisation of some of the known sources, which provides crucial information about the nature of galaxies where these events originate as well as their nearby progenitor environments. 
The repeating FRB 121102 was the first to be localised to within $\sim0.1\arcmin\arcmin$ resolution with the Jansky VLA \citep{Chatterjee17} and is found to be associated to a dwarf star-forming host galaxy at z=0.19273 \citep{Tendulkar17} with a steady radio source at a separation of $\lesssim 0.01\arcmin\arcmin$ determined with the European VLBI \citep{Marcote17}. 
Recent localisations of three apparently non-repeating FRBs and repeating FRB 180916.J0158+65 indicate that the host galaxies and progenitor environments for these bursts differ substantially from those of FRB 121102 \citep{Bannister19,Prochaska19,Ravi19,Marcote20}. 
The Canadian Hydrogen Intensity Mapping Experiment (CHIME) detected FRB 200428 which is spatially coincident with the galactic Soft Gamma-ray Repeater (SGR) 1935+2154, thus reporting the first possible Galactic origin of these bursts \citep{ScholzCHIME20}.

\begin{table*}
%\label{Table1}
\begin{center}
\caption{Observed and inferred parameters for non-repeating FRBs reported with Parkes and with total DM exceeding $500\ {\rm pc\ cm^{-3}}$. We use a DM cutoff to minimize the error in the estimation of inferred FRB parameters due to the host galaxy DM assumption. For each reported FRB, we select the observation with largest $S/N$ from the \href{http://www.astronomy.swin.edu.au/pulsar/frbcat/}{FRB catalogue}.
%FRB 141113 and 170107 were detected by Arecibo and ASKAP, respectively, and FRBs 160317, 160410 and 160608 were detected by UTMOST, while the rest of the FRBs were detected by Parkes. 
We also exclude the FRBs with unresolved/imaginary intrinsic widths from our analysis (see equation \ref{wint}). The definitions of all burst parameters are discussed in Section \ref{sec2}.
}
\label{Table1}
\bgroup
\def\arraystretch{1.0}
\begin{tabular}{| c | c | c | c | c | c | c | c | c |}
\hline
\hline
\centering

% FRB, Reference, Speak, DMtot, DM_MW, z, L_int1/2, w_obs, w_int1/2 
FRB & Reference & $S_{{\rm peak,obs}}$ & $DM_{{\rm tot}}$ & $DM_{{\rm MW}}$ & $z$ & $L_{{\rm int1/2}}$ & $w_{{\rm obs}}$ & $w_{{\rm int1/2}}$ \\
& & (Jy) & (${\rm pc\ cm^{-3}}$) & (${\rm pc\ cm^{-3}}$) & & ($10^{44}$ erg/s) & (ms) & (ms)\\
\hline \hline
010125 & \citet{Petroff16} & 0.54 & 790.3 & 110 & 0.82 & 2.06/2.06 & 10.6 & $4.06/4.05$ \\ \hline
010312 & \citet{Keane19} & 0.15 & 1187 & 51 & 1.42 & 1.66/1.66 & 37.0 & $14.56/14.55$ \\ \hline
010621 & \citet{Petroff16} & 0.53 & 748.0 & 523 & 0.20 & 0.11/0.11 & 8.0 & $2.93/2.93$ \\ \hline
%010724 & 1.57 & 31.48 & 375.0 & 0.35 & 0.53/0.53 & 0.78 & 20.0 & $2.60\times10^{-4}/5.19\times10^{-2}$ & $14.53/14.53$ \\ \hline
090625 & \citet{Petroff16} & 1.14 & 899.55 & 31.69 & 1.06 & 7.36/7.57 & 1.92 & $0.73/0.71$ \\ \hline
110220 & \citet{Petroff16} & 1.11 & 944.38 & 34.77 & 1.12 & 6.52/6.53 & 6.59 & $3.06/3.05$ \\ \hline
%%110523 & GBT & \citet{Masui15} & 0.60 & 623.30 & 0.69 & 1.34/2.70 & 1.73 & $0.77/0.38$ \\ \hline
110626 & \citet{Petroff16} & 0.63 & 723.0 & 47.46 & 0.81 & 2.22/2.26 & 1.41 & $0.58/0.56$ \\ \hline
110703 & \citet{Petroff16} & 0.45 & 1103.6 & 32.33 & 1.33 & 4.32/4.37 & 3.90 & $1.55/1.53$ \\ \hline
120127 & \citet{Petroff16} & 0.62 & 553.3 & 31.82 & 0.61 & 1.00/1.00 & 1.21 & $0.60/0.60$ \\ \hline
121002 & \citet{Petroff16} & 0.43 & 1629.18 & 74.27 & 2.00 & 11.49/11.86 & 5.44 & $1.66/1.61$ \\ \hline
130626 & \citet{Petroff16} & 0.74 & 952.4 & 66.87 & 1.09 & 5.14/5.30 & 1.98 & $0.74/0.71$ \\ \hline
%130628 & 1.91 & 1.22 & 469.88 & 0.47 & 4.99/5.94 & 0.06 & 0.64 & $1.02\times10^{-3}/8.89\times10^{-2}$ & $0.11/0.09$ \\ \hline
130729 & \citet{Petroff16} & 0.22 & 861 & 31 & 1.01 & 1.00/1.00 & 15.61 & $7.73/7.73$ \\ \hline
131104 & \citet{Petroff16} & 1.16 & 779 & 71.1 & 0.85 & 3.80/3.82 & 2.37 & $1.15/1.15$ \\ \hline
140514 & \citet{Petroff16} & 0.47 & 562.7 & 34.9 & 0.62 & 0.64/0.65 & 2.82 & $1.68/1.68$ \\ \hline
%141113 & 0.04 & 0.08 & 400.3 & 0.18 & 0.003/0.003 & 0.0005 & 2.00 & $1.62\times10^{-5}/1.79\times10^{-2}$ & $1.65/1.65$ \\ \hline
150215 & \citet{Petroff17} & 0.70 & 1105.6 & 427.2 & 0.82 & 2.14/2.14 & 2.88 & $1.37/1.37$ \\ \hline
%151206 & Parkes & \citet{Bhandari18} & 0.30 & 1909.80 & 2.28 & 20.63/20.13 & 3.0 & $0.44/0.45$ \\ \hline
151230 & \citet{Bhandari18} & 0.42 & 960.4 & 38.0 & 1.13 & 2.70/2.71 & 4.4 & $1.98/1.97$ \\ \hline
171209 & \citet{Oslowski19} & 1.48 & 1457.4 & 13.00 & 1.84 & 29.90/30.47 & 2.5 & $0.87/0.85$ \\ \hline
%%180309 & \citet{Oslowski19} & 20.8 & 263.47 & 44.69 & 0.19 & 1.76/1.76 & 0.58 & $0.48/0.48$ \\ \hline
180311 & \citet{Oslowski19} & 0.15 & 1570.9 & 45.20 & 1.95 & 3.64/3.64 & 13.4 & $4.53/4.53$ \\ \hline
180714 & \citet{Oslowski19} & 0.60 & 1467.9 & 257.0 & 1.52 & 7.93/8.02 & 2.9 & $1.15/1.14$ \\ \hline
%%160317 & UTMOST & \citet{Caleb17} & 3.0 & 1165 & 1.03 & 14.58/14.65 & 21.0 & $9.82/9.76$ \\ \hline
%160410 & 7.0 & 28.00 & 278 & 0.19 & 0.65/0.65 & 0.20 & 4.0 & $1.61\times10^{-4}/1.32\times10^{-1}$ & $3.04/3.04$ \\ \hline
%%160608 & UTMOST & \citet{Caleb17} & 4.3 & 682 & 0.51 & 3.79/3.80 & 9.0 & $5.40/5.39$ \\ \hline
%%170107 & ASKAP & \citet{Bannister17} & 22.3 & 609.5 & 0.68 & 163.88/171.71 & 2.6 & $0.34/0.33$ \\ \hline
%%170416 & ASKAP & \citet{Shannon18} & 19.4 & 523.2 & 0.56 & 22.06/22.07 & 5.0 & $2.86/2.86$ \\ \hline
%%170428 & ASKAP & \citet{Shannon18} & 7.7 & 991.7 & 1.17 & 96.79/98.14 & 4.4 & $1.02/1.00$ \\ \hline
%%171116 & ASKAP & \citet{Shannon18} & 19.6 & 618.5 & 0.69 & 60.58/60.93 & 3.2 & $1.04/1.04$ \\ \hline
%%180110 & ASKAP & \citet{Shannon18} & 128.1 & 715.7 & 0.82 & 964.88/988.45 & 3.2 & $0.61/0.60$ \\ \hline
%%180131 & ASKAP & \citet{Shannon18} & 22.2 & 657.7 & 0.74 & 56.54/56.64 & 4.5 & $2.03/2.02$ \\ \hline
\hline
\end{tabular}
\egroup
\end{center}
\vspace{-0.25cm}
\end{table*}
%\vspace{-2cm}

The cosmological origin of FRBs is strongly suggested by their large dispersion measures (integrated electron column density along the line of sight, DM = $\int n_e dl \sim 10^3 \ {\rm pc\ cm^{-3}}$), which typically exceeds the expected Galactic interstellar medium (ISM) contribution by almost an order of magnitude \citep{CL02}. Assuming that most of the excess DM is due to the ionized intergalactic medium (IGM) contribution \citep{Ioka03,Inoue04}, the inferred redshifts are in the range $z \sim 0.2-2$ with a significant isotropic energy release of $\sim 10^{38}-10^{40}\ {\rm erg}$ \citep{Thornton13,KP15,Champion16}. Due to their possible cosmological origin, FRBs can also be potentially used as a probe to study the distribution of free electrons in the IGM and cosmology \citep{Gao14,Zheng14}. The $\sim$ms pulse duration constrains the FRB source size, thereby implying high radio brightness temperatures and coherent emission.
% \citep{Katz14,LG14}. 

Although the all-sky isotropic event rate for FRBs above fluence $\sim$1 Jy ms is relatively high $\sim 10^3 - 10^4\ {\rm day}^{-1}$ \citep{Thornton13,Champion16,Rane16,Lawrence17}, a large majority of FRBs appear to be one-off events despite dedicated follow-up efforts spanning several hundreds of hours \citep{Petroff15,Ravi15,Shannon18}.
%none of the FRBs except FRB 121102 \citep{Scholz16,Spitler16} and FRB 180814.J0422+73 \citep{CHIME19}
%have been observed to repeat yet despite dedicated follow-up efforts . 
This might be due to two possible reasons: (1) two different classes of FRB progenitors (non-repeating and repeating bursts) as suggested by \citet{Keane16}, or (2) observational bias due to the finer localization and higher sensitivity of Arecibo and CHIME relative to Parkes. 
It is also possible that low/clustered repetition rates \citep{Opp18} or unfavourable luminosity functions \citep{CP18,Caleb19} hinder the detection of subsequent bursts from repeating sources for telescopes with given sensitivity and observing time.  
While the repeating FRB 180916.J0158+65 has been found to exhibit a periodicity of $\sim$16 days in its repetition activity \citep{CHIME20}, it is also possible that the first repeating source FRB 121102 shows a periodicity of $\sim$160 days in its repetition behaviour \citep{Rajwade20}.

Several progenitor models, including both cataclysmic and non-cataclysmic scenarios, have been proposed in the FRB literature: 
collapsing supermassive neutron stars (NSs; \citealt{FR14,Zhang14}), 
compact binary mergers \citep{Piro12,Kashiyama13,Totani13}, 
galactic flaring stars \citep{Loeb14}, 
radio emission from pulsar companions \citep{MZ14}, 
magnetar giant flares \citep{PP10,Kulkarni14,Lyubarsky14,Katz16}, %Thornton et al. 2013, Pen \& Connor 2015,  
supergiant pulses from young pulsars \citep{Connor16,CW16,Lyutikov16}, %Katz 2016, 
young rapidly spinning magnetars \citep{Kashiyama17,Metzger17}
and plasma stream interacting with NS magnetosphere \citep{Zhang17}. 
%and coherent curvature radiation \citep{Kumar17,LK18}. 
While most of the aforementioned models are primarily based on timescales and energetics considerations, \citet{Kumar17} and \citet{LK18} have recently discussed the detailed calculations for the emission conditions and the plasma properties for coherent curvature radiation model.

In this work, we concentrate only on the cosmological origin of FRBs 
%as supported by the localisation of FRB 121102 at $z=0.19273$. We 
and consider the bursts with total DM $\geq 500\ {\rm pc\ cm^{-3}}$ that were detected with Parkes.  %published until February 2019 
We place a lower DM cutoff for the FRBs considered in our analysis here to reduce the error in the estimates of the inferred parameters which are based on the assumption of the host galaxy DM contribution. 
Previous population synthesis efforts have focussed on investigating the underlying FRB source classes \citep{Gardenier19}, volumetric densities of events \citep{Caleb16,Niino18}, fluence distribution of FRBs \citep{James19a}, effect of instrumental parameters on observed properties \citep{Connor19}, brightness distributions and their relation to the spatial density of FRB sources \citep{Vedantham16,ME18,MB19} and the fraction of repeating sources \citep{Caleb19}. 
Unlike the previous studies, here we derive the physical properties of FRB sources directly from the current observed distributions. 
%repeated captures of asteroids by NSs (Geng \& Huang 2015, Dai et al. 2016), 
% annihilating BHs (Keane et al. 2012)
%dark matter induced collapse of NS (Fuller \& Ott 2015)

In this study, we first present a method to estimate the true properties of the non-repeating and repeating burst populations from the current observations. 
We then perform Monte Carlo (MC) simulations based on specific model distributions of the true properties to constrain the scattering properties of the intervening IGM and ISM, FRB spatial density as a function of $z$, host galaxy DM and the spectral index of the assumed power-law FRB energy density. 
We also discuss whether the repeating FRB 121102 is representative of the entire FRB population based on its repeating behaviour and the follow-up observations for the non-repeating FRBs. This paper is organised as follows. 
In Section \ref{sec2}, we estimate the distances and intrinsic widths for the FRBs using appropriate models for the IGM and ISM properties, and further evaluate the energies and luminosities for these bursts. We then describe our MC code in Section \ref{sec3} and discuss the simulation results in Section \ref{sec4}. We constrain the model parameters for the FRB population using the current observations. In Section \ref{sec5}, we investigate whether the repeating FRB 121102 is representative of all FRBs, and finally present our summary and conclusions in Section \ref{sec6}.

%\vspace{0.1 in}
%\EntryGap 
%\SmallEntryGap
\section{FRB intrinsic parameters from observables}
\label{sec2}
In this section, we first estimate the distances to the observed FRBs from their total dispersion measure ($DM_{{\rm tot}}$) by assuming a fixed host galaxy DM contribution ($DM_{{\rm host}}$). Throughout this work, we only consider the Parkes FRBs with $DM_{\rm tot}$ values exceeding $500\ {\rm pc\ cm^{-3}}$ to minimize the error in the inferred FRB source properties (such as luminosity, energy, etc.) that are based on the estimates of FRB distances. 
The inferred FRB distances are determined by the assumptions about the host galaxy properties and the source location inside it, with the error in $z$ expected to be larger for a larger contribution from the relatively uncertain $DM_{{\rm host}}$ to $DM_{{\rm tot}}$. 
It should be noted that restricting the Parkes FRB sample up to a certain $DM_{\rm tot}$ cutoff can introduce some bias through sample selection effects which will influence the results from population modelling and KS analysis of these events. However, a large majority ($\gtrsim 80\%$) of the bursts currently detected with Parkes satisfy $DM_{\rm tot} \geq 500\ {\rm pc\ cm^{-3}}$ and are therefore included in our analysis. Here we assume that the bias due to the uncertainty in the $DM_{\rm host}$ contribution significantly overrides the bias introduced by restricting our sample.

While the $DM_{\rm host}$ contribution to $DM_{\rm tot}$ is generally stochastic with more variability for nearby FRBs, the contributions from the free electrons in the galactic halo to $DM_{\rm host}$ and density inhomogeneities (such as cosmic voids and strong filaments) to $DM_{\rm IGM}$ are both expected to be small. 
We then obtain the intrinsic pulse widths ($w_{{\rm int}}$) from the observed FRB widths ($w_{{\rm obs}}$) using scattering models for the pulse temporal broadening due to the multipath propagation through the ionized ISM and IGM. The burst luminosities and energies are calculated for a flat energy spectrum from the peak flux density, distance and the frequency range for FRB radio emission. 
We also estimate the error in the inferred parameters due to the various assumptions that are involved in our models. %\vspace{-0.3cm} 
%We study the correlation between the intrinsic parameters and the relative contributions from different width components. 

\begin{table*}
%\label{Table1}
\begin{center}
\caption{DM components of the five FRB sources that were localised to their host galaxies until April 2020. While FRB 121102 and FRB 180916 are the first two repeating sources that were reported, three non-repeating sources FRB 190523, FRB 180924 and FRB 181112 were localised upon detection. The $DM_{\rm MW}$ estimates are obtained from the Milky Way disk and halo contributions based on the NE2001 model. We obtain the $DM_{\rm host}$ contribution by subtracting $DM_{\rm IGM}$ (equation \ref{DMigm}) from $DM_{\rm Ex} = DM_{\rm tot} - DM_{\rm MW}$.}
\label{Table1DM}
\bgroup
\def\arraystretch{1.0}
\begin{tabular}{| c | c | c | c | c | c | c | c |} % c | c | c |
\hline
\hline
\centering
FRB & Telescope & Reference & $DM_{\rm tot}$ & $DM_{\rm MW}$ & $z$ & $DM_{\rm host}/DM_{\rm tot}$ & $DM_{\rm MW}/DM_{\rm tot}$ \\ % & Telescope & $DM_{\rm host}/DM_{\rm tot}$ & $DM_{\rm MW}/DM_{\rm tot}$
&  &  & (${\rm pc\ cm^{-3}}$) & (${\rm pc\ cm^{-3}}$) & & & \\ \hline \hline % & & & 
121102 & Arecibo & \citet{Tendulkar17} & 558 & 188 & 0.19 & 0.42 & 0.34\\ \hline % & 0.42 & 0.34 & Arecibo 
180916 & CHIME &\citet{Marcote20} & 349 & 199 & 0.03 & 0.36 & 0.57\\ \hline % & 0.36 & 0.57 & CHIME 
190523 & DSA-10 & \citet{Ravi19b} & 761 & 102 & 0.66 & 0.21 & 0.13\\ \hline %  & 0.21 & 0.13 & DSA-10 
180924 & ASKAP & \citet{Bannister19} & 361 & 70 & 0.32 & 0.16 & 0.19\\ \hline %  & 0.16 & 0.19 & ASKAP 
181112 & ASKAP & \citet{Prochaska19} & 589 & 182 & 0.47 & 0.09 & 0.31\\ \hline % & 0.09 & 0.31 & ASKAP 
\hline
\end{tabular}
\egroup
\end{center}
\vspace{-0.1cm}
\end{table*}
%\vspace{-2cm}

%\vspace{-0.1cm}
\subsection{Distance and width estimates}
The total DM for any FRB has contributions from the IGM ($DM_{\rm IGM}$), the Milky Way (MW) ISM ($DM_{\rm MW}$) and the host galaxy ISM. Including the cosmological expansion factor for the host galaxy contribution gives
\begin{equation}
DM_{\rm tot} = DM_{\rm IGM} + DM_{\rm MW} + \frac{DM_{\rm host}}{(1+z)}.  
\label{DMtot}
\end{equation}
The IGM contribution increases with the source redshift as \citep{Ioka03,Inoue04,DZ14}
\begin{eqnarray}
%\begin{aligned}
DM_{\rm IGM} = \frac{c}{H_0}\int_{0}^{z}\frac{f_{\rm IGM}n_{\rm e}(z^{\prime})x(z^{\prime})dz^{\prime}}{(1 + z^{\prime})^{2} [\Omega_{\rm m}(1+z^{\prime})^3 + \Omega_{\Lambda}]^{0.5}}\nonumber\\
 = (1294.9\ {\rm pc\ cm^{-3}}) \int_{0}^{z}\frac{(1+z^{\prime})dz^{\prime}}{\sqrt{(1+z^{\prime})^3 + 2.7}}
%\end{aligned}
\label{DMigm}
\end{eqnarray}
where $f_{\rm IGM} = 0.83$ is the fraction of baryon mass in the IGM, $n_e(z) = 2.1\times10^{-7} (1+z)^3 \ {\rm cm^{-3}}$ is the number density of free electrons, $x(z) \approx 7/8$ is the ionization fraction with cosmological parameters as $H_{0} = 68 {\rm \ km\ s^{-1}}{\rm Mpc^{-1}}$, $\Omega_{\rm m} = 0.27$ and $\Omega_{\Lambda} = 0.73$ \citep{Planck14}. 

The DM contribution from the Galactic ISM along the FRB source line of sight is obtained from the NE2001 model \citep{CL02}. The host galaxy DM contribution depends on the type of the galaxy, location of the FRB source within the galaxy as well as our viewing angle relative to the galaxy. 
The recent localisations of the non-repeating sources FRB 180924, FRB 181112 and FRB 190523 along with that of the repeating FRB 180916 can be particularly useful in order to constrain the relative DM contributions from the IGM, host galaxy and MW. In Table \ref{Table1DM}, we list the DM contributions from these components for the three apparently non-repeating and two repeating FRB sources. 
While the repeating FRB 121102 and FRB 180916 were localised close to the star-forming regions inside dwarf and spiral galaxy respectively \citep{Tendulkar17,Marcote20}, the non-repeating FRBs 180924, 181112 and 190523 were found to be located inside more massive galaxies \citep{Bannister19,Prochaska19,Ravi19b}. Recent works have investigated the baryonic content of the Universe from the DM and also the properties of the host galaxies for the FRBs that were localised by ASKAP within $0.11 < z < 0.52$ \citep{Macquart2020,Bhandari20}. 
\citet{Chittidi20} and \citet{Simha20} have studied the contributions to the FRB 190608 observed pulse from the local environment of the burst as well as the galaxies that are present in the foreground. 

%Although the host galaxy for the repeating FRB 121102 has been identified to be a dwarf star-forming galaxy \citep{Tendulkar17}, there is still no information about the host galaxies for the other sources. 
Due to the general uncertainties associated with the host galaxy properties and the FRB source location inside them, here we assume that the host galaxy has a free electron density distribution similar to that of the MW with a typical contribution of $DM_{\rm host} \approx 100\ {\rm pc\ cm^{-3}}$. 
It can be seen that for the five localised FRB sources that are known at present, the $DM_{\rm host}$ component does not outweigh the IGM DM contribution and is also comparable to the MW contribution. 
Therefore, it is reasonable to assume here a fixed $DM_{\rm host}$ contribution in order to derive the burst distances. Later, we consider a more general scenario whereby the host galaxy DM is similar to that for the MW from the NE2001 model (see Section 3). 
With the values of $DM_{\rm tot}$, $DM_{\rm MW}$ and $DM_{\rm host}$ known, we solve for the redshifts of the non-repeating bursts from equation (\ref{DMtot}). As the host galaxy for the repeating FRB 121102 is localized at z=0.19273 \citep{Tendulkar17}, the $DM_{\rm IGM}$ value is precisely known and equation (\ref{DMtot}) further gives $DM_{\rm host} \approx 281\ {\rm pc\ cm^{-3}}$. Once $z$ is estimated, the comoving distance to the source is obtained as $D(z) = (8.49\ {\rm Gpc}) \int_{0}^{z} [(1+z^{\prime})^3 + 2.7]^{-0.5} dz^{\prime}$ with a luminosity distance $D_{L}(z) = (1+z)D(z)$.

\begin{table*}
%\label{Table1}
\begin{center}
\caption{Observed parameters for the repeating FRB 121102 sub-bursts that were published until February 2019. In our analysis here, we have excluded the sub-bursts with unresolved/imaginary intrinsic widths.}
\label{Table1r}
\bgroup
\def\arraystretch{1.0}
\begin{tabular}{| c | c | c | c | c | c ||}
\hline
\hline
\centering
MJD & Telescope & Pulses detected & $S_{{\rm peak,obs}}$ & $w_{{\rm obs}}$ & Reference \\
%FRB & Telescope & Reference & $S_{{\rm peak,obs}}$ & $DM_{{\rm tot}}$ & $z$ & $L_{{\rm int1/2}}$ & $w_{{\rm obs}}$ & $w_{{\rm int1/2}}$ \\
& & & (Jy) & (ms) & \\
%& & & (Jy) & (${\rm pc\ cm^{-3}}$) & & ($10^{44}$ erg/s) & (ms) & (ms)\\
\hline \hline
56233.283 -- 57175.748 & Arecibo (1.4 GHz) & 10 & 0.02 -- 0.31 & 2.80 -- 8.70 & \citet{Spitler16} \\ \hline
57339.356 -- 57345.462 & GBT (2.0 GHz) & 4 & 0.04 -- 0.09 &  5.97 -- 6.73 & \citet{Scholz16} \\ \hline
57364.205 & Arecibo (1.4 GHz) & 1 & 0.03 & 2.50 & \citet{Scholz16} \\ \hline
57647.232 -- 57765.136 & GBT (2.0 GHz) & 9 & 0.08 -- 0.56 & 1.36 -- 3.45 & \citet{Scholz17} \\ \hline
57765.101 -- 57765.143 & Arecibo (1.4 GHz) & 4 & 0.02 -- 0.09 & 3.66 -- 4.34 & \citet{Scholz17} \\ \hline
57623.744 -- 57649.452 & VLA (3.0 GHz) & 9 & 0.13 -- 3.34 & 1.1 -- 2.5 & \citet{Law17} \\ \hline
57769.688 -- 57803.693 & Effelsberg (1.4 GHz) & 13 & 0.11 -- 0.80 & 1.80 -- 5.10 & \citet{Hardy17} \\ \hline
57747.130 -- 57772.129 & Arecibo (4.5 GHz) & 15 & 0.1 -- 1.2 & 0.15 -- 1.92 & \citet{Michilli18} \\ \hline
57991.580 -- 57991.583 & GBT (6.0 GHz) & 2 & 0.4 -- 0.9 & 0.27 -- 0.59 & \citet{Michilli18} \\ \hline
57991.410 -- 57991.448 & GBT (6.0 GHz) & 18 & 0.05 -- 0.70 & 0.18 -- 1.74 & \citet{Gajjar18} \\ \hline
57620.392 -- 57620.400 & Effelsberg (5.0 GHz) & 3 & 0.1 -- 0.3 & 0.5 -- 1.7 & \citet{Spitler18} \\ \hline
\hline
\end{tabular}
\egroup
\end{center}
\vspace{-0.1cm}
\end{table*}
%\vspace{-2cm}

The intrinsic width of a cosmological FRB source is broadened due to both propagation and telescope effects. Excluding the pulse broadening components from the observed width gives 
\begin{equation}
w_{\rm int}^{2} = \frac{w_{\rm obs}^{2} - (w_{\rm DM}^2 + w_{\rm samp}^2 + w_{\rm IGM}^2 + w_{\rm ISM, MW}^2)}{(1+z)^2} - w_{\rm ISM, host}^2 
\label{wint}
\end{equation}
where, $w_{\rm DM} = 8.3\times10^6 (DM_{\rm tot}\: \Delta\nu/\nu_{0}^{3})\ {\rm ms}$, is the dispersive smearing across single frequency channels with $\Delta\nu$ and $\nu_{0}$ being the channel bandwidth and the central frequency of observation in MHz, respectively. While $w_{\rm samp}$ is the sampling time of the observation, $w_{\rm IGM}/w_{\rm ISM,MW}/w_{\rm ISM,host}$ denotes the pulse temporal broadening due to scattering in the IGM/MW ISM/host galaxy ISM and $(1+z)$ is the cosmic expansion factor. 

It should be noted that the distribution of the pulse observed/intrinsic width $w_{\rm obs/int}$ is directly affected by the temporal resolution of the telescope/survey used. 
\citet{Connor19} discuss in detail the effect of time and frequency resolution of detection instruments on the distribution of pulse width as well as other observed properties. 
While most non-repeating FRBs detected with Parkes were at a fine resolution of $\sim 0.1\ {\rm ms}$, the other events detected by GBT, UTMOST and ASKAP were at a coarse temporal resolution of $\sim 1.0\ {\rm ms}$. 
In our analysis here, we include the instrument time resolution $\sim w_{\rm samp}$ for each burst while computing $w_{\rm int}$ from $w_{\rm obs}$ using equation (\ref{wint}) and discuss the associated bias $\Delta w_{\rm int}$ later in this section. While a telescope with a coarse time resolution is less likely to detect a pulse with a relatively small $w_{\rm obs}$ due to the associated instrumental noise, there will also be an observing bias against the events with fairly large $w_{\rm obs}$ as the sensitivity reduces gradually. While \citet{Shannon18} have shown that the luminosity distributions for the Parkes FRBs at finer resolution and ASKAP FRBs at coarser resolution are fairly similar, the observing bias tends to select more events at relatively smaller distances. 

As the radio pulses propagate through the ionised plasma in the intervening IGM and ISM, they are scattered due to the inhomogeneities in the electron density along the line of sight resulting in multipath propagation and thereby scatter broadening. Although detailed constraints on scattering timescales are difficult to obtain directly, it is expected that most of the scattering is primarily due to IGM turbulence \citep{Williamson72,MK13} whereas the Galactic scattering contribution along such lines of sight is comparatively smaller. 
%The reported scattering timescales are significantly larger compared to the scattering timescales expected from the Galactic turbulence along such lines of sight and 
%it is expected that most of the scattering is predominantly due to the IGM \citep{Williamson72,MK13}. 
It should be noted that the majority of the 13 FRBs that were initially reported with CHIME were found to be significantly scattered \citep{CHIME19b}, which is in contrast to the ASKAP FRBs that showed minimal scattering contribution to the pulse width \citep{Shannon18}.   
%CHIME found a majority (8/13) of events that were significantly scattered, but ASKAP found evidence for scattering in just 3/20 events (Shannon et al. 2018). 
Furthermore, the apparent correlation between the scattering widths and DM was found to be relatively weak. 
%7/13 CHIME bursts show temporal scattering and they report a weak correlation between scattering time and DM. 
Due to the absence of sufficient information on the FRB scattering timescales, we consider two models to evaluate the ISM and IGM scattering broadening timescales for each FRB as discussed below.

\begin{figure*}
%\vspace{-10em}
\gridline{\fig{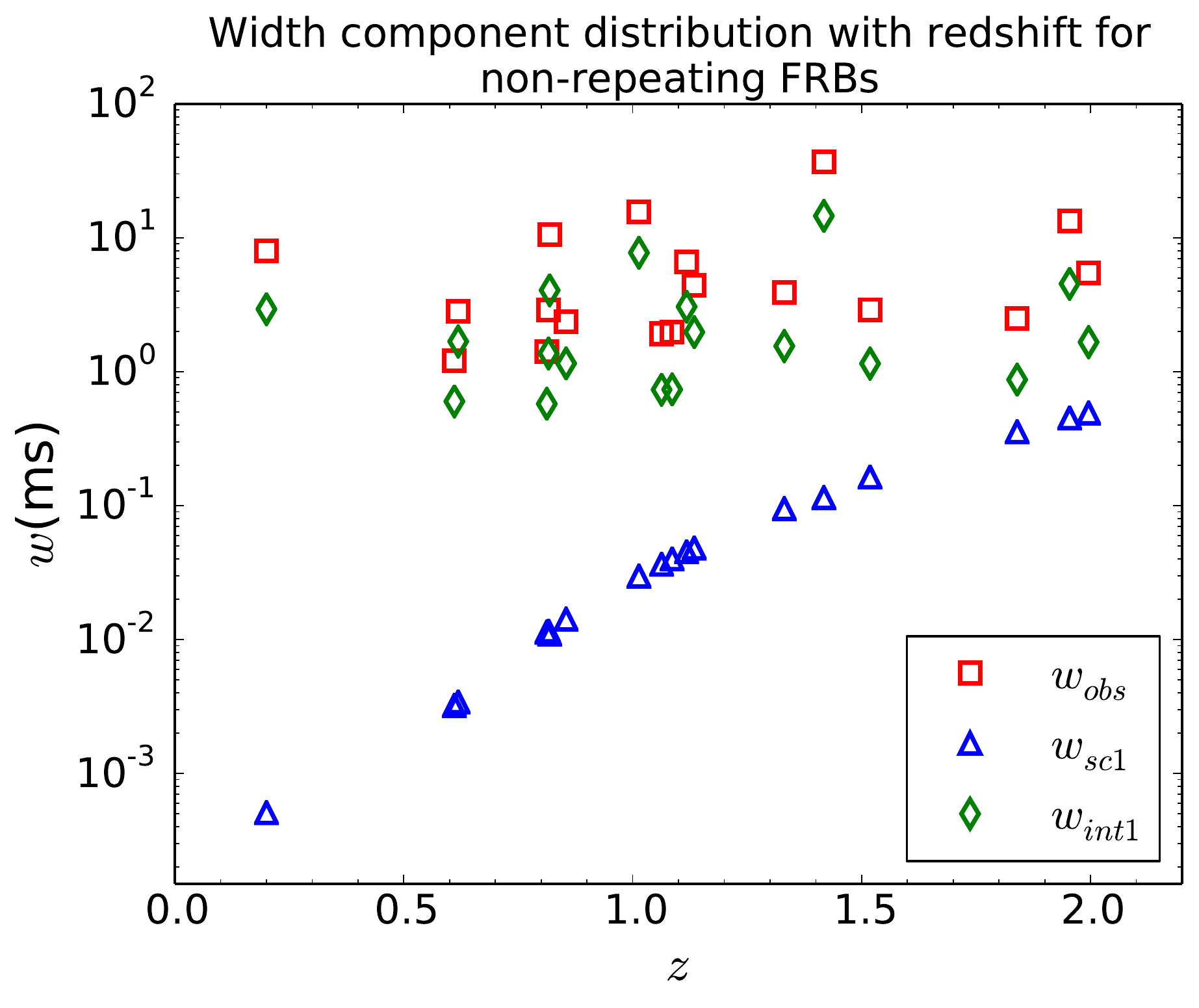}{0.46\textwidth}{}
          \fig{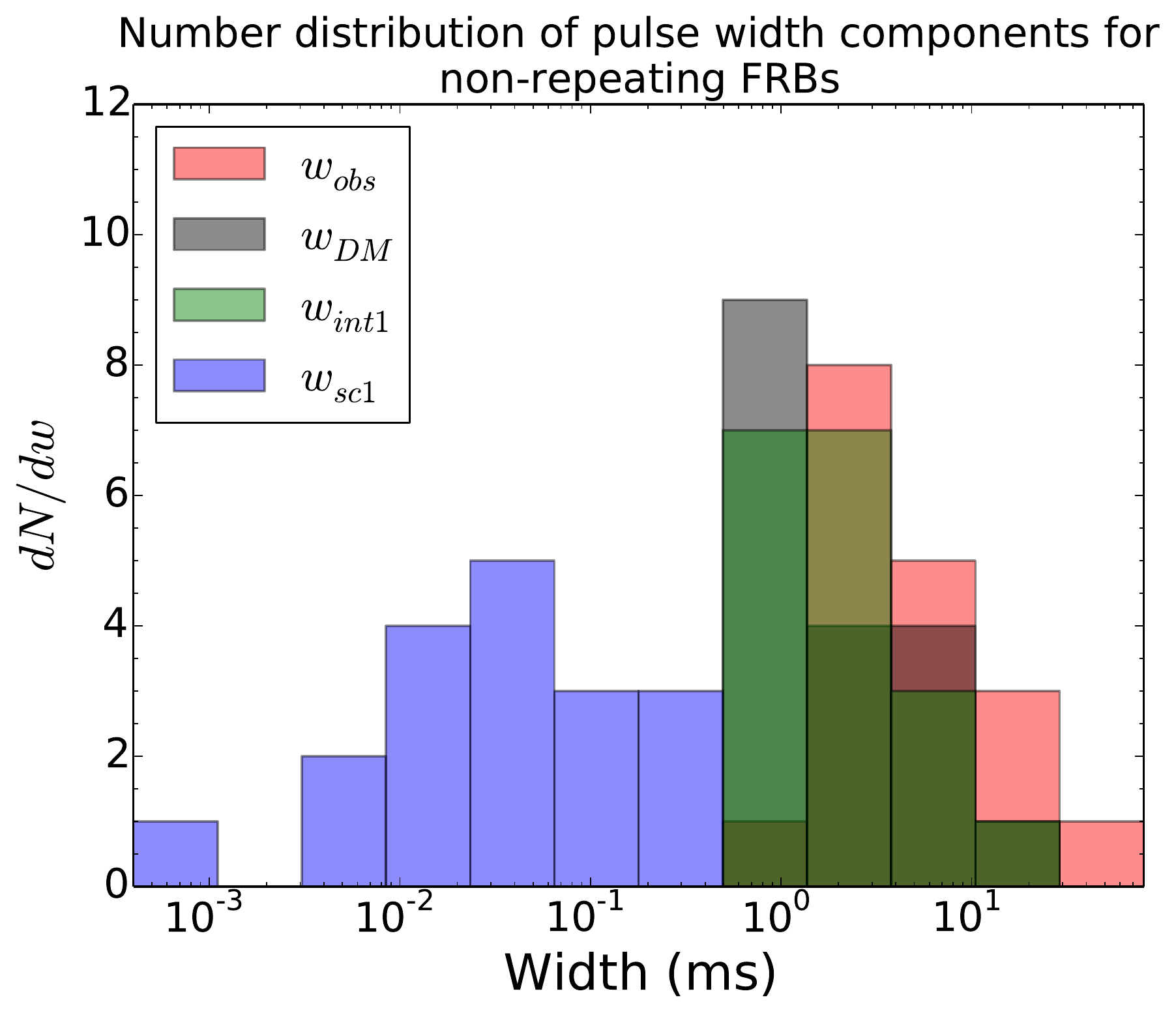}{0.46\textwidth}{}          
          }  \vspace{-2.5em}        
\gridline{\fig{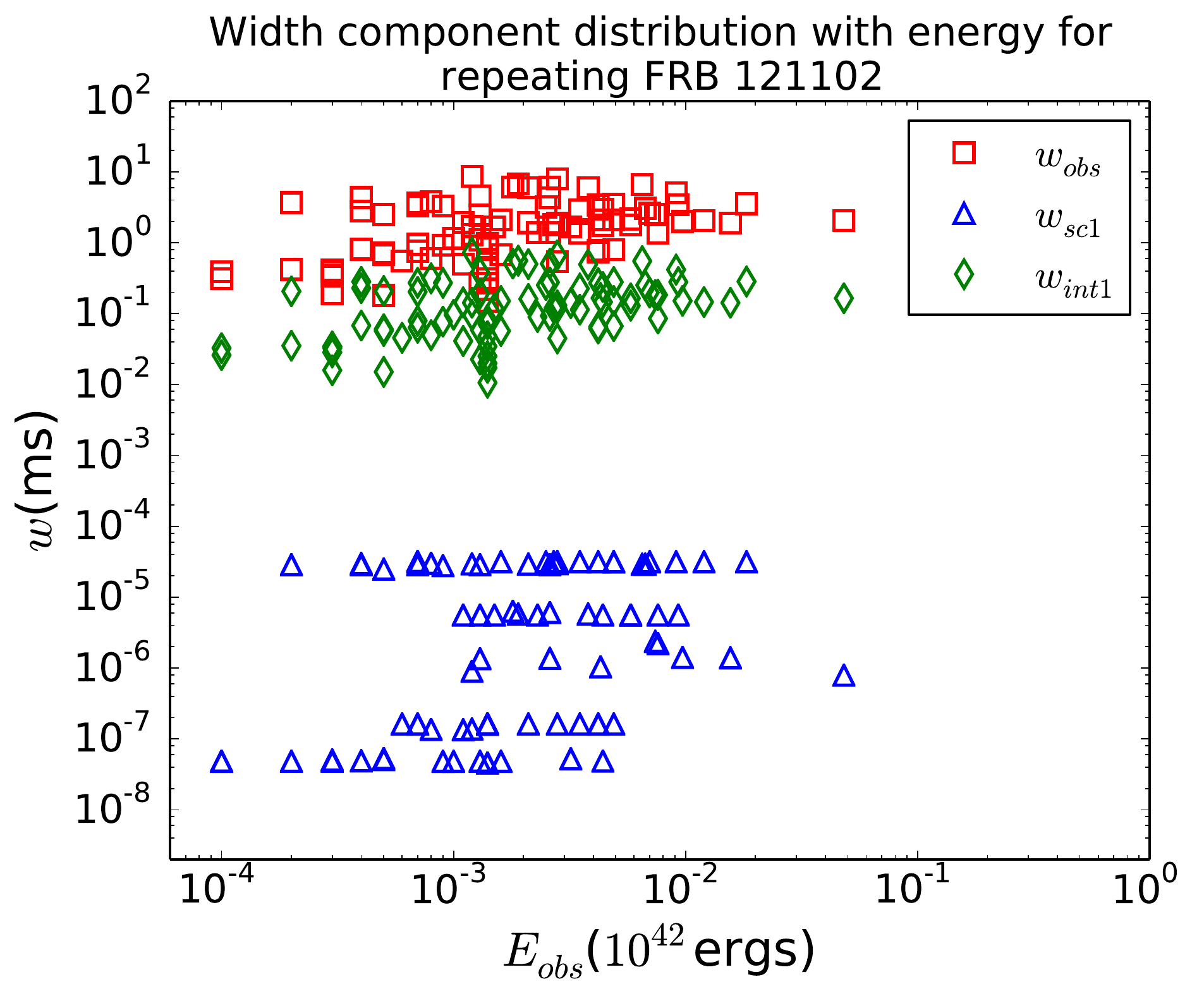}{0.46\textwidth}{}
          \fig{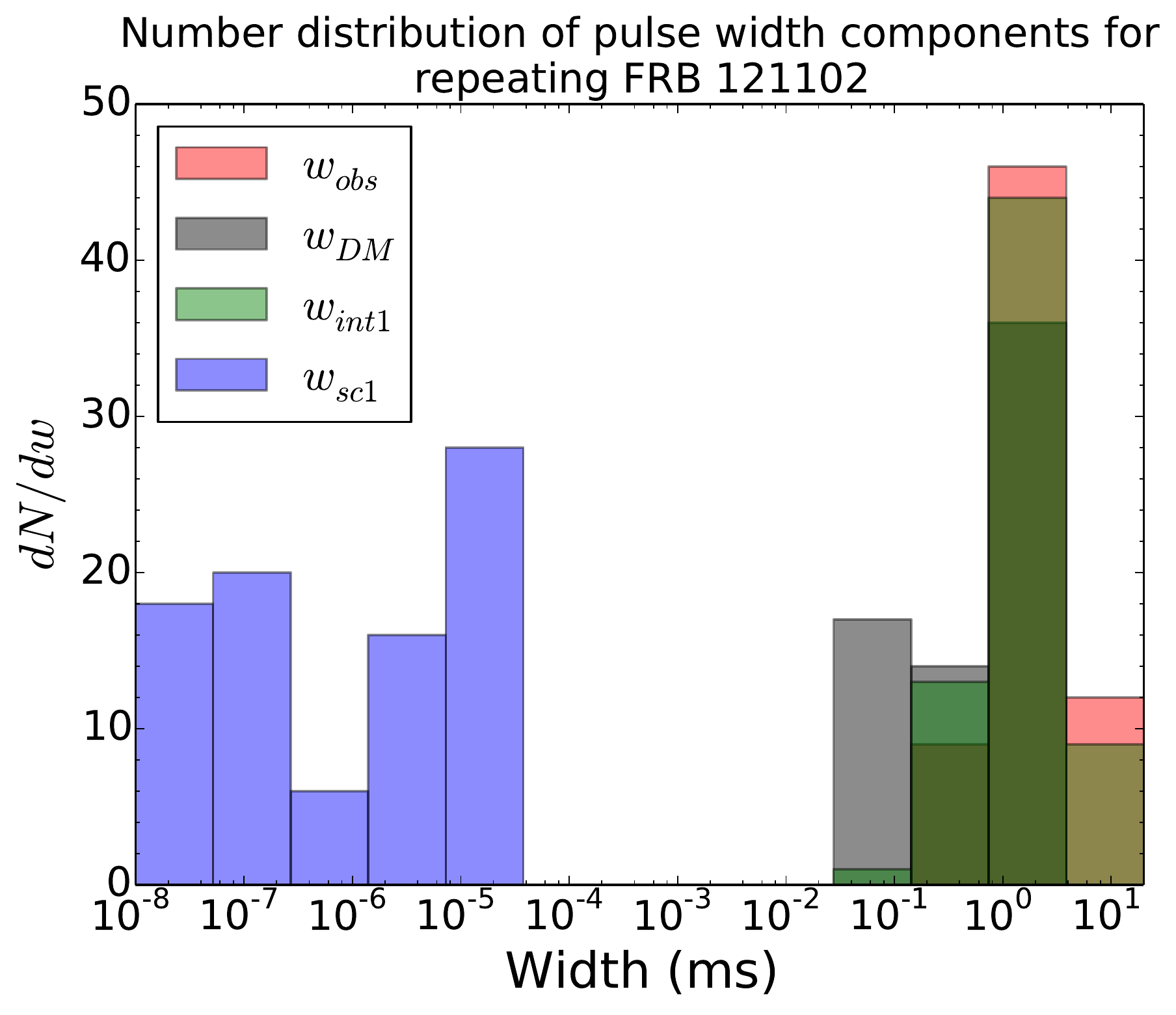}{0.46\textwidth}{}
          }  \vspace{-2.5em}        
  \caption{\emph{Distribution of pulse width components for Parkes FRBs and FRB 121102:} 	
	{\it Top-left panel:} Variation of $w_{\rm obs}$, $w_{\rm sc1}$ and $w_{\rm int1}$ with distance for Parkes FRBs,	
	{\it Top-right panel:} Number distribution of $w_{\rm obs}$, $w_{\rm DM}$, $w_{\rm int1}$ and $w_{\rm sc1}$ for Parkes FRBs, 
	{\it Bottom-left panel:} Variation of $w_{\rm obs}$, $w_{\rm sc1}$ and $w_{\rm int1}$ with burst energy for FRB 121102 sub-bursts,
	{\it Bottom-right panel:} Number distribution of $w_{\rm obs}$, $w_{\rm DM}$, $w_{\rm int1}$ and $w_{\rm sc1}$ for FRB 121102 sub-bursts.	
}%
	\label{wcompNRR} 
\end{figure*}

%\vspace{-0.2cm}
\begin{enumerate}[leftmargin=*]
{\setlength \itemindent{0pt} \item \emph{Scattering model 1:} We assume that the temporal broadening $w_{\rm ISM,host/MW}$ due to scattering in the host galaxy/MW ISM is related to $DM_{\rm host/MW}$ as given by the empirical fit obtained by \citet{Krishnakumar15},
\begin{eqnarray}
\begin{aligned}
\label{wISMhostMW}
&w_{\rm ISM,host/MW} = (4.1\times10^{-8}\ {\rm ms})\ 4f(1-f) \\
&\times(1.00+1.94\times10^{-3} DM_{\rm host/MW}^{2.0})\frac{DM_{\rm host/MW}^{2.2}}{\nu_{\rm 0,GHz}^{4.4}}
\end{aligned}
\end{eqnarray}
where $\nu_{\rm 0,GHz} = \nu_{0}/10^3$ is the central frequency in GHz and $4f(1-f)$ is the lever-arm factor by which $w_{\rm ISM,host/MW}$ is suppressed. We use $f$ = 25 kpc/$D_{L}$, where $D_{L}$ is the luminosity distance (in kpc) from the source to the observer and 25 kpc is the typical extent of a MW-like galaxy. Although most of the scattering material is present in the ISM of the host galaxy or the MW, their contribution to scatter broadening is expected to be significantly suppressed by a factor $4f(1-f) \sim 10^{-4}$ due to the asymmetric placement of the scattering screens relative to the source and the observer \citep{Williamson72,Vandenberg76,Lorimer13}. As the average electron density fluctuations in the IGM are expected to be less significant compared to the ISM, we assume that the scatter broadening due to IGM can be obtained by rescaling the ISM contribution by three orders of magnitude \citep{Lorimer13,Caleb16},
\begin{eqnarray}
%\begin{aligned}
\label{wIGM}
w_{\rm IGM} = (4.1\times10^{-11}\ {\rm ms})\ (1.00+1.94\times10^{-3} DM_{\rm IGM}^{2.0})\nonumber \\
\times\frac{DM_{\rm IGM}^{2.2}}{\nu_{\rm 0,GHz}^{4.4}}.
%\end{aligned}
\end{eqnarray}
As opposed to the host galaxy/MW ISM scattering, IGM scattering is unaffected by geometrical effects such as the lever-arm effect.} %\\
%\vspace{-0.2cm}

{\setlength\itemindent{0pt} \item \emph{Scattering model 2:}} In this model, we assume that the ISM scatter broadening contribution from the host galaxy and the MW are still given by the $w_{\rm ISM}-DM$ relation in equation (\ref{wISMhostMW}). However, instead of rescaling $w_{\rm ISM,host/MW}$ to obtain $w_{\rm IGM}$, we use the theoretical temporal smearing expression for IGM turbulence as obtained by \citet{MK13},
\begin{eqnarray}
w_{\rm IGM}(z) = \frac{k_{\rm IGM}}{\nu_{\rm 0,GHz}^{4}Z_{L}}\int^{z}_{0}\frac{dz^{\prime}}{\left[\Omega_{m}(1+z^{\prime})^3 + \Omega_{\Lambda}\right]^{0.5}} \nonumber \\
\times \int^{z}_{0}\frac{(1+z^{\prime})^{3}}{\left[\Omega_{m}(1+z^{\prime})^3 + \Omega_{\Lambda}\right]^{0.5}}dz^{\prime}
\label{wIGM2}
\end{eqnarray}
where $Z_{L} = 1 + (1/2)z$ and $k_{\rm IGM}$ is the normalisation factor (see Appendix \ref{AppendixA} for a detailed derivation of equation \ref{wIGM2}). We fix the value of $k_{\rm IGM}= 7.35\times10^{11}\ {\rm ms\ MHz^{4}}$ from equation (\ref{wint}) such that $w_{\rm int} \leq \sqrt{w_{\rm obs}^{2} - (w_{\rm DM}^2 + w_{\rm samp}^2 + w_{\rm IGM}^2)}/(1+z)$ is a real quantity for all the resolved FRBs in Table \ref{Table1}. 
\end{enumerate}

While $w_{\rm IGM}$ from model 1 is based on the assumption that the nature of IGM turbulence is similar to that of the Galactic ISM and can be estimated with an observationally established empirical fit, $w_{\rm IGM}$ from model 2 is based on a completely theoretical model for IGM turbulence which has not been observationally verified. Previous FRB population studies \citep{Bera16,Caleb16} have used $w_{\rm ISM}-DM$ relation for pulsars in the MW ISM from \citet{Bhat04} in order to estimate the IGM and ISM scatter broadening widths. However, it has already been shown by \citet{Hassall13} and \citet{Lorimer13} that the scatter broadening of known FRBs is significantly smaller compared to that estimated from the \citet{Bhat04} model. The pulse scattering width estimated from the \citet{Bhat04} model increases considerably beyond $z \sim 0.5$ (see Figure 1 of \citealt{Bera16}) and typically exceeds the observed pulse widths for the known FRBs at high redshifts (see Table \ref{Table1}). For $DM_{\rm host/MW} \lesssim 100\ {\rm pc\ cm^{-3}} \ll DM_{\rm IGM}$, $w_{\rm ISM,host/MW}$ obtained from equation (\ref{wISMhostMW}) is considerably smaller compared to $w_{\rm IGM}$ and other width components in equation (\ref{wint}). The values of all the observed burst and telescope parameters in equations (\ref{DMtot}-\ref{wIGM2}) are obtained from the \href{http://www.astronomy.swin.edu.au/pulsar/frbcat/}{FRB catalogue}. 

The modelled scatter broadening width for a given FRB source can be written as, $w_{\rm sc} = [w_{\rm ISM,MW}^2 + w_{\rm ISM,host}^2 (1+z)^2 + w_{\rm IGM}^2]^{0.5}$. 
For Parkes FRBs and scattering model 1, we show the variation of the width components $w_{\rm obs}$, $w_{\rm sc}$ and $w_{\rm int}$ with distance in the top-left panel of Figure \ref{wcompNRR}, while the number distribution of width components $w_{\rm obs}$, $w_{\rm DM}$, $w_{\rm int}$ and $w_{sc}$ for model 1 are shown in the top-right panel of Figure \ref{wcompNRR}. 
The ISM broadening contributions obtained from the scattering models are found to be very small with $w_{\rm ISM,MW} \lesssim 10^{-3}\ {\rm ms}$ and $w_{\rm ISM,host} \lesssim 10^{-6}\ {\rm ms}$, which is expected as the ISM contribution is suppressed relative to the IGM contribution by the geometrical factor $4f(1-f) \sim 10^{-4}$. While the width broadening due to IGM turbulence is larger at least by an order of magnitude for scattering model 2 at smaller redshifts $z \lesssim 1$, the IGM contributions for both scattering models are roughly equal for $z \gtrsim 2$ as $w_{\rm IGM}$ increases faster with distance for model 1. The dispersive smearing $w_{\rm DM}$ is approximately of the same order of magnitude as $w_{\rm obs}$ and $w_{\rm int}$ for a given FRB, with $10^{-1}\ {\rm ms} \lesssim w_{\rm DM} \lesssim 10^{1}\ {\rm ms}$ for all bursts. The smallest contribution to the pulse width broadening is from scattering with $w_{\rm sc} \lesssim 1\ {\rm ms}$ for almost all bursts. As $w_{\rm obs} \sim w_{\rm DM} \gg w_{\rm sc}$ for most FRBs, $w_{\rm DM}$ is the dominant contribution to the temporal broadening. Even though the intrinsic pulse width varies considerably with $10^{-1}\ {\rm ms} \lesssim w_{\rm int} \lesssim 10\ {\rm ms}$, the two scattering models are essentially indistinguishable due to the small scatter broadening contributions with $w_{\rm int,1} \approx w_{\rm int,2}$ and therefore we only show the results for scattering model 1 in Figure \ref{wcompNRR}.

The bottom-left panel of Figure \ref{wcompNRR} shows the variation of width components with $E_{\rm obs}$, while the bottom-right panel shows the number distribution of width components for the sub-bursts of FRB 121102. 
The observed parameters for the FRB 121102 sub-bursts that we consider in our analysis here are listed in Table \ref{Table1r}. For each dataset, we mention the telescope/frequency, number of detected pulses along with the range in $S_{peak,obs}$ and $w_{obs}$. 
It should be noted that many intrinsically shorter FRB 121102 sub-bursts have been detected due to high time-resolution systems such as the Breakthrough Listen observations at 4-8 GHz with GBT \citep{Gajjar18}.
Similar to the non-repeating FRBs, the sub-bursts of FRB 121102 were also detected by instruments with time-resolution varying over a wide range: $\sim 0.1\ {\rm ms}$ for Arecibo and Effelsberg to $\sim 1.0\ {\rm ms}$ for GBT. This affects the distributions of the true properties for FRB 121102 sub-bursts such as $w_{\rm int}$ and $L_{\rm int}$. We compute $w_{\rm int}$ for sub-bursts from a given survey by including the associated time resolution $\sim w_{\rm samp}$ along with the uncertainty. However, we do not consider the effect of different sensitivity thresholds for different searches of FRB 121102 in our simplistic study here.

We find that the modelled ISM broadening contributions from both host galaxy and MW are very small with $w_{\rm ISM,host/MW} \lesssim 10^{-4}\ {\rm ms}$ for all sub-bursts. As the redshift $z \approx 0.19273$ is relatively small for FRB 121102, the width broadening due to IGM turbulence is much more significant for scattering model 2 relative to model 1. The dispersive smearing is found to be smaller compared to most Parkes FRBs with $10^{-1}\ {\rm ms} \lesssim w_{\rm DM} \lesssim 1\ {\rm ms}$. While $w_{\rm obs}$ and $w_{\rm int}$ are approximately of the same order of magnitude, $w_{\rm DM}$ is about one order of magnitude smaller. 
As the dispersive smearing contribution directly relates to the instrumental detection parameters such as the channel bandwidth $\Delta \nu$ and the central frequency $\nu_0$, we evaluate $w_{\rm DM}$ separately for bursts that are detected with different telescopes in our FRB sample. 
It should also be noted that $DM_{\rm tot} \geq 500\ {\rm pc\ cm^{-3}}$ for Parkes FRBs considered here is expected to result in larger dispersive smearing on average as compared to the FRB 121102 sub-bursts.

We find that scatter broadening $w_{\rm sc} \lesssim 2\times10^{-2}\ {\rm ms}$ is the smallest contribution to the width broadening. Even though $w_{\rm DM}$ is the dominant contribution to the pulse broadening with $w_{\rm obs} > w_{\rm DM} \gg w_{\rm sc}$, $w_{\rm DM}$ for FRB 121102 sub-bursts are considerably smaller compared to that for the Parkes FRBs due to the relatively small $DM_{\rm tot}$ for FRB 121102. The intrinsic width varies considerably within $1\ {\rm ms} \lesssim w_{\rm int} \lesssim 10\ {\rm ms}$ with $w_{\rm int} \approx w_{\rm obs}$, implying that a considerable fraction of $w_{\rm obs}$ for FRB 121102 is from $w_{\rm int}$ and not due to the dispersive smearing or scatter broadening of the pulse. Although $w_{\rm sc1} \ll w_{\rm sc2}$, the scattering models are still indistinguishable with $w_{\rm int,1} \approx w_{\rm int,2}$ due to the minimal IGM and ISM scatter broadening contributions. 

For both Parkes bursts and FRB 121102, we find that most of the pulse temporal broadening is due to dispersive smearing and not IGM or ISM scattering. The contribution from $w_{\rm DM}$ to the width broadening is found to be considerably larger for the Parkes FRBs in comparison to the FRB 121102 sub-bursts, which is expected due to the larger $DM_{\rm tot}$ values for the Parkes FRBs. 
While the variation in $w_{\rm DM}$ for a given $z$ is only due to a combination of different observation frequencies $\Delta \nu$ and $\nu_0$ across various telescopes for the repeating FRB 121102, 
%(bottom-left panel of Figure \ref{wcompNRR}), 
different galactic contributions $DM_{\rm host/MW}$ also play a major role in determining the magnitude of $w_{\rm DM}$ for the Parkes FRBs. %(top-left panel of Figure \ref{wcompNRR}).

The IGM pulse broadening obtained from scattering models is the dominant contribution to $w_{\rm sc}$ for both classes of FRBs while the $w_{\rm ISM}$ contributions are significantly smaller due to the geometrical lever-arm effect. The intrinsic width for both FRB classes is found to be largely scattering model-independent. Moreover, there is a considerable spread in the $w_{\rm int}$ values within a range of $\sim 37\ {\rm ms}$/$\sim 8\ {\rm ms}$ for the Parkes/FRB 121102 bursts with most FRBs having $w_{\rm int} \lesssim 5\ {\rm ms}$/$w_{\rm int} \lesssim 3\ {\rm ms}$. We estimate the average relative broadening of the intrinsic width, $\Delta w_{\rm int}/w_{\rm int} = (w_{\rm obs} - w_{\rm int})/w_{\rm int}$, to be $\sim 150\%$/$\sim 20\%$ for the Parkes/FRB 121102 bursts. 
It should be noted that larger $w_{\rm obs}$ corresponds to a lower instrument sensitivity, thereby resulting in a observing bias against bursts that are smeared over longer duration and/or have larger intrinsic width.

\begin{figure*}
%\vspace{-10em}
\gridline{\fig{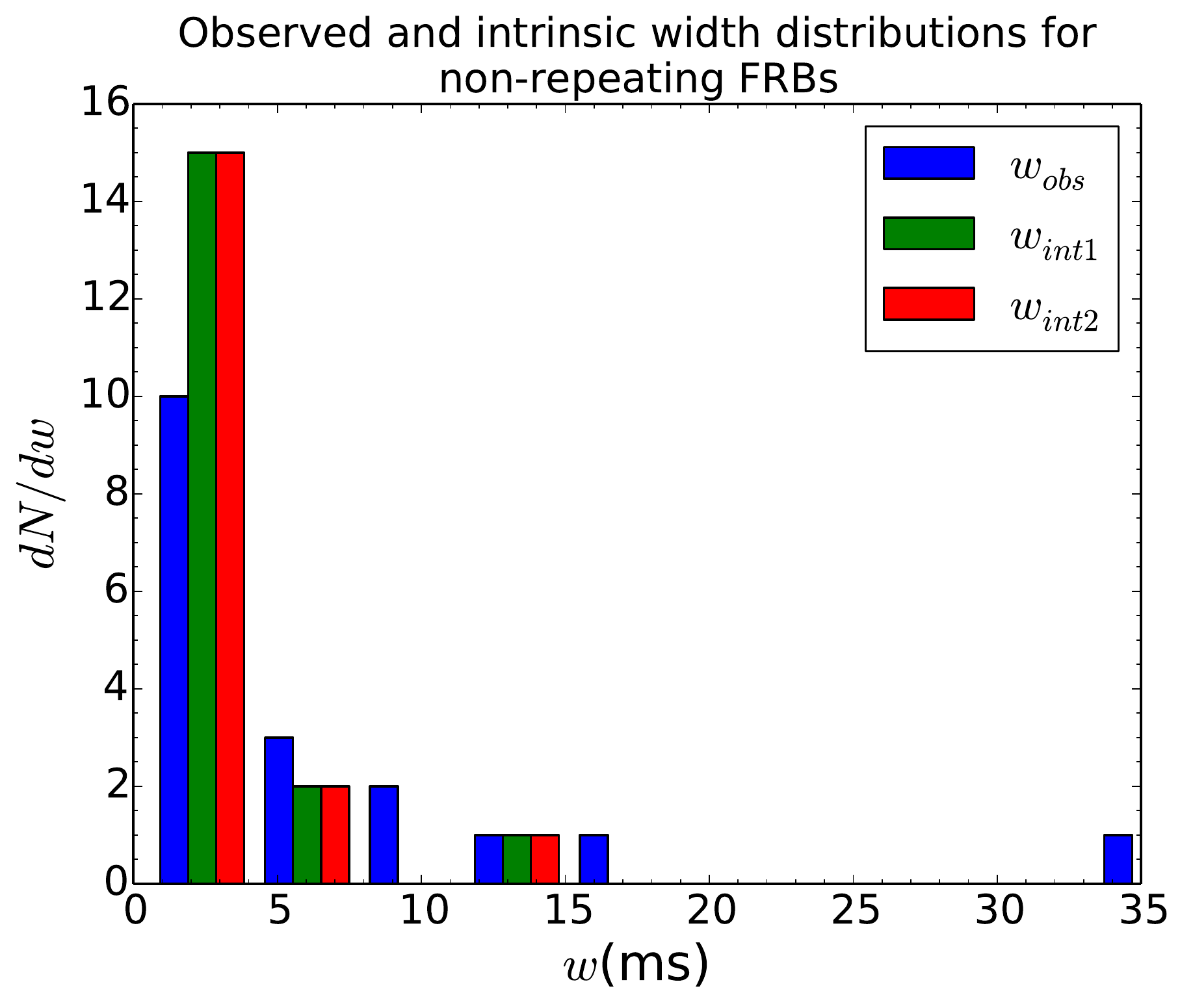}{0.46\textwidth}{}
          \fig{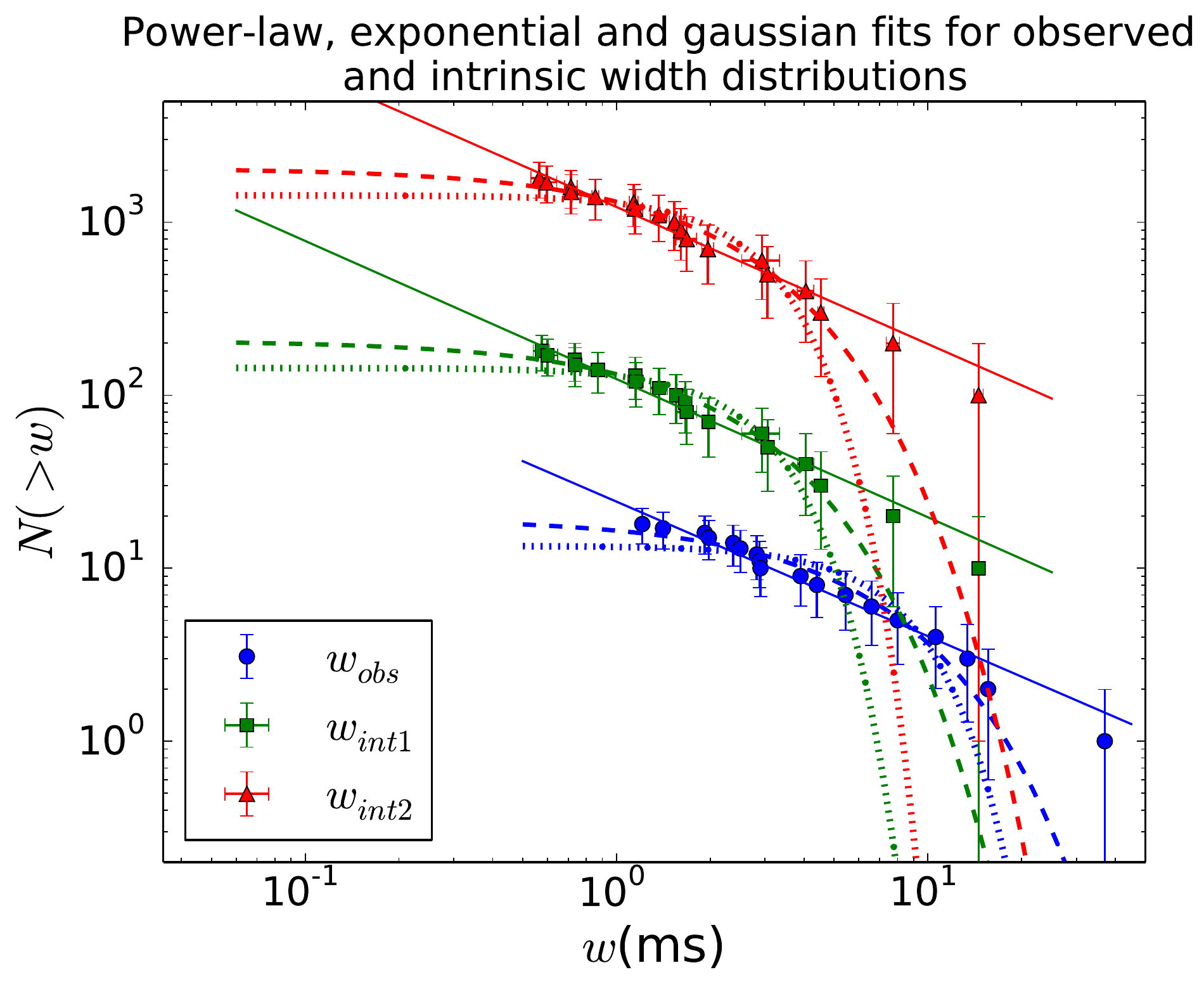}{0.46\textwidth}{}
          }  \vspace{-2.5em}        
\gridline{\fig{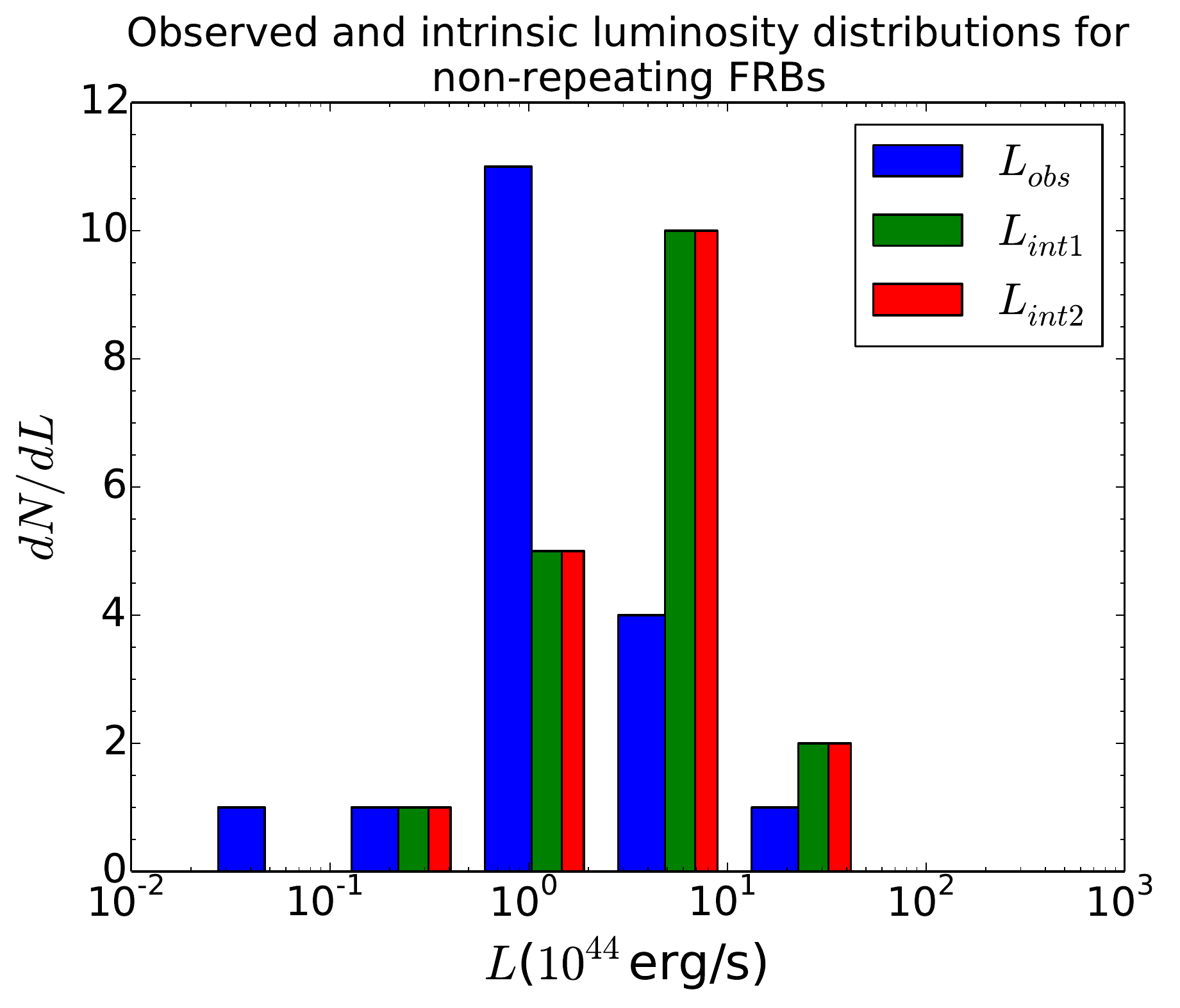}{0.46\textwidth}{}
          \fig{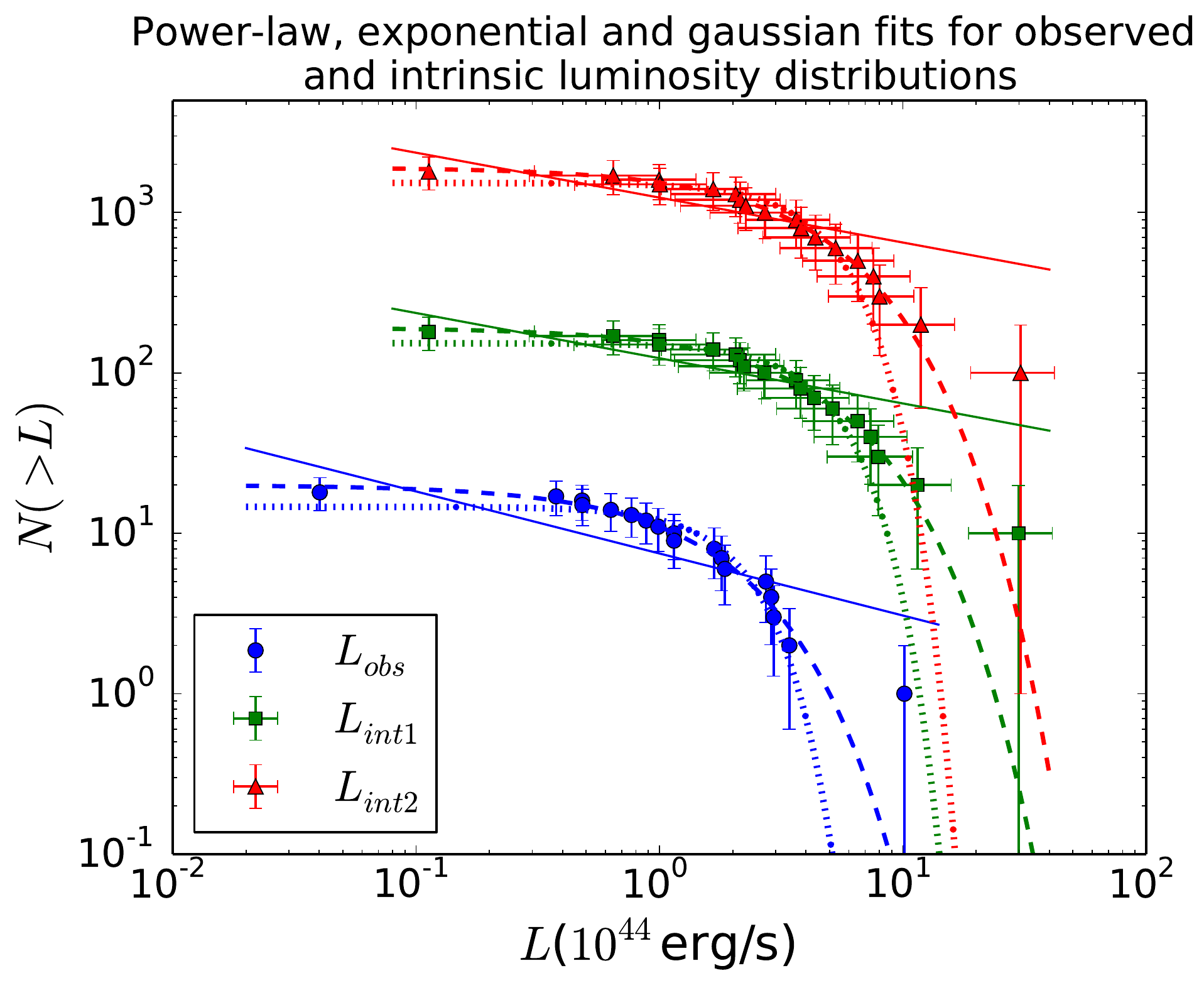}{0.46\textwidth}{}
          }  \vspace{-2.5em}        
  \caption{\emph{Pulse width and luminosity distributions for Parkes bursts:} 
	{\it Top-left panel:} Histograms for widths $w_{\rm obs}$, $w_{\rm int1}$ and $w_{\rm int2}$, 
	{\it Top-right panel:} Chi-squared fits for cumulative distributions of $w_{\rm obs}$, $w_{\rm int1}$ and $w_{\rm int2}$,
	%{\it Center-left panel:} Histograms of peak flux densities $S_{peak,obs}$, $S_{peak,int1}$ and $S_{peak,int2}$,
	%{\it Center-right panel:} Chi-squared fits for cumulative distributions of $S_{peak,obs}$, $S_{peak,int1}$ and $S_{peak,int2}$,	
	{\it Bottom-left panel:} Histograms of luminosities $L_{\rm obs}$, $L_{\rm int1}$ and $L_{\rm int2}$,
	{\it Bottom-right panel:} Chi-squared fits for cumulative distributions of $L_{\rm obs}$, $L_{\rm int1}$ and $L_{\rm int2}$.
	The index 1/2 for the burst parameters denotes scattering model 1/2. 
	The power-law, exponential and gaussian cumulative distributions are shown using solid, dashed and dotted lines, respectively (see Table \ref{Table2}). 
	%The functional forms used for the chi-squared fits of the cumulative distributions are power-law, exponential and gaussian (see Table \ref{Table2}).		
}%
	\label{distwSpeakLnr} 
\end{figure*}

\begin{figure*}
%\vspace{-10em}
\gridline{\fig{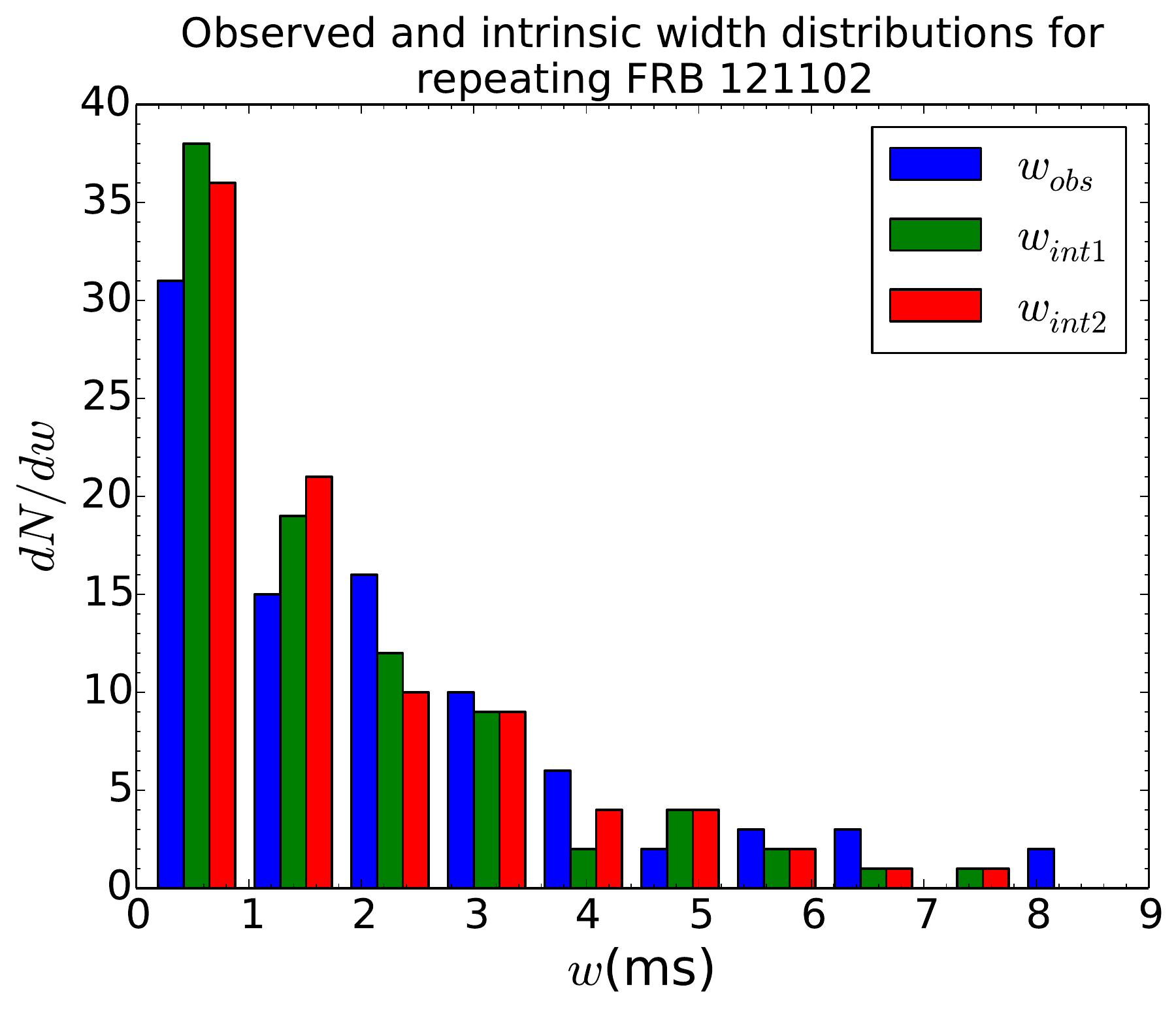}{0.46\textwidth}{}
          \fig{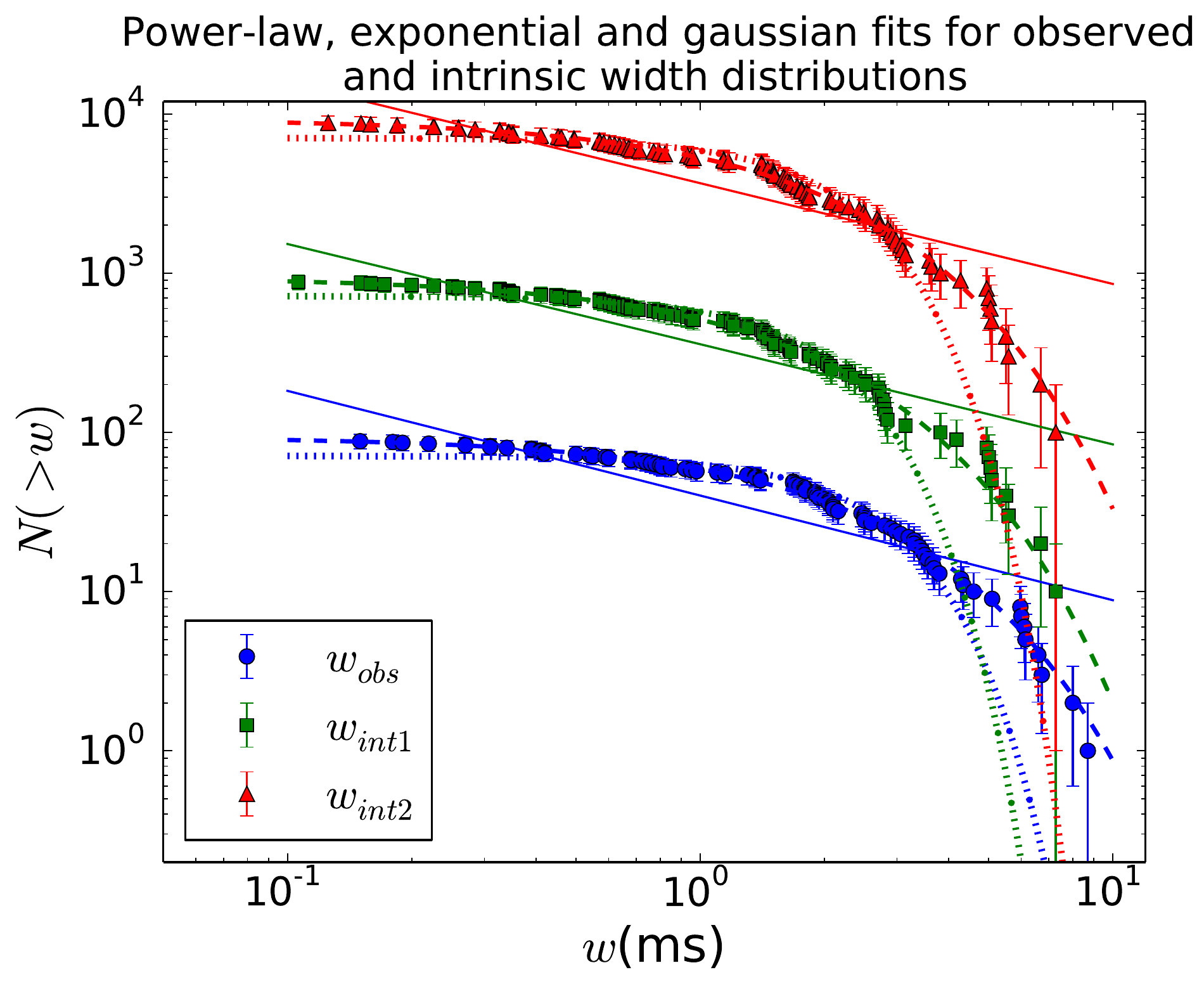}{0.46\textwidth}{}
          }  \vspace{-2.5em}        
\gridline{\fig{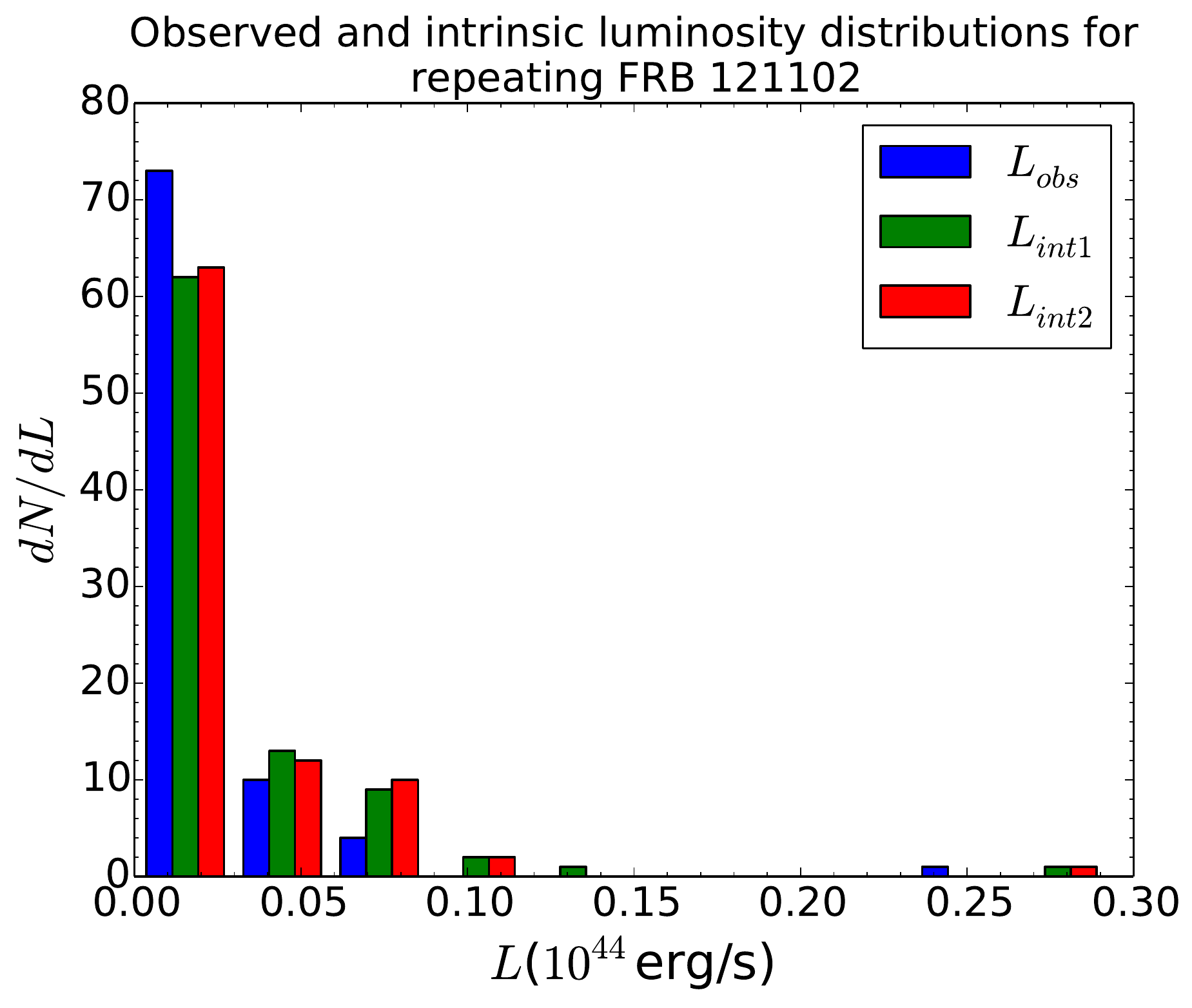}{0.46\textwidth}{}
          \fig{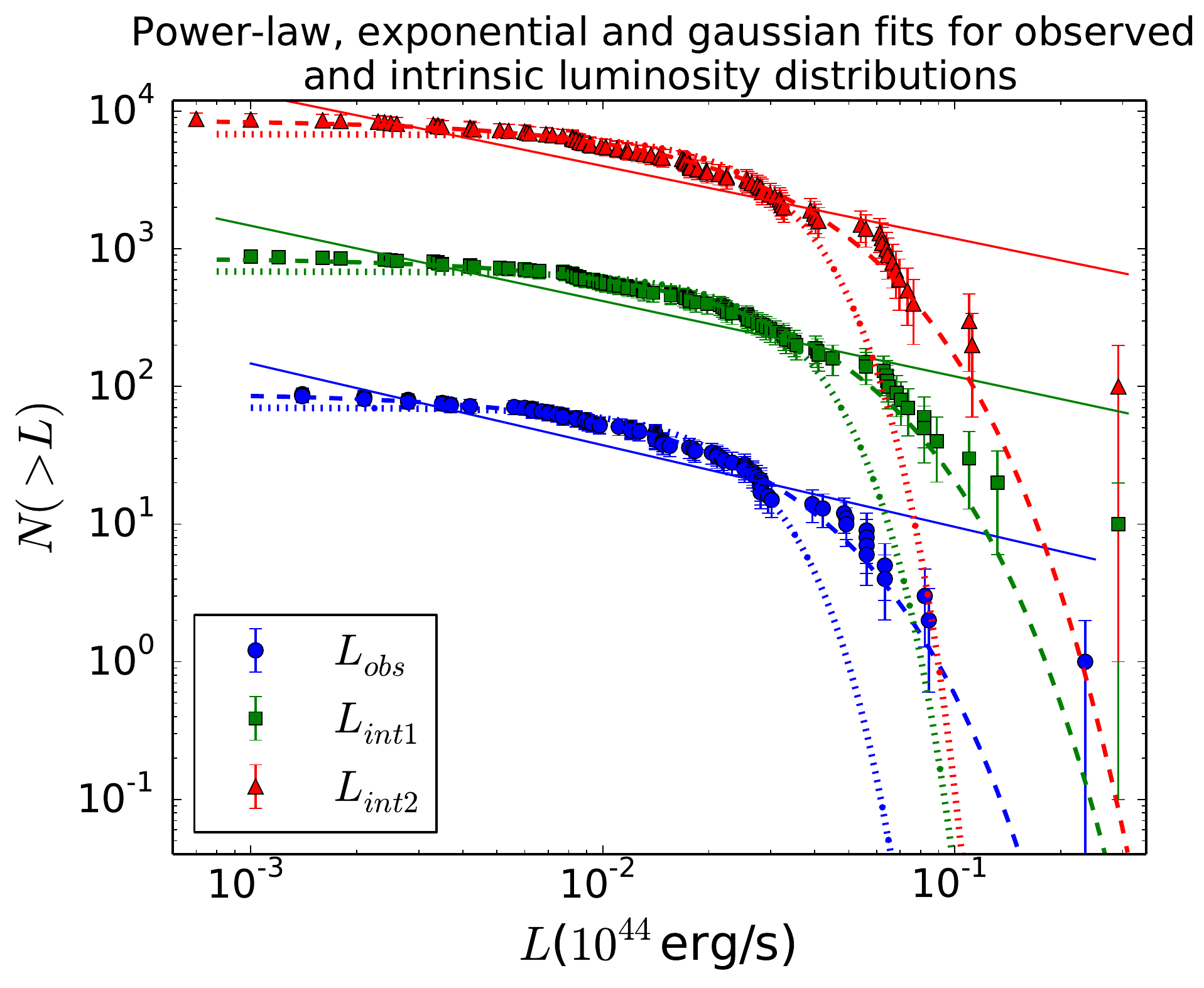}{0.46\textwidth}{}
          }  \vspace{-2.5em}        
  \caption{\emph{Pulse width and luminosity distributions for FRB 121102 bursts:} 
	{\it Top-left panel:} Histograms for widths $w_{\rm obs}$, $w_{\rm int1}$ and $w_{\rm int2}$, 
	{\it Top-right panel:} Chi-squared fits for cumulative distributions of $w_{\rm obs}$, $w_{\rm int1}$ and $w_{\rm int2}$,
	{\it Bottom-left panel:} Histograms of luminosities $L_{\rm obs}$, $L_{\rm int1}$ and $L_{\rm int2}$,
	{\it Bottom-right panel:} Chi-squared fits for cumulative distributions of $L_{\rm obs}$, $L_{\rm int1}$ and $L_{\rm int2}$.
	The index 1/2 for the burst parameters denotes scattering model 1/2. 
        The burst parameters for the repeating FRB 121102 are obtained from \citet{Spitler16}, \citet{Scholz16}, \citet{Scholz17}, \citet{Law17}, \citet{Hardy17}, \citet{Michilli18}, \citet{Gajjar18} and \citet{Spitler18}. 
        The power-law, exponential and gaussian cumulative distributions are shown using solid, dashed and dotted lines, respectively (see Table \ref{Table3}).
	%The functional forms used for the chi-squared fits of the cumulative distributions are power-law, exponential and gaussian (see Table \ref{Table3}).	
}%
	\label{distwSpeakLr} 
\end{figure*}

%\vspace{-0.1cm}
\subsection{Luminosity and energy estimates}
As the width of a radio pulse gets broadened by scattering in the turbulent plasma, the pulse is smeared across a longer time interval, thereby reducing its peak flux density. The fluence $\mathcal{F}_{\rm obs}$, proportional to the total emitted energy of the pulse, is assumed to be unaffected by the scatter broadening for each burst. Once $w_{\rm int}$ for a given FRB is obtained from equation (\ref{wint}), the corresponding intrinsic peak flux density can be estimated from the fluence as $S_{\rm peak,int} = \mathcal{F}_{\rm obs}/w_{\rm int}$. 

For a power-law energy distribution of the FRB source, the bolometric luminosity and energy for the burst are given by \citep{Lorimer13},
\begin{eqnarray}
L = \frac{4\pi D^{2}(z) (\nu_{\rm max}^{\prime \alpha+1} - \nu_{\rm min}^{\prime \alpha+1}) (\nu_2 - \nu_1)}{(1+z)^{\alpha-1} (\nu_{2}^{\alpha+1} - \nu_{1}^{\alpha+1})} S_{\rm peak}  \nonumber \\ %\left(\frac{S_{\rm peak} w_{\rm obs}}{w_{int}}\right)
E = \frac{4\pi D^{2}(z) (\nu_{\rm max}^{\prime \alpha+1} - \nu_{\rm min}^{\prime \alpha+1})(\nu_2 - \nu_1)}{(1+z)^{\alpha-1}(\nu_{2}^{\alpha+1} - \nu_{1}^{\alpha+1})} \mathcal{F}_{\rm obs} %\left(\frac{\mathcal{F}_{\rm obs}w_{\rm obs}}{w_{int}}\right)
\label{EdistplLE}
\end{eqnarray}
where $\nu_{\rm min/max}^{\prime}$ is the minimum/maximum source emission frequency in the FRB comoving frame, $\alpha$ is the spectral index and $\nu_{1/2}$ is the lowest/highest frequency in the observing band of the telescope. 
%{\bf Here we account for the reduction in the pulse luminosity and energy due to propagation effects such as scattering and dispersive smearing.}
In order to evaluate the intrinsic luminosity and energy distributions for the observed bursts, we assume a flat energy spectrum ($\alpha \approx 0$) to obtain: $L _{\rm int}= 4\pi S_{\rm peak,int} D^{2}(z) (\nu_{\rm max}^{\prime} - \nu_{\rm min}^{\prime})(1+z)$ and $E _{\rm obs}= 4\pi \mathcal{F}_{\rm obs} D^{2}(z) (\nu_{\rm max}^{\prime} - \nu_{\rm min}^{\prime})(1+z)$. Here we use $\nu_{\rm min}^{\prime} = 300\ {\rm MHz}$ and $\nu_{\rm max}^{\prime} = 8\ {\rm GHz}$ that are consistent with the current observed FRB population (\href{http://www.astronomy.swin.edu.au/pulsar/frbcat/}{FRB catalogue}). The assumption of a flat energy spectrum is reasonable as the FRB emission spectrum is poorly constrained at present with the spectral indices varying within a wide range. 

The left-half panels in Figure \ref{distwSpeakLnr} show the histograms for the distributions of pulse width 
%, peak flux density 
and luminosity of the 18 Parkes bursts from Table \ref{Table1}, while the right-half panels show three functional fits for the cumulative distributions of the corresponding quantities. We obtain chi-squared fits for the cumulative distributions of the Parkes burst parameters using three different functional forms: power-law, exponential and gaussian with zero mean, where the error for each data point is quantified with Poisson fluctuations (see Table \ref{Table2} in Appendix \ref{AppendixC} for the parameter details). We find that $w_{\rm int}$ for most bursts is a factor of $\sim 2-3$ smaller compared to $w_{\rm obs}$ and is within a relatively broad range of $\sim 0.3-10\ {\rm ms}$. While most of the bursts have $w_{\rm int} \lesssim 5\ {\rm ms}$, there is considerable spread in the width values suggesting that they are not peaked around $w_{\rm int} \approx 1\ {\rm ms}$ as assumed for previous MC simulations \citep{Bera16,Caleb16}. 

We find that the cumulative distribution of $w_{\rm int}$ for Parkes bursts is best fitted with an exponential distribution, with a cutoff around $w_{\rm int} \sim 2\ {\rm ms}$ that is about a third of the $w_{\rm obs}$ exponential cutoff. 
%The intrinsic peak flux density $S_{peak,int}$ varies within a large range of $\sim 10^{-2}-10^{2}\ {\rm Jy}$ with a distribution that is roughly symmetric around the peak $\sim 1\ {\rm Jy}$. Similar to $S_{peak,int}$, 
$L_{\rm int}$ varies by almost three orders of magnitude from $\sim 10^{43}\ {\rm erg/s}$ to $\sim 10^{46}\ {\rm erg/s}$ with a peak luminosity around $\sim 5\times10^{44}\ {\rm erg/s}$. As the inferred $L_{\rm int}$ values of the observed Parkes bursts vary within a wide range, FRBs are significantly unlikely to be standard candles. We find that the cumulative distribution of $L_{\rm int}$ is best fitted with an exponential distribution for the Parkes bursts, with a $L_{\rm int}$ cutoff around $\sim 4\times10^{44}\ {\rm erg/s}$. Furthermore, there is no significant difference in the width and luminosity distribution fit parameters obtained by changing the scattering models. 

The left-half panels in Figure \ref{distwSpeakLr} show the histograms for the width and luminosity distributions of the repeating FRB 121102 sub-bursts obtained from \citet{Spitler16}, \citet{Scholz16}, \citet{Scholz17}, \citet{Law17}, \citet{Hardy17}, \citet{Michilli18}, \citet{Gajjar18} and \citet{Spitler18}. The right-half panels show the chi-squared fits (with the same functional forms as the Parkes bursts) for the cumulative width and luminosity distributions, with the parameter details listed in Table \ref{Table3} of Appendix \ref{AppendixC}. Even though the average $w_{\rm int}$ for repeating FRB 121102 is smaller compared to that for the Parkes bursts, it still varies by almost two orders of magnitude from $\sim 0.1-8\ {\rm ms}$ with most bursts having $w_{\rm int} \lesssim 3\ {\rm ms}$. The fractional pulse broadening $\Delta w_{\rm int}/w_{\rm int} = (w_{\rm obs} - w_{\rm int})/w_{\rm int}$ for FRB 121102 is also found to be smaller compared to the Parkes bursts, which is expected due to its smaller distance, and thereby lesser scatter and dispersion broadening. \citet{Hessels19} have shown that the intrinsic duration of bursts from repeating FRB 121102 are in the range $\sim$1-5 ms while finer burst temporal structures are found up to 0.1 ms.

To study the width and luminosity distributions for a general population of non-repeating and repeating FRBs, the fluence completeness threshold should ideally be considered at low energies as the observational data can be incomplete due to the telescope sensitivity. \citet{Gourdji19} have recently used Arecibo data for 41 low-energy bursts from FRB 121102 to show that the power-law index for the burst energy distribution can vary considerably for different thresholds. In particular, they obtain a steeper power-law energy distribution compared to the previously reported results, by omitting bursts that are below the energy threshold $E_{th} = 2\times10^{37}\ {\rm erg}$ from their analysis. \citet{Oos19} have shown using 30 FRB 121102 bursts that were detected with WSRT/Apertif that a single power-law function may not fit the complete range of burst energy distribution for this repeater.

Here we find that the cumulative $w_{\rm int}$ distribution is best fitted with an exponential distribution with a cutoff $w_{\rm int} \sim 1.6\ {\rm ms}$ that is slightly smaller than the corresponding cutoff for $w_{\rm obs} \sim 2.1\ {\rm ms}$. The luminosity varies in a considerably smaller range, $L_{\rm int} \sim 10^{41}-10^{43}\ {\rm erg/s}$, compared to the non-repeaters. We find that the cumulative $L_{\rm int}$ distribution is best fitted with an exponential distribution, with a cutoff $L_{\rm int} \sim 2.7\times10^{42}\ {\rm erg/s}$ that is slightly larger compared to the $L_{\rm obs}$ cutoff. Similar to the non-repeating bursts, the difference between the scattering models is almost negligible.

\subsection{Error estimates for inferred parameters}
We estimated the FRB distances from the observed $DM_{\rm tot}$ values by assigning $DM_{\rm host} \approx 100\ {\rm pc\ cm^{-3}}$ as the fixed DM contribution from the host galaxy. However, the actual value of $DM_{\rm host}$ can vary over a significantly broader range depending on the type of the host galaxy and the location of the FRB source within it, thereby affecting our estimate for the inferred $z$. We also used two scattering models for the pulse temporal broadening in the turbulent ISM and IGM in order to compute $w_{\rm int}$ from the observable $w_{\rm obs}$. The assumptions used in the scattering models considered can affect the intrinsic width obtained. Furthermore, in addition to the pulse temporal broadening from propagation across turbulent plasma, the peak flux density $S_{\rm peak,int}$ is also reduced due to the finite size of the telescope beams. The inferred burst luminosities and energies are then directly affected by these modified $S_{\rm peak,int}$ values as well as the assumptions about the FRB energy density spectrum which we assume to be a power-law distribution in this work. Below we provide some rough estimates for each of these uncertainties in the inferred parameters. %that were computed earlier.

\begin{enumerate}[leftmargin=*]
{\setlength\itemindent{0pt} \item \emph{Error in the $z$ estimate from $DM_{\rm host}$ contribution:}} We previously assumed that the host galaxy has a free electron density structure that is similar to the MW and it provides a typical contribution of $DM_{\rm host} \approx 100\ {\rm pc\ cm^{-3}}$ to $DM_{\rm tot}$. As the values for $DM_{\rm tot}$ and $DM_{\rm MW}$ are known for a given FRB line of sight, the estimate for the burst distance is primarily based on the assumption about the host galaxy DM contribution. In general, $DM_{\rm host}$ can have considerable spread $\Delta DM_{\rm host} \sim 100\ {\rm pc\ cm^{-3}}$ due to the unknown location of the FRB source and our viewing angle relative to the galaxy. With $DM_{\rm IGM} \approx 750z\ {\rm pc\ cm^{-3}}$ for the FRBs listed in Table 1, we have $DM_{\rm Ex} \approx DM_{\rm host}/(1+z) + 750z$ from equation (\ref{DMtot}). Substituting $\Delta DM_{\rm host} \sim DM_{\rm host} \approx 100\ {\rm pc\ cm^{-3}}$ gives the redshift error to be $\Delta z \approx 0.2/(1+z)$ assuming $\Delta z \sim z \sim 1$ for most FRBs. The value for $\Delta z$ gradually decreases with increasing distance which is expected as the relative contribution from the uncertain $DM_{\rm host}$ to $DM_{\rm tot}$ decreases. \SmallEntryGap

{\setlength\itemindent{0pt} \item \emph{Error in the estimate for $w_{\rm int}$:}} We find that the ISM broadening contributions from both scattering models are suppressed significantly due to the geometrical lever-arm factor of $\sim$$10^{-4}$ and are very small with $w_{\rm ISM,MW/host} \lesssim 10^{-3}\ {\rm ms}$ for all reported FRBs. Even though $w_{\rm IGM} \sim 10^{-2}-10^{-1}\ {\rm ms}$ is considerably larger, the modelled IGM temporal broadening contributions are still typically smaller by at least an order of magnitude compared to the dispersive smearing. Using equation (\ref{wint}), the intrinsic width can therefore be approximately written as $w_{\rm int}^{2} \approx (w_{\rm obs}^{2} - w_{\rm DM}^{2})/(1+z)^{2}$ for all reported bursts. As $w_{\rm obs}$ and $w_{\rm DM} \propto DM_{\rm tot}$ are directly determined from the observations, the relative error in the intrinsic width is primarily due to $\Delta z$ and is given as $\Delta w_{\rm int}/w_{\rm int} = 0.2/(1+z)^{2}$.%\\

For scattering model 1, the IGM/ISM pulse width broadening is based on the DM contribution with $w_{\rm IGM/ISM} \propto DM_{\rm IGM/ISM}^{4.2}$ for sufficiently large values of DM. Although density fluctuations in the ISM/IGM can increase $DM_{\rm ISM/IGM}$ by the factor of a few which can then increase $w_{\rm IGM/ISM}$ by almost two orders of magnitude, the corresponding change in $w_{\rm int}$ is found to be negligible due to the modelled scatter broadening widths being very small (see Figure \ref{wcompNRR}). In case of scattering model 2, the IGM pulse broadening is determined by both $z$ and $k_{\rm IGM}$. From equation (\ref{wIGM2}), $w_{\rm IGM}$ increases by almost an order of magnitude for the $z$ range of FRBs in Table \ref{Table1} with the error in $w_{\rm IGM}$ due to $\Delta z$ being smaller for larger $z$. The value of the normalization constant $k_{\rm IGM}$ is fixed using the width parameters of a single FRB and can vary by a factor of few in general. However, as $w_{\rm IGM} \lesssim 10^{-2}\ w_{\rm obs}$ for most FRBs that we consider in our study, the dependence of $w_{\rm int}$ on the specific value of $k_{\rm IGM}$ is relatively weak. \SmallEntryGap
 
{\setlength\itemindent{0pt} \item \emph{Effect of beam shape on $S_{\rm peak}$:}} While the intrinsic peak flux density $S_{\rm peak,int}$ is diminished due to the pulse broadening from multipath propagation, the actual observed flux can be even smaller due to the finite size of the telescope beam used for detection. For a typical Gaussian beam profile, the observed flux density can be written as $S_{\rm peak,obs} \approx S_{\rm peak,int}\ {\rm exp}(-r^{\prime 2}/r_{\rm beam}^{2})$, where $r_{\rm beam}$ is the beam radius and $r^{\prime}$ is the radial distance from the beam center. The flux density averaged over an entire beam area is then $\langle S_{\rm peak,obs} \rangle = (S_{\rm peak,int}/\pi r_{\rm beam}^{2}) \int_{0}^{r_{\rm beam}} {\rm exp}(-r^{\prime 2}/r_{\rm beam}^{2}) 2\pi r^{\prime} dr^{\prime} = S_{\rm peak}(1 - e^{-1})$. Therefore, the average relative uncertainty in the $S_{\rm peak,int}$ values from the telescope and propagation effects is $\Delta S_{\rm peak,int}/S_{\rm peak,int} = \sqrt{(1/e)^{2} + (\Delta w_{\rm int}/w_{\rm int})^{2}} = \sqrt{(1/e)^{2} + 0.04/(1+z)^{4}} \approx 1/e$. \SmallEntryGap

{\setlength\itemindent{0pt} \item \emph{Effect on the inferred luminosity and energy:}} The inferred bolometric luminosity and energy are obtained using equation (\ref {EdistplLE}) from $S_{\rm peak}$ and $\mathcal{F}_{\rm obs}$, respectively, for a given burst. The measured values for $\mathcal{F}_{\rm obs}$ and $S_{\rm peak}$ are affected by the finite telescope beam size and $\Delta w_{\rm int}$. For our calculations, we assume the simple case of a flat FRB energy spectrum with $\alpha \approx 0$ for coherent source emission from 300 MHz to 8 GHz and a telescope detection bandwidth $\nu_{\rm bw} = \nu_{2} - \nu_{1}$. However, the value of $\alpha$ is highly uncertain from the current observations. As $D(z) \propto z$, in terms of the inferred parameters we have $L \propto S_{\rm peak} f(\alpha) z^{2}$ and $E \propto \mathcal{F}_{\rm obs} f(\alpha) z^{2}$, where $f(\alpha) = (1+z)^{1-\alpha} (\nu_{max}^{\prime \alpha + 1} - \nu_{min}^{\prime \alpha + 1})/(\nu_{2}^{\alpha + 1} - \nu_{1}^{\alpha + 1})$. For Parkes $\nu_{\rm bw} = 0.34\ {\rm GHz}$ and a typical FRB $z \sim 1$, the difference in the values of $f(\alpha)$ for $\alpha = 0$ and Kolmogorov spectral index $\alpha = -1.4$ is found to be very small $\lesssim 1/15$ and hence $\Delta f(\alpha)/f(\alpha) \ll 1$. The relative uncertainty in the inferred luminosity and energy values are then obtained with $\Delta L/L = \sqrt{(\Delta S_{\rm peak}/S_{\rm peak})^{2} + 4(\Delta z/z)^{2}}$ and $\Delta E/E = \sqrt{(\Delta \mathcal{F}_{\rm obs}/\mathcal{F}_{\rm obs})^{2} + 4(\Delta z/z)^{2}} \approx 1/e$.

\end{enumerate}

We have included the relative uncertainties in both the derived parameters $z$ and $w_{\rm int}$ as well as the inferred parameters $L$ and $E$ for our analysis. We evaluate $\Delta w_{\rm int}(w_{\rm int},z)$ and $\Delta L(z, S_{\rm peak}, L)$ for each burst in the non-repeating FRBs/FRB 121102 sample, and plot them as error bars for the cumulative distributions that are shown in Figures \ref{distwSpeakLnr} and \ref{distwSpeakLr}. The chi-squared fits for the cumulative $w_{\rm int}$ and $L$ distributions (see Tables \ref{Table2} and \ref{Table3}) are obtained after assigning proportionate weights to these uncertainties. 
In addition to the uncertainties involved with the derived burst parameters, some additional bias can also result from restricting the Parkes FRB sample to a smaller sub-sample with $DM_{\rm tot} \geq 500\ {\rm pc\ cm^{-3}}$. However, the selection bias introduced from applying the $DM_{\rm tot}$ cutoff is not expected to be significant as the vast majority $\gtrsim 80\%$ of Parkes FRBs currently have $DM_{\rm tot}$ values exceeding  $500\ {\rm pc\ cm^{-3}}$. Furthermore, we expect the bias from the uncertainty in the $DM_{\rm host}$ contribution for $DM_{\rm tot} \leq 500\ {\rm pc\ cm^{-3}}$ bursts to significantly outweigh that from the sample selection.

\begin{table*}
\begin{center}
\caption{System parameters for the Parkes multibeam (MB) receiver are obtained from \citet{Stav96} and those for the Arecibo L-band feed array (ALFA) are obtained from \href{http://www.naic.edu/alfa/}{http://www.naic.edu/alfa/}. %\citet{Thorn13}, \citet{Spitler14}. 
}
\label{Table4}
\bgroup
\def\arraystretch{1.0}
\begin{tabular}{| c | c | c |}
\hline
\hline
\centering
Parameter & Parkes MB & Arecibo ALFA \\ \hline \hline
Digitization factor ($\delta$) & 1.07 & 1.16 \\ \hline
System temperature ($T_{\rm sys}$) & 30 & 30 \\ \hline
Central frequency in MHz ($\nu_0$) & 1352 & 1375 \\ \hline
Frequency bandwidth in MHz ($\nu_{\rm bw}$) & 338 & 323 \\ \hline
Channel bandwidth in MHz ($\Delta \nu_0$) & 0.390 & 0.336 \\ \hline
Sampling width in ms ($w_{\rm samp}$) & 0.0640 & 0.0655 \\ \hline
\hline
\end{tabular}
\egroup
\end{center}
\end{table*}
%\vspace{-0.4cm}

%\EntryGap
%\SmallEntryGap
\section{Monte Carlo simulations}
\label{sec3}
In the previous section, we presented a method in order to estimate the true properties of the FRBs such as $w_{\rm int}$, $S_{\rm peak,int}$, $L_{\rm int}$ and $E_{\rm obs}$, from the observables for both non-repeating FRBs and FRB 121102 sub-bursts. Here we describe our MC code with which we constrain the various properties of the FRB source, its host galaxy and the intervening turbulent plasma from the observed properties of the reported Parkes/Arecibo FRBs. We first discuss the initial parameters and distributions used in the MC code. Next, we briefly describe the algorithm of our MC code.

\subsection{Input parameters}
The input parameters used for our MC simulations are:
\begin{itemize}[leftmargin=*]
\item \emph{Burst type:} We categorise all FRBs into two different classes of bursts: non-repeating and repeating FRBs. We model the population of the non-repeating/repeating FRBs found at the Parkes MB/Arecibo ALFA (see Table \ref{Table4} for the system parameters of these surveys). Parkes MB/Arecibo ALFA has 13/7 beams with different beam center gains $G_{\rm beam}$ and beam radii $r_{\rm beam}$. For Parkes MB, $r_{\rm beam} = 7.0^{\prime}\ (7.05^{\prime})\ [7.25^{\prime}]$ and $G_{\rm beam} = 0.731\ (0.690)\ [0.581]\ {\rm K\ Jy^{-1}}$ for beam 1\ (2-7)\ [8-13], while Arecibo ALFA has $r_{\rm beam} = 3.35^{\prime}\ (3.35^{\prime})$ and $G_{\rm beam} = 10.4\ (8.2)\ {\rm K\ Jy^{-1}}$ for beam 1\ (2-7). %\\

\item \emph{Scattering model:} As the distributions of the true properties derived from the current observations are found to be practically similar for both scattering models (see Section 2.1), here we only consider model 2 to determine the scatter broadening of the intrinsic pulse width for each FRB due to propagation through the turbulent ISM and IGM. %\\
%, we consider either scattering model 1 or model 2 .\\

\item \emph{FRB intrinsic width and luminosity distributions:} We consider two separate model distributions for both $w_{\rm int}$ and $L_{\rm int}$ in our MC simulations. As the observed pulse widths for non-repeating bursts/FRB 121102 vary within a broad range of $\sim 0.6-37.0\ {\rm ms}$/$\sim 0.15-8.70\ {\rm ms}$ and are peaked around $\sim 3.0\ {\rm ms}$/$\sim 1.8\ {\rm ms}$, we use model distributions for $w_{\rm int}$ as: (a) W1: lognormal distribution with mean $\mu_1 = 0$ and standard deviation $\sigma_1 = 0.25$, and (b) W2: lognormal distribution with mean $\mu_2 = 0$ and standard deviation $\sigma_2 = 0.50$. For the known $w_{\rm obs}$ distribution, the W1/W2 intrinsic width distribution physically corresponds to larger/smaller pulse temporal broadening due to scattering in the intervening plasma. 

The inferred luminosities for the non-repeating FRBs/FRB 121102 vary within a wide range 
%of $\sim 10^{43}-10^{47}\ {\rm erg/s}$/$\sim 10^{41}-10^{43}\ {\rm erg/s}$ 
with fewer bursts detected at larger luminosities. Here we assume power-law (PL) intrinsic luminosities for both classes of FRBs with: (a) L1: PL distribution with index $\alpha_{L_{\rm int}} = -1.3$, and (b) L2: PL distribution with index $\alpha_{L_{\rm int}} = -1.8$. 
We use $L_{\rm min} = 10^{43}\ {\rm erg/s}$ and $L_{\rm max} = 10^{45}\ {\rm erg/s}$ for non-repeating FRBs whereas the corresponding quantities are $10^{41}\ {\rm erg/s}$ and $10^{43}\ {\rm erg/s}$ for repeating FRB 121102.  
Here the broader luminosity distribution L1 physically corresponds to larger pulse scatter broadening due to multipath propagation. %\\

\item \emph{FRB source spatial density $n(z)$:} In order to estimate the number of FRB sources in a given comoving volume, we consider three different spatial density distributions:

\begin{enumerate}[leftmargin=*]
\item \emph{Non-evolving (NE) population:} The number of FRB progenitors increases linearly with the comoving volume for a non-evolving population. From Table \ref{Table1}, we know that the maximum inferred redshift value for the reported FRBs is $z \approx 2.3$. We generate FRBs up to a maximum redshift $z_{\rm max} = 3.0$, corresponding to a maximum comoving volume $V_{\rm c,max} \approx 1286\ {\rm Gpc}^3$ for typical cosmological parameters. The comoving distances to the FRBs are obtained as $D_c = (3 \zeta_1 V_{\rm c,max}/4\pi)^{1/3}$, where $\zeta_1$ is a uniform random number between 0 and 1. The associated $z$ is then obtained by inverting $D(z) = (8.49\ {\rm Gpc}) \int_{0}^{z} [(1+z^{\prime})^3 + 2.7]^{-0.5} dz^{\prime}$. %\\

\item \emph{Tracking cosmic star formation history (SFH):} As the majority of the FRB progenitor models suggested (including both cataclysmic and non-cataclysmic scenarios) involve young neutron stars, the spatial distribution of FRBs is expected to track the cosmic SFH. Furthermore, few reported FRBs have inferred distances exceeding $z \approx 2.0$, in which case the FRB spatial density can be significantly different compared to a NE population. We consider the cosmic SFH functional fit suggested by \citet{MD14},
\begin{equation}
\psi(z) = (0.015\ {\rm M_{\odot} yr^{-1} Mpc^{-3}}) \frac{(1+z)^{2.7}}{1 + [(1+z)/2.9]^{5.6}}.
\label{SFReqn}
\end{equation} 
The FRB redshifts are generated by inverting $\zeta_2 = \int_0^z \psi(z^{\prime})dz^{\prime}/\int_0^3 \psi(z^{\prime})dz^{\prime}$, where $\zeta_2$ is a uniform random number between 0 and 1. We then obtain $z = 12.05\zeta_2 - 57.16\zeta_2^2 + 167.97\zeta_2^3 - 259.28\zeta_2^4 + 199.04\zeta_2^5 - 59.63\zeta_2^6$. %\\

\item \emph{Power-law distribution:} We also consider a broken power-law FRB spatial density given by, 
\begin{equation}
n(z) = n_0\left\{
\begin{array}{ll}
(1 + z)^{\alpha_l}, & z_{\rm min} \leq z < z_{\rm crit}\\
(1 + z)^{\alpha_u}, & z_{\rm crit} \leq z \leq z_{\rm max}\\
\end{array}
\right. 
\label{nz_pl}
\end{equation}
where $n_0$ is a constant, $\alpha_l/\alpha_u$ is the lower/upper power-law index, $z_{\rm min}/z_{\rm max}$ is the minimum/maximum redshift and $z_{\rm crit}$ is the redshift at which $n(z)$ peaks. We try to constrain the distribution parameters $\alpha_l$, $\alpha_u$ and $z_{\rm crit}$ from the observed FRB population. 
We have chosen the power-law indices to be $\alpha_l \approx 2.7$ and $\alpha_u \approx -2.9$, with $z_{\rm crit} \approx 1.85$ as the peak of the $n(z)$ distribution. The PL indices are motivated from the cosmic SFH distribution (see equation \ref{SFReqn}), with $\psi(z \ll 1) \propto (1+z)^{2.7}$ at low $z$ and $\psi(z \gg 1) \propto (1+z)^{-2.9}$ at high $z$. We assume that the peak of the spatial density distribution is similar to that of the cosmic SFH with $z_{crit} \approx 1.85$. We also consider more general cases with varying $(\alpha_l, \alpha_u)$ or varying $z_{crit}$ later in our analysis (see Section \ref{PL_nz}). 
Based on the inferred $z$ obtained from $DM_{tot}$ in the previous section (see Table \ref{Table1}), here we consider FRB sources distributed within $z_{\rm min} = 0$ and $z_{\rm max} = 3.0$.
%From the cosmic SFH fit (equation \ref{SFReqn}), we have $\alpha_l \approx 2.7$ at low $z$, $\alpha_u \approx -2.9$ at high $z$ and $z_{\rm crit} \approx 1.85$ from the peak of the distribution. 
%We use $z_{\rm min} = 0$ and $z_{\rm max} = 3.0$ to generate the FRB redshifts from the power-law $n(z)$ distribution. %\\
\end{enumerate}

\item \emph{$\beta$-parameter for $DM_{\rm host}$:} As opposed to a constant $DM_{\rm host}$ contribution along all lines of sight, we assume that the free electron density distribution in the host galaxy is similar to that of the MW and can be obtained using the NE2001 model. We estimate the DM contribution due to the host galaxy ISM along the FRB source line of sight as, $DM_{\rm host} = \beta DM_{\rm NE2001}$, where $\beta$ is the parameter that accounts for the size of the FRB source host galaxy relative to the MW and $DM_{\rm NE2001}$ is the DM value predicted by the NE2001 model. In this work, we consider $\beta = 0.1$, 1.0 and 10.0 for the non-repeating FRBs. For the repeating FRB 121102, we use $DM_{\rm MW} = 188\ {\rm pc\ cm^{-3}}$ and $DM_{\rm host} = 281\ {\rm pc\ cm^{-3}}$ for all generated FRBs. 
We assume that all bursts are located at the position of the Solar system in the host galaxy and all lines of site are weighted equally, as the location of the FRB source within its host galaxy is highly uncertain at present. This also eliminates any possible bias from the specific choice of a volumetric function for the distribution of source locations within the host galaxy. As $DM_{\rm tot}$ is expected to be significantly larger compared to $DM_{\rm host}$ (constrained by the plasma frequency and further diminished by the cosmological expansion factor), the assumption about the FRB source location is not expected to affect the results. 

For our simple analysis here, we ignore the contribution to the DM from free electrons in the halos of any intervening galaxies and assume a diffuse homogeneous structure for both IGM and ISM. Recently, \citet{PZ19} have shown that the  contribution to $DM_{\rm Ex}$ from $DM_{\rm host}$ is essentially stochastic with the relative scatter in $DM_{\rm Ex}$ values being largest for nearby FRBs due to a larger relative uncertainty involved from $DM_{\rm host}$. For host galaxies that are similar to the MW, the free electrons in the galactic halo are expected to have a relatively small contribution with $DM_{\rm host} \lesssim 30\ {\rm pc\ cm^{-3}}$ \citep{Dolag15}. While $DM_{\rm IGM}$ can vary for FRBs at a similar $z$ depending on the density inhomogeneities along the line of sight \citep{McQuinn14}, the DM variability from cosmic web voids and strong filaments is found to be typically small $\lesssim 10\ {\rm pc\ cm^{-3}}$ \citep{Smith11,SD18,Ravi19}. %\\
 
\item \emph{FRB energy density spectral index $\alpha$:} Instead of using a flat FRB energy spectrum, we assume that the FRB energy spectrum can be modelled using a power-law, $E_{\nu^{\prime}} = k \nu^{\prime \alpha}$, where $\nu^{\prime}$ is the frequency in the source frame and $\alpha$ is the spectral index. The FRB bolometric luminosity and energy are then given by equation (\ref{EdistplLE}). Even though the coherent emission mechanism for FRBs suggests a negative spectral index, the spectral indices for some of the reported bursts vary within a wide range. 
\citet{Macquart2019} recently obtained the best-fit value $\alpha \approx -1.5$ from the spectra of 23 FRBs detected with ASKAP \citep{Bannister17,Shannon18}, which is similar to that of the Galactic pulsar population.
For completeness, here we consider $\alpha = -3.0$, -1.5 and 2.0.  

\end{itemize}

Recent works have shown that the distributions of observables can statistically constrain FRB source properties such as the luminosity function and the evolution history of the cosmic rate density \citep{Vedantham16,ME18,Niino18,MB19}. 
%It is expected that $DM_{Ex}$ anti-correlates with the burst fluence as expected for luminosity functions for cosmological FRB sources \citep{Shannon18}. 
\citet{ME18} showed that in order to obtain a cosmological distribution of FRBs that is dominated by the brighter events, the FRB luminosity function should have a power-law index that is flatter than $\alpha_{L_{\rm int}} = -1.5$. The intrinsic power-law luminosity distribution for non-repeating FRBs was shown to have $\alpha_{L_{\rm int}} = -0.6 \pm 0.3$ for ASKAP bursts \citep{LP19} and $-1.2 \leq \alpha_{L_{\rm int}} \leq -0.5$ for Parkes detections \citep{LK16}. These distributions agree well with the intrinsic luminosity models that we consider here. Although the shape of the luminosity function still has to be studied in more detail, recent works have shown that Schechter functions are preferred over power-law or normal distributions \citep{Luo18,Caleb16}.

\begin{figure*}
%\vspace{-10em}
\gridline{\fig{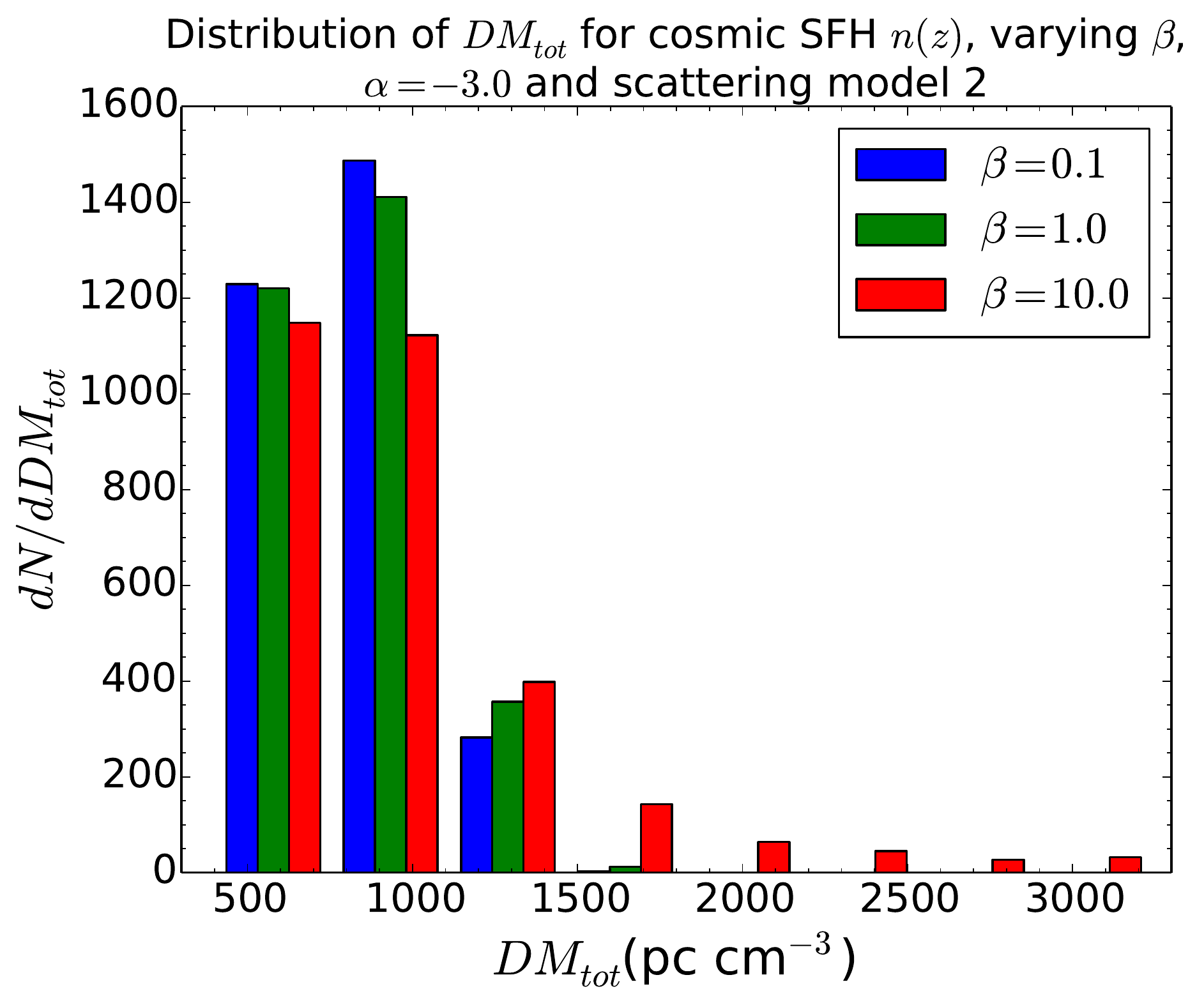}{0.46\textwidth}{}
          \fig{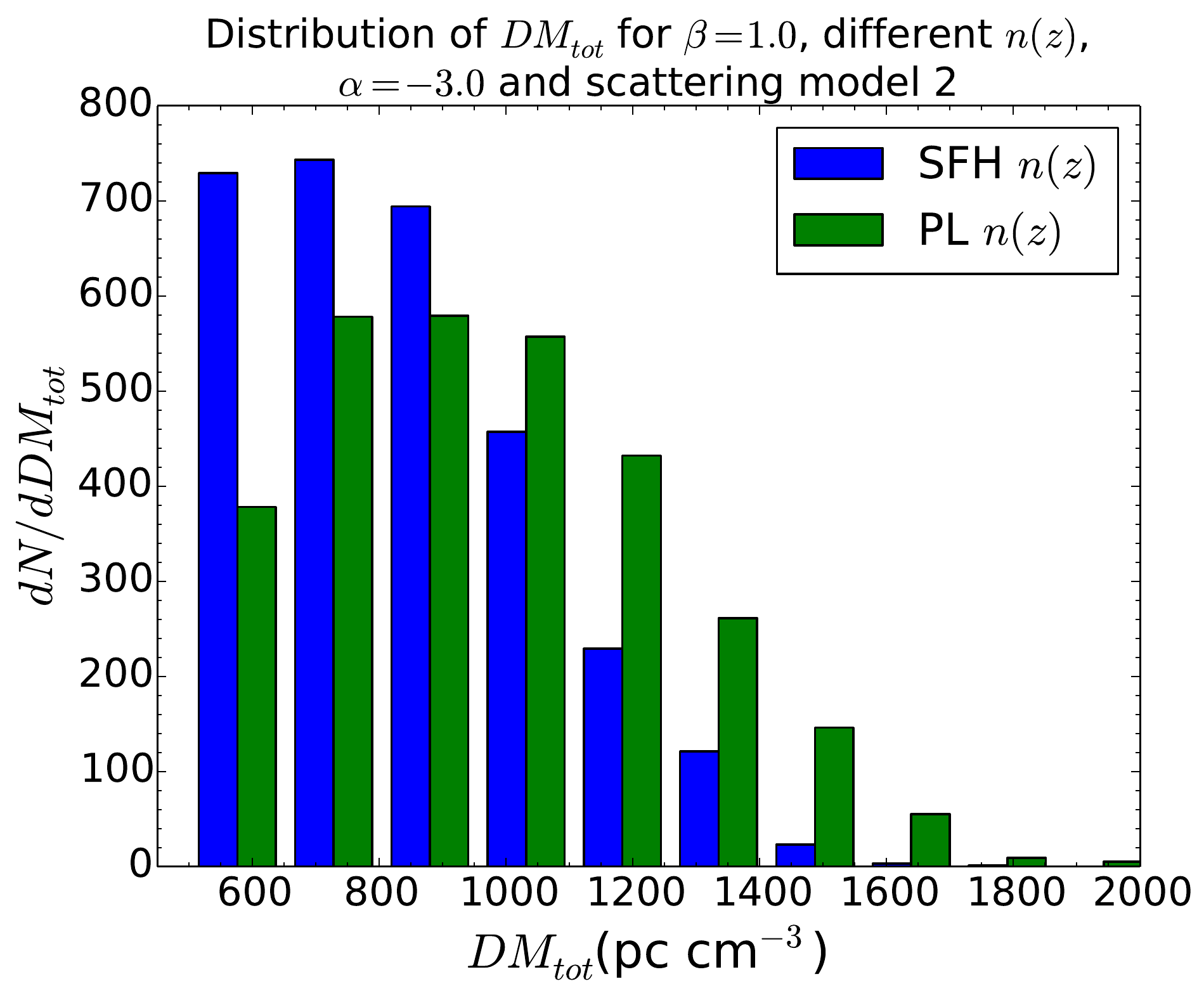}{0.46\textwidth}{}
          }  \vspace{-2.5em}        
\gridline{\fig{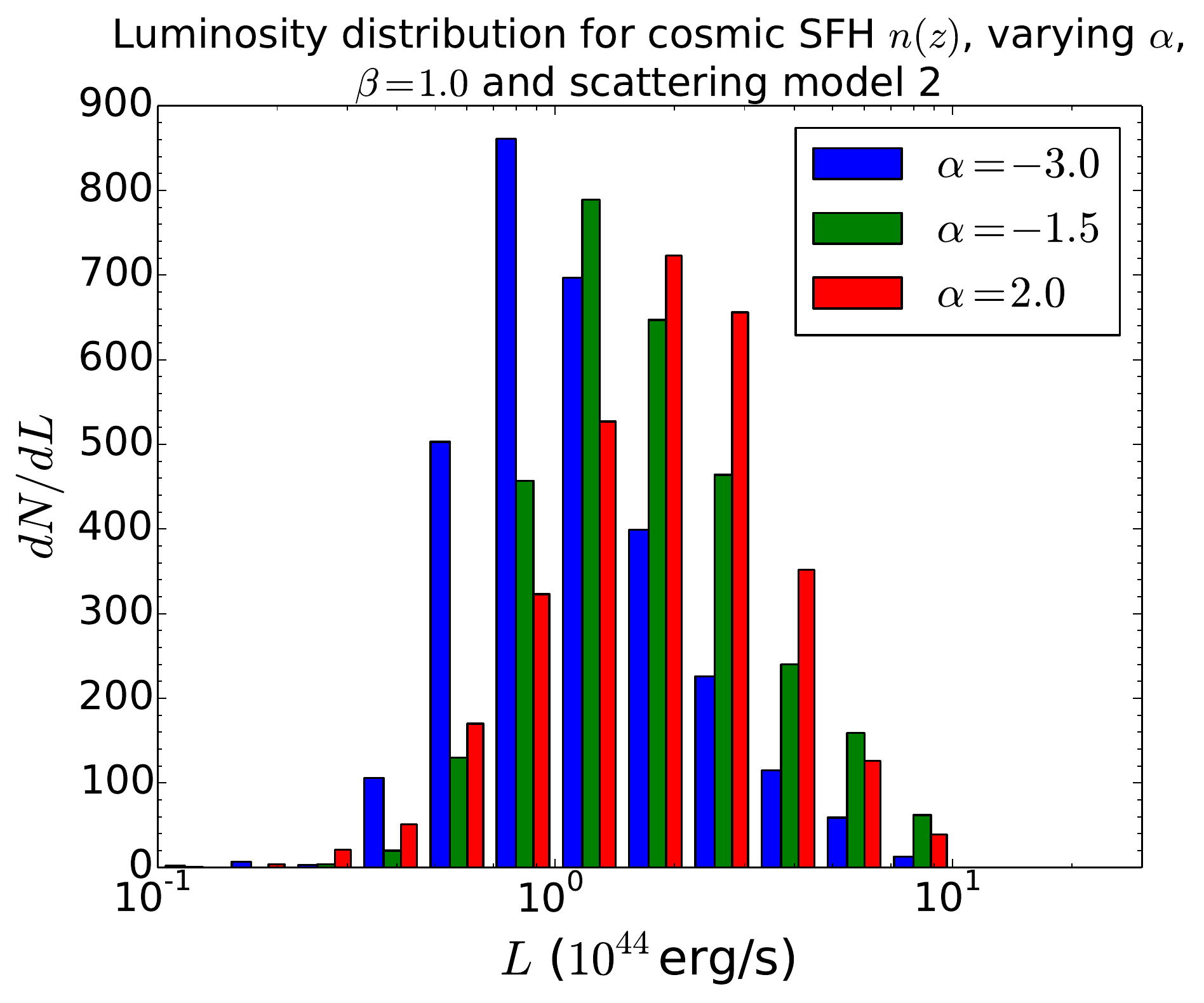}{0.46\textwidth}{}
          \fig{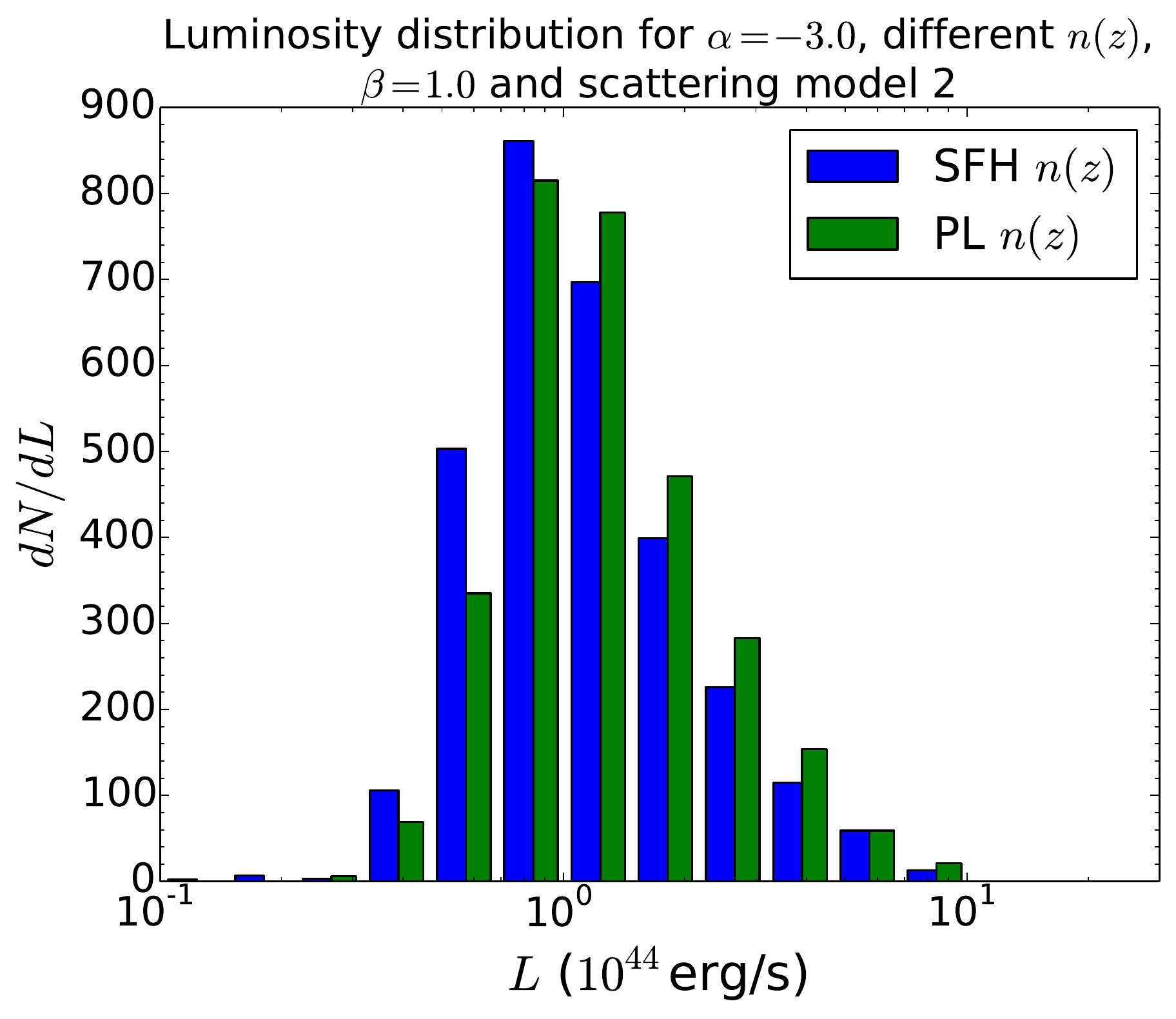}{0.46\textwidth}{}
          }  \vspace{-2.5em}        
  \caption{\emph{Dispersion measure and luminosity distributions for 3000 simulated non-repeating Parkes FRBs:} 	
	{\it Top-left panel:} $DM_{\rm tot}$ distribution for cosmic SFH spatial density and varying $\beta=0.1,\ 1.0,\ 10.0$,  
	{\it Top-right panel:} $DM_{\rm tot}$ distribution for $\beta=1.0$ and cosmic SFH/PL spatial density ($z_{\rm crit}=1.85$, $\alpha_l=3$, $\alpha_u=0$), 	
	{\it Bottom-left panel:} Luminosity distribution for cosmic SFH spatial density and varying $\alpha=-3.0,\ -1.5,\ 2.0$, 	
	{\it Bottom-right panel:} Luminosity distribution for $\alpha=-3.0$ and cosmic SFH/PL spatial density ($z_{\rm crit}=1.85$, $\alpha_l=3$, $\alpha_u=0$). All the simulations are done for scattering model 2, intrinsic width distribution w1 and intrinsic luminosity distribution L1.}%
	\label{sim_checks} 
\end{figure*}

\subsection{MC code algorithm}
At the start of the simulation, we select a specific burst type for the FRB events to be generated. 
% a specific scattering model and burst type for the FRB events to be generated. 
We then draw the intrinsic pulse width $w_{\rm int}$ and luminosity $L_{\rm int}$ of the first event from the corresponding model distributions (W1/W2 for width and L1/L2 for luminosity) and compute the burst energy $E_{\rm int} = w_{\rm int}L_{\rm int}$.
% best fit chi-squared distributions ({\bf Tables \ref{Table2} and \ref{Table3} in Appendix \ref{AppendixC}})
Next, we draw $D(z)$ for the burst and evaluate the corresponding $z$ for the chosen FRB spatial density model $n(z)$. 
Once the distance to the FRB is known, we further estimate $DM_{\rm IGM}(z)$ and $w_{\rm IGM}(z)$ from scattering model 2. 
We then draw a random line of sight in the MW and get $DM_{\rm MW}$ from the NE2001 model. For the host galaxy ISM contribution, we choose a random line of sight in the host galaxy to evaluate $DM_{\rm host} = \beta DM_{\rm NE2001}$ for the $\beta$ value selected. Once the $DM_{\rm MW}$ and $DM_{\rm host}$ contributions are known, we obtain the corresponding width broadening components $w_{\rm ISM,MW}$ and $w_{\rm ISM,host}$. We include the IGM contribution to the DM to estimate $DM_{\rm tot}$ and the pulse dispersive smearing $w_{\rm DM}$. The observed pulse width $w_{\rm obs}$ is obtained by adding $w_{\rm samp}$ to the previously estimated width components (see equation \ref{wint}). 

For a power-law FRB energy density with spectral index $\alpha$ chosen earlier, the peak flux density at the beam center $S_{\rm peak,bc}$ can be obtained from $L_{\rm int}$ for an observation in the frequency band between $\nu_1 = \nu_0 - (1/2)\nu_{\rm bw}$ and $\nu_2 = \nu_0 + (1/2)\nu_{\rm bw}$ as
\begin{equation}
S_{\rm peak,bc} = \frac{L_{\rm int} (1+z)^{\alpha - 1}}{4\pi D(z)^{2} (\nu_{\rm max}^{\prime \alpha+1} - \nu_{\rm min}^{\prime \alpha+1})} \frac{w_{\rm int}}{w_{\rm obs}} \left(\frac{\nu_2^{\alpha+1} - \nu_1^{\alpha+1}}{\nu_2 - \nu_1}\right)
\label{Speak_bc}
\end{equation}
where the factor $w_{\rm int}/w_{\rm obs}$ accounts for the reduction in $L_{\rm int}$ due to pulse scattering and/or dispersive smearing.
We assume FRB coherent emission to be in the frequency range between $\nu_{\rm min}^{\prime} = 300\ {\rm MHz}$ and $\nu_{\rm max}^{\prime} = 8\ {\rm GHz}$, in agreement with the current observations. However, the observed flux $S_{\rm peak,obs}$ can be significantly smaller compared to $S_{\rm peak,bc}$ due to the finite telescope beam size and for a Gaussian beam profile is given by
\begin{equation}
S_{\rm peak,obs} = S_{\rm peak,bc}\ {\rm exp}\left[-(2\ {\rm ln}2) \frac{r^{\prime 2}}{r_{\rm beam}^2}\right].
\label{Speak_obs}
\end{equation} 
While the probability of a particular beam detecting the FRB event is proportional to its area $\sim \pi r_{\rm beam}^2$, the radial distance $r^{\prime}$ from the beam center is chosen randomly. For pulses detected in single-pulse searches, the search trial width $w_{\rm trial}$ has to be closest to the observed pulse width $w_{\rm obs}$. In our simulation, we generate $w_{\rm trial}$ in powers of two starting from $w_{\rm samp}$ in order to find $w_{\rm trial}$ nearest to $w_{\rm obs}$. Once $w_{\rm trial}$ is determined, the signal-to-noise ratio $S/N$ for optimal detection is obtained from $S_{\rm peak,obs}$ and the other telescope parameters as
\begin{equation}
\frac{S}{N} = \frac{S_{\rm peak,obs}}{T_{\rm sys} \delta} G_{\rm beam} \sqrt{2 \nu_{\rm bw} w_{\rm trial}}.
\label{SNR}
\end{equation}
The FRB event is detected only if $S/N \geq 9\ (5)$ for Parkes MB (Arecibo ALFA). In addition, we discard all the events for which $DM_{\rm tot} < 500\ {\rm pc\ cm^{-3}}$ is obtained.
The simulation continues with the aforementioned algorithm until $N_{\rm det} = 3000$ events have been detected. From $S_{\rm peak,obs}$ of a given FRB, we further obtain the observed fluence $\mathcal{F}_{\rm obs} = S_{\rm peak,obs}w_{\rm obs}$, luminosity $L_{\rm obs} = 4\pi D(z)^2 (1+z) S_{\rm peak,obs} (\nu_{\rm max}^{\prime} - \nu_{\rm min}^{\prime})$ and energy $E_{\rm obs} = 4\pi D(z)^2 (1+z) \mathcal{F}_{\rm peak,obs} (\nu_{\rm max}^{\prime} - \nu_{\rm min}^{\prime})$. The observed ($S_{\rm peak,obs}$, $\mathcal{F}_{\rm obs}$, $DM_{\rm tot}$, $w_{\rm obs}$) and inferred ($z$, $L_{\rm obs}$, $E_{\rm obs}$) properties for every detected burst are stored for comparison with the observed FRB population in order to constrain the input parameters. %\\ %\\

%\SmallEntryGap
%\EntryGap
\vspace{0.15cm} 
\section{Simulation results and parametric constraints}
\label{sec4}
In the previous section, we described the MC code algorithm and discussed the input parameters for our code. Here we present the simulation results for the non-repeating FRBs and further compare them with observations in order to constrain the model parameters of the FRB population. 
%We consider NE, SFH and PL spatial densities for the non-repeating FRBs, while the parameters $n(z)$ and $\beta$ are kept fixed for the repeating population based on the singular detection FRB 121102.
%\subsection{Non-repeating FRBs}
We discuss the simulation results of the NE and SFH $n(z)$ distributions for non-repeating FRBs that are detectable with Parkes, and use Kolmogorov-Smirnov (KS) test to compare them with the current observations. 
For NE and SFH $n(z)$, we perform MC simulations with all four combinations of the model distributions for $w_{\rm int}$ and $L_{\rm int}$ - we hereby denote them as w$i$L$i$, where $i = 1,2$ corresponds to the specific distribution (see Section 3.1). 
We identify the parametric spaces ($\beta$,$\alpha$) where the p-value $\gamma$ obtained from the KS test is maximised for a given spatial density 
%choice of the scattering model and $n(z)$ 
distribution. 
Then, we specifically consider the cases where $\gamma$ is maximised to further constrain the parameters of the PL $n(z)$ distribution and identify ($\alpha_l$,$\alpha_u$,$z_{\rm crit}$) favored by the observations. 
The p-values obtained from the KS test comparison of the simulated and observed parameter distributions are listed in Table \ref{Table5} of Appendix \ref{AppendixC} for all the cases considered. We perform all KS tests under the null hypothesis that the two samples were drawn from the same distribution unless the p-value $\gamma < 0.05$.

We first perform some validity tests in order to check the robustness of our MC population synthesis algorithm. We show the results for the simulated $DM_{\rm tot}$ and $L$ distributions of 3000 Parkes FRBs considering different source properties. In the top-half panels of Figure \ref{sim_checks}, we compare the $DM_{\rm tot}$ distributions for varying host galaxy DM contributions $\beta=0.1,\ 1.0,\ 10.0$ as well as different spatial density models (SFH and PL). Here we consider a PL spatial density distribution with model parameters $z_{\rm crit}=1.85$, $\alpha_l=3.0$ and $\alpha_u=0$. As opposed to cosmic SFH, the PL spatial density does not fall off significantly beyond $z_{\rm crit}=1.85$ thereby resulting in more bursts with larger $DM_{\rm tot}$ values (see top-right panel of Figure \ref{sim_checks}). Similarly, a larger host galaxy DM contribution with $\beta = 10.0$ results in more FRBs with larger observed $DM_{\rm tot}$ as compared to $\beta \sim 0.1-1$, as shown in the left-half panel of Figure \ref{sim_checks}). We then compare the luminosity distributions for varying FRB energy density spectral index $\alpha=-3.0,\ -1.5,\ 2.0$ as well as SFH and PL spatial densities, as shown in the bottom-half panels of Figure \ref{sim_checks}. As expected for a fixed $\alpha=-3.0$, the PL $n(z)$ model leads to more FRBs with larger luminosities as compared to the cosmic SFH $n(z)$ due to more FRBs that are located at relatively larger distances (see the bottom-right panel of Figure \ref{sim_checks}). For SFH FRB spatial density model, we then obtain the luminosity distribution by considering different spectral indices $\alpha$. Here, as expected, a more negative $\alpha$ leads to more bursts with smaller energies and thereby smaller luminosities as shown in bottom-left panel of Figure \ref{sim_checks}.

\begin{figure*}
%\vspace{-10em}
\gridline{\fig{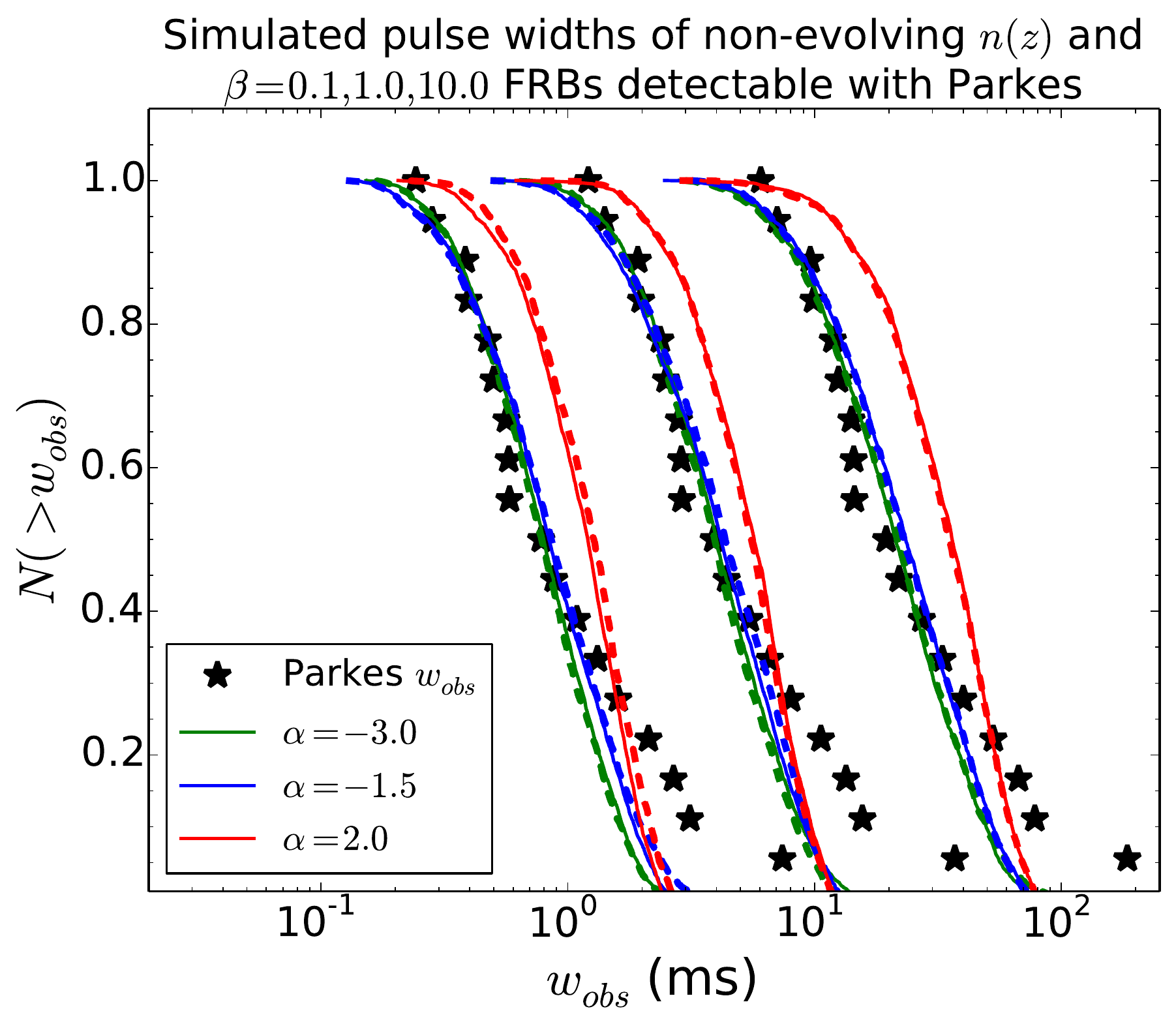}{0.43\textwidth}{}
          \fig{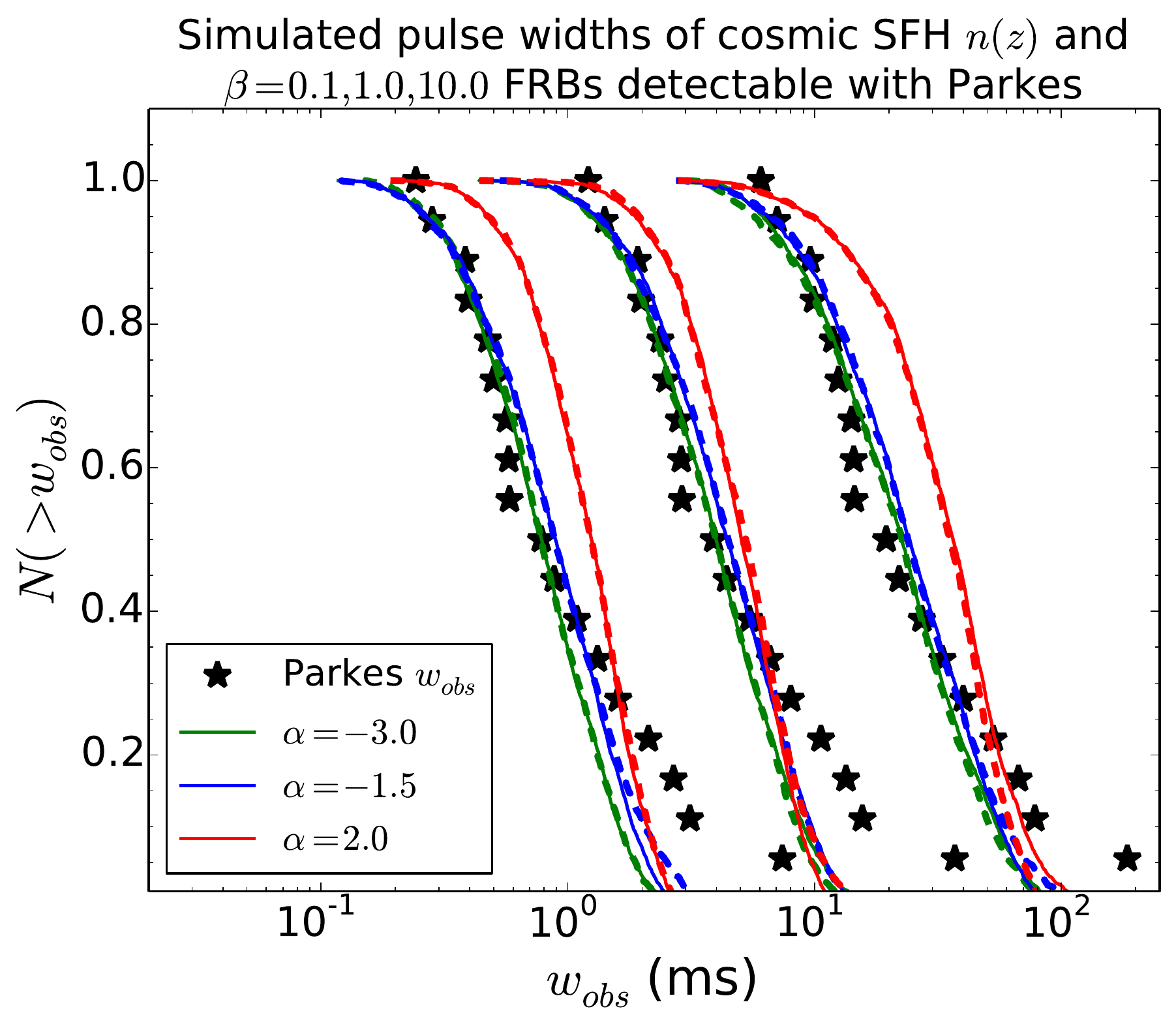}{0.43\textwidth}{}
          }  \vspace{-3.2em}        
\gridline{\fig{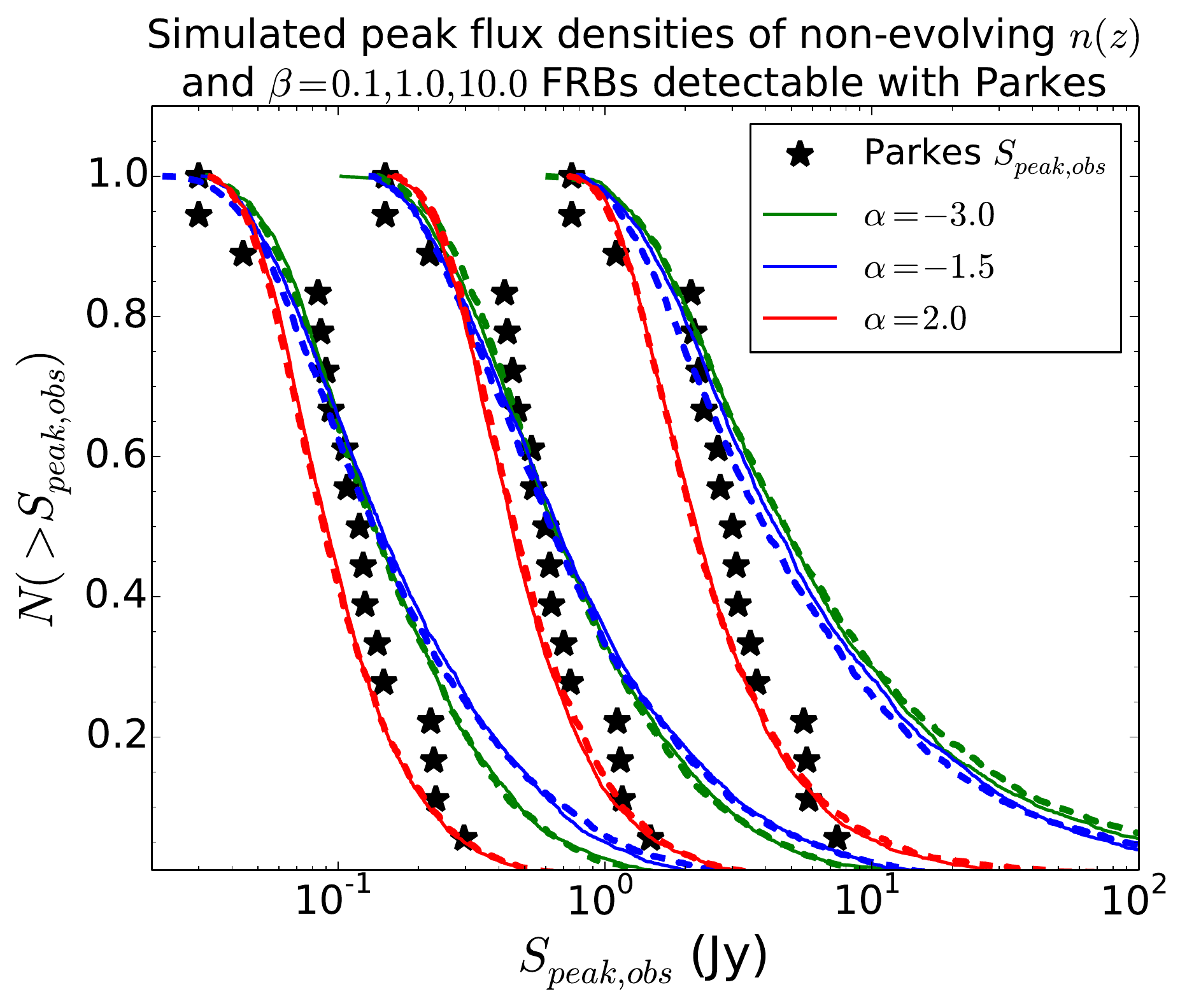}{0.43\textwidth}{}
          \fig{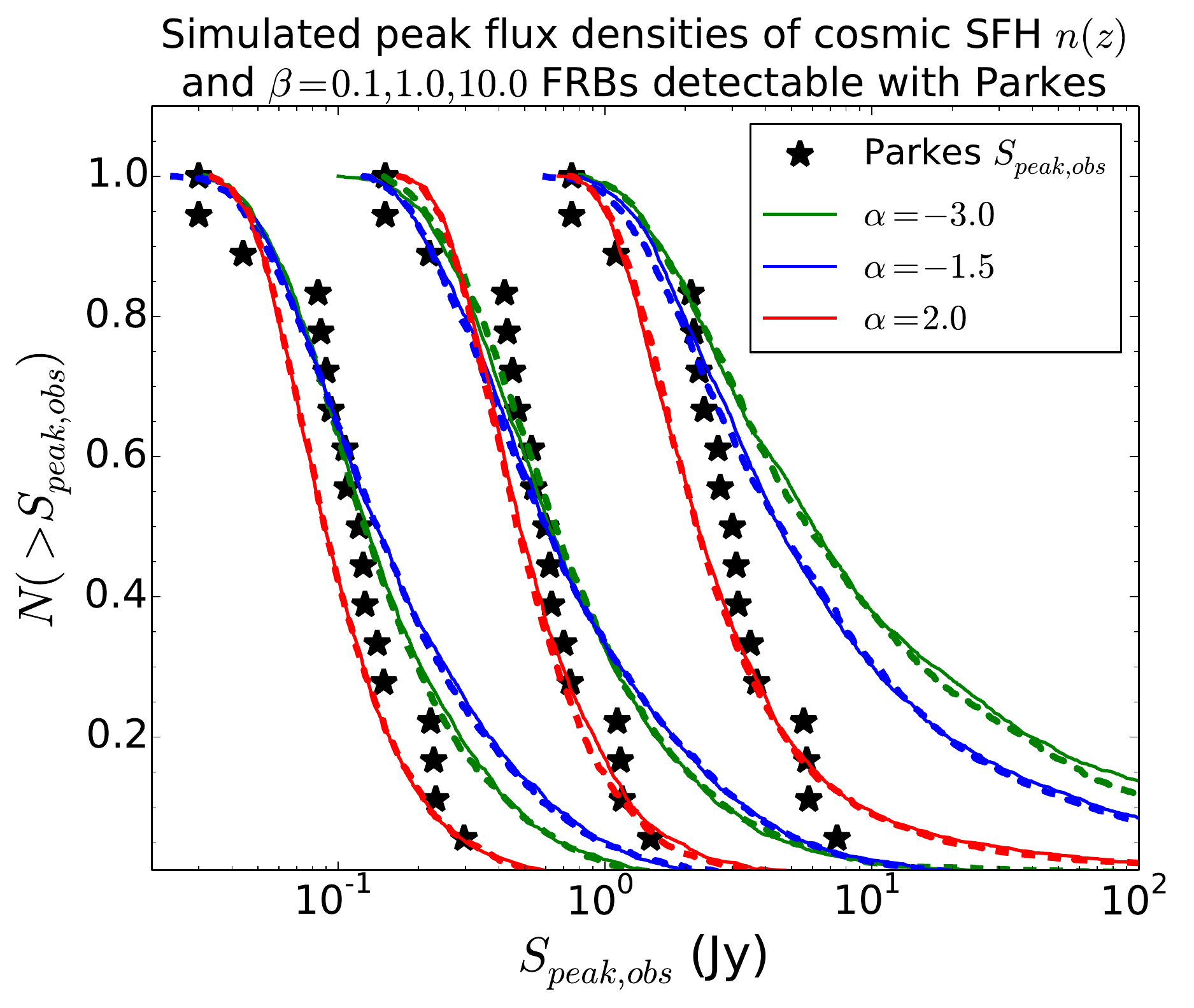}{0.43\textwidth}{}
          }  \vspace{-3.2em}
\gridline{\fig{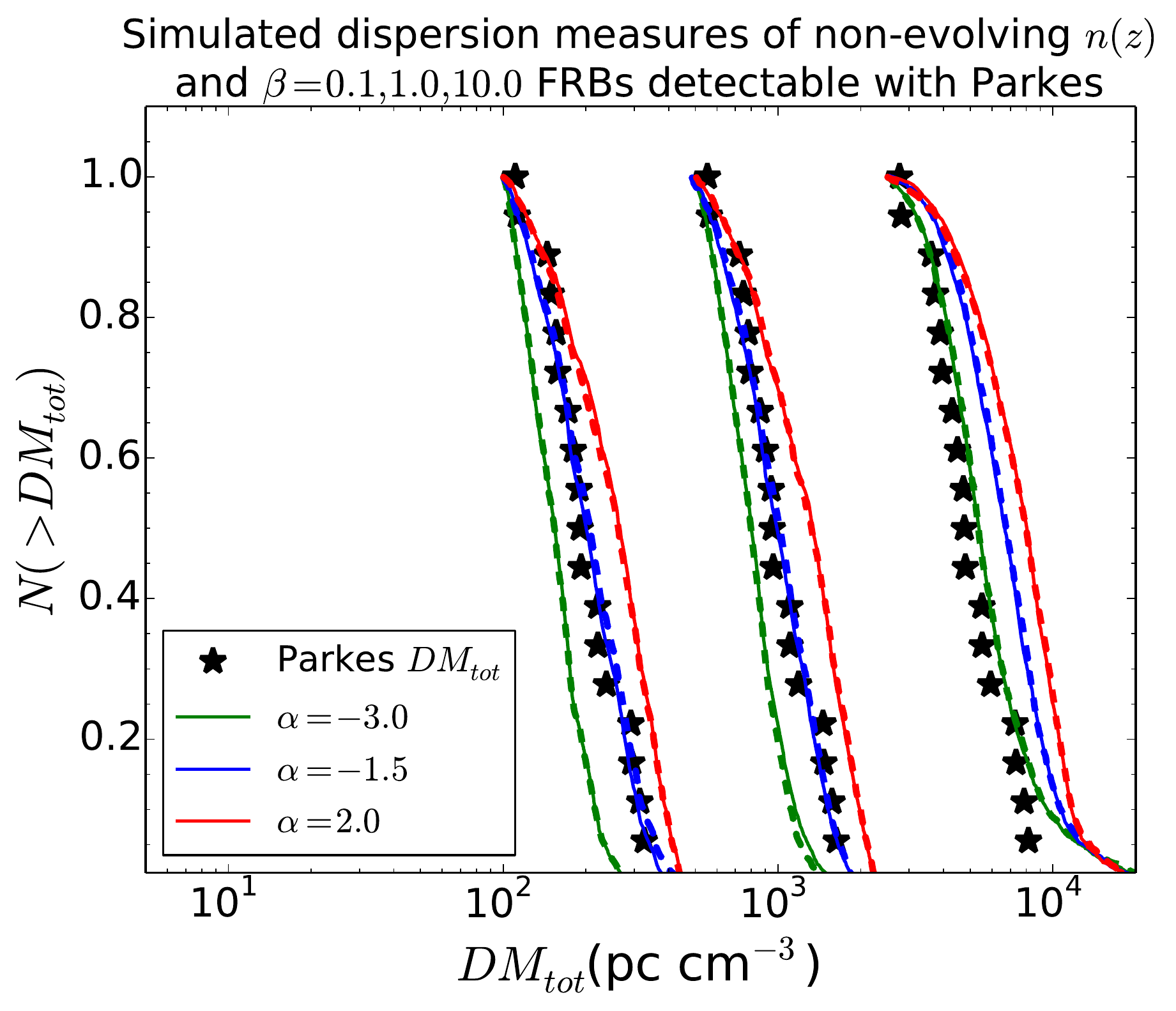}{0.43\textwidth}{}
          \fig{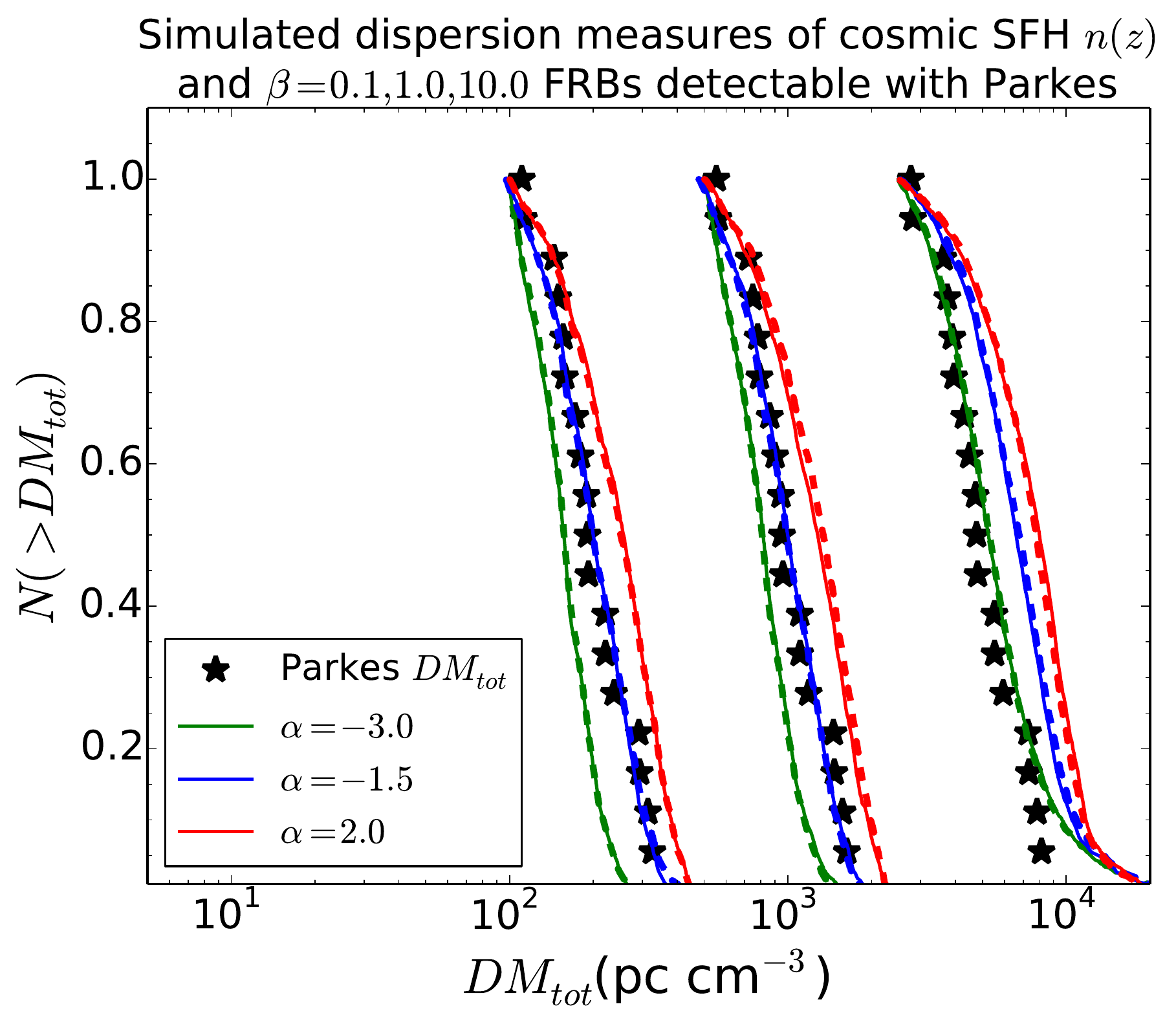}{0.43\textwidth}{}
          }  \vspace{-2.8em}        
  \caption{\emph{Comparison of $w_{\rm obs}$, $S_{\rm peak,obs}$ and $DM_{\rm tot}$ for simulated NE (left-half panels) and SFH (right-half panels) FRBs with non-repeating FRBs detected by Parkes:}
	 In each panel, the MC simulation results are shown for $DM_{\rm host}$ parameter $\beta=0.1,1.0,10.0$, spectral index $\alpha = -3.0,-1.5,2.0$, scattering model 2, and intrinsic models w1L1 and w1L2. 
	 The cumulative parameter distributions for intrinsic model w1L1/w1L2 are denoted by the dashed/solid lines and the corresponding p-values are presented in the top-half of Table \ref{Table5}.
	 The simulation results for the intrinsic width model W2 are not shown here as they are similar to those for $w_{\rm int}$ model W1. 
	 The values for the burst parameters are scaled from their actual values by a factor of 0.2/5.0 for $\beta=0.1/10.0$ to avoid overlap. 
	{\it Top-left panel:} Simulation results for $w_{\rm obs}$ of NE FRBs,
	{\it Top-right panel:} Simulation results for $w_{\rm obs}$ of SFH FRBs,
	{\it Center-left panel:} Simulation results for $S_{\rm peak,obs}$ of NE FRBs,
	{\it Center-right panel:} Simulation results for $S_{\rm peak,obs}$ of SFH FRBs,
	{\it Bottom-left panel:} Simulation results for $DM_{\rm tot}$ of NE FRBs,
	{\it Bottom-right panel:} Simulation results for $DM_{\rm tot}$ of SFH FRBs.
%The index 1/2 used to denote the KS values signify scattering model 1/2. 
}%
\label{fig5} 
\end{figure*}

\begin{figure*}
%\vspace{-10em}
\gridline{\fig{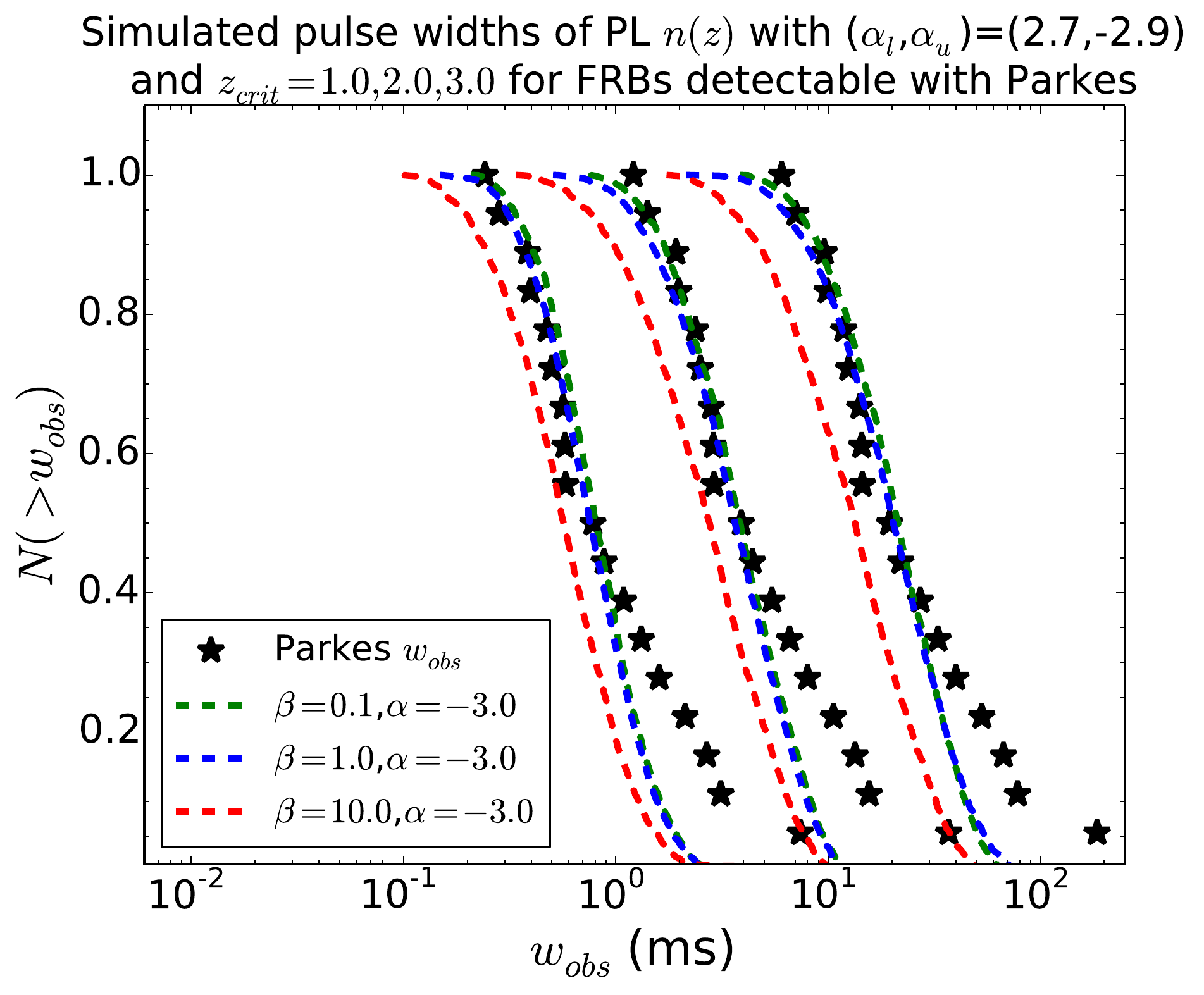}{0.43\textwidth}{}
          \fig{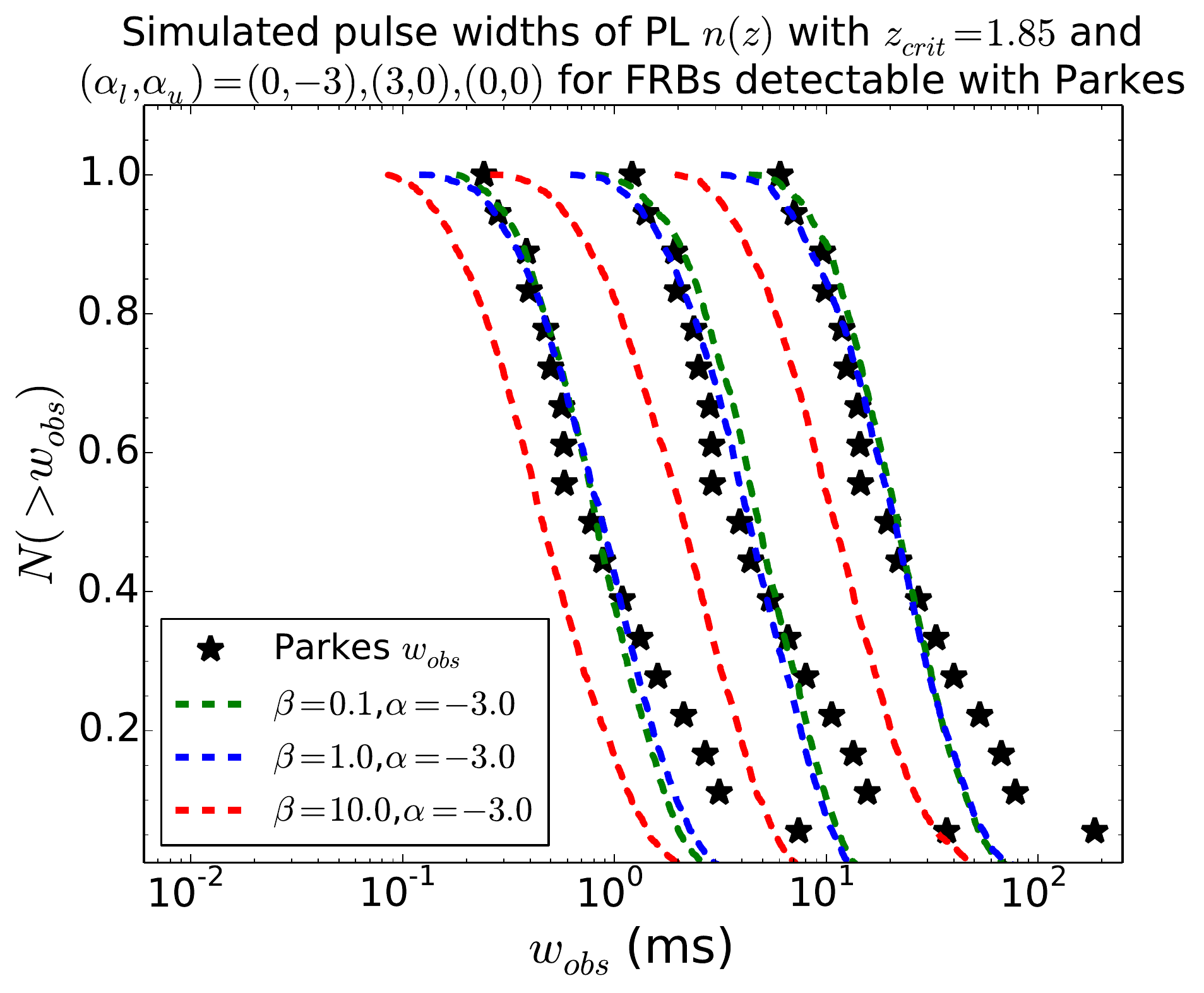}{0.43\textwidth}{}
          }  \vspace{-3.2em}        
\gridline{\fig{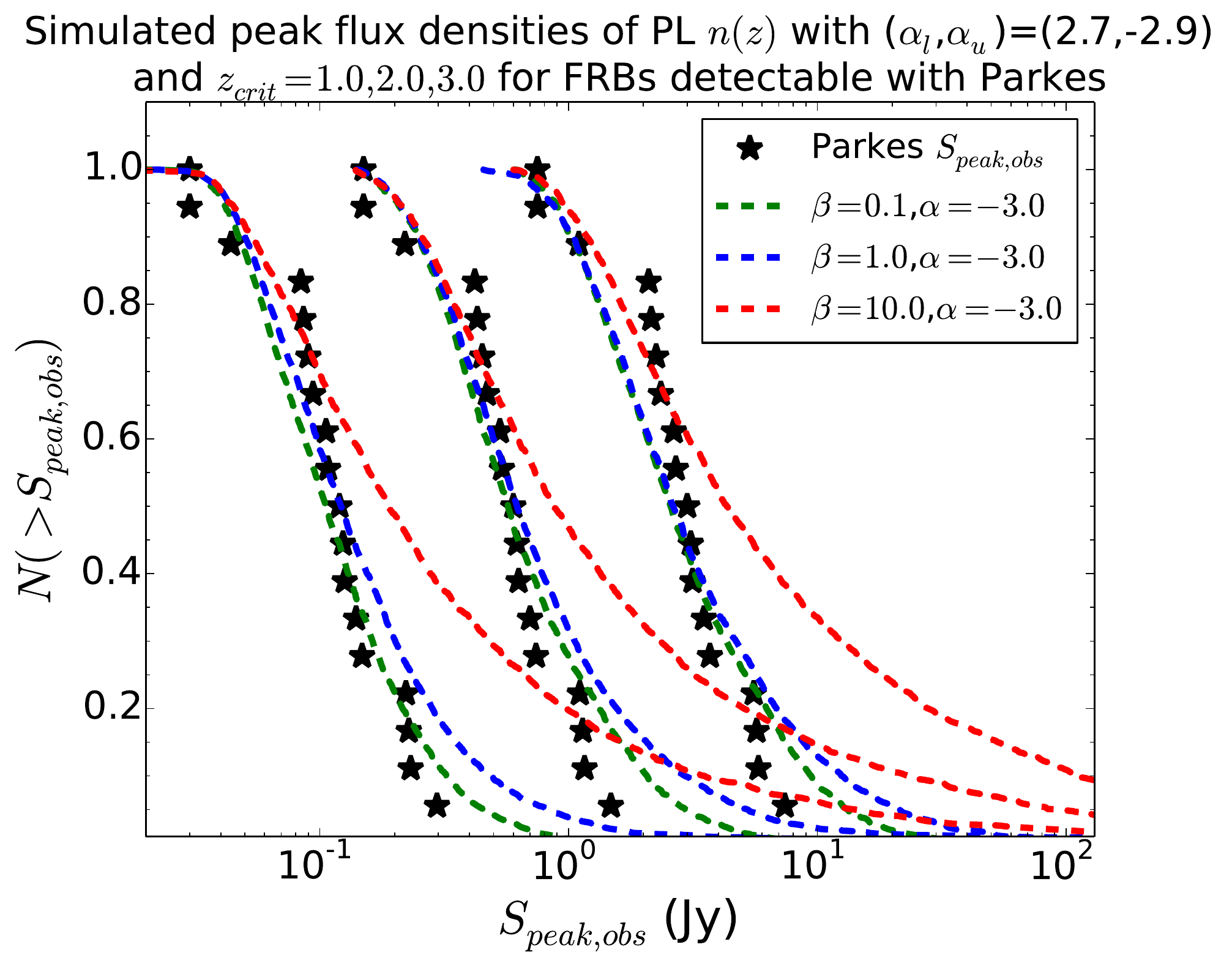}{0.43\textwidth}{}
          \fig{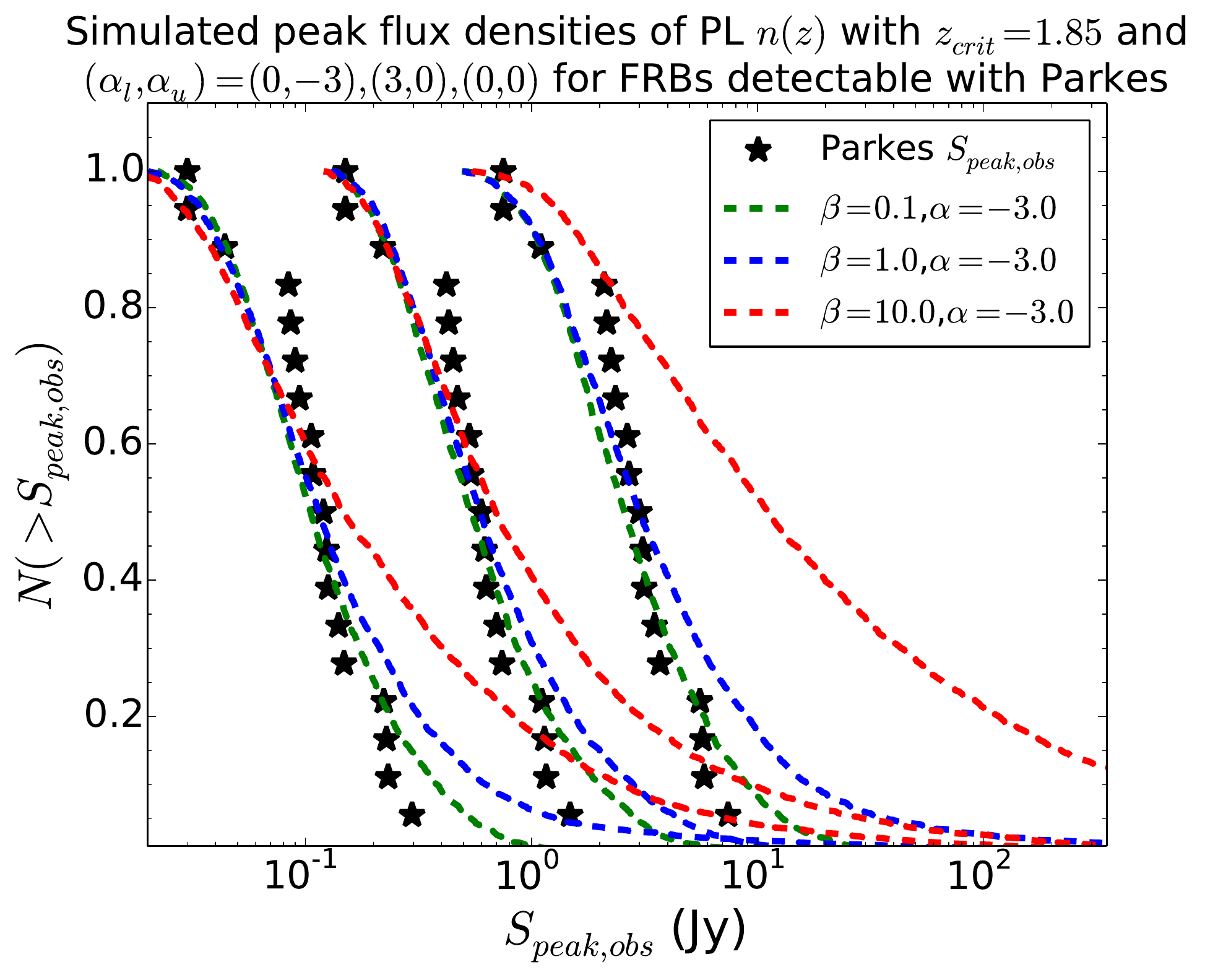}{0.43\textwidth}{}
          }  \vspace{-3.2em}
\gridline{\fig{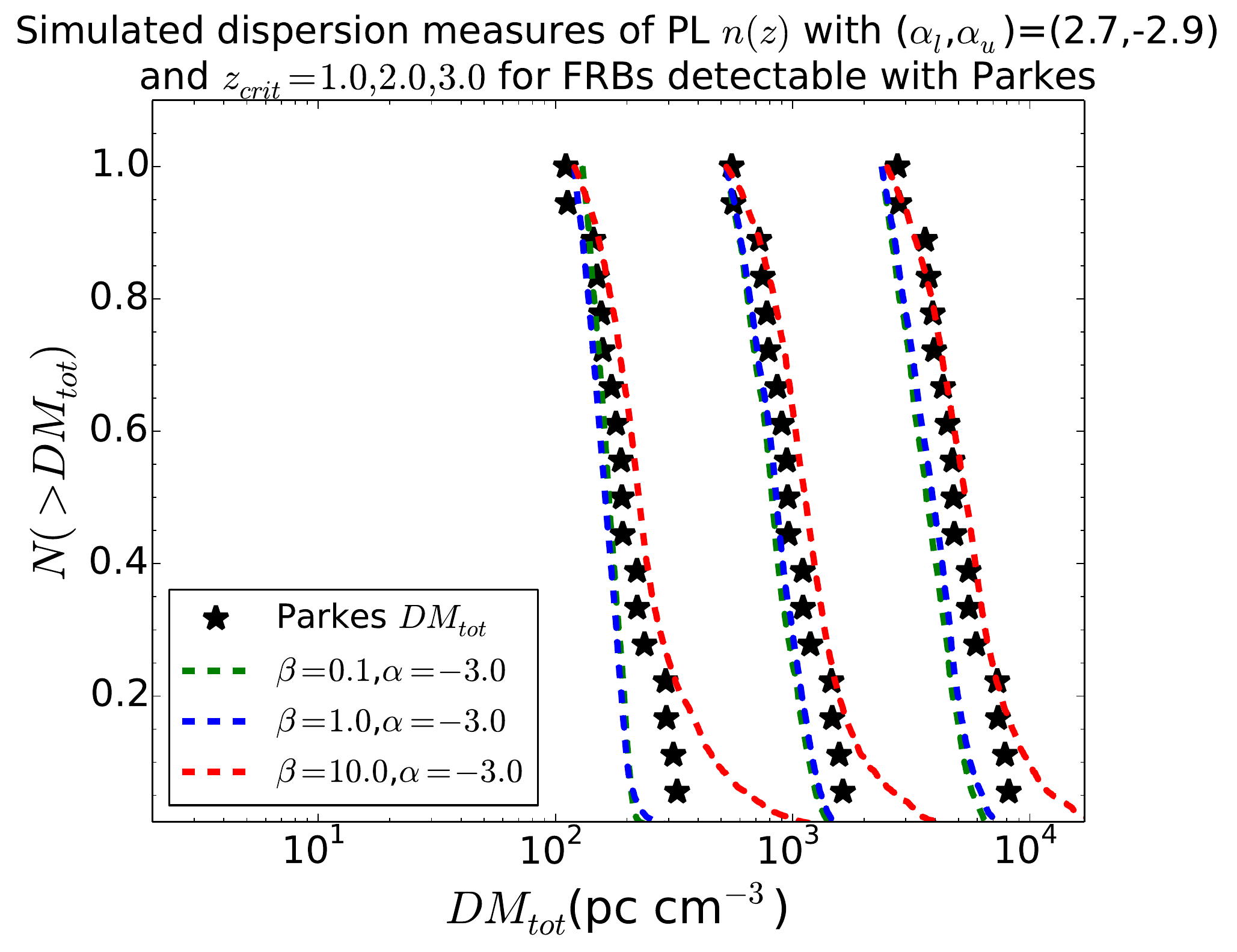}{0.43\textwidth}{}
          \fig{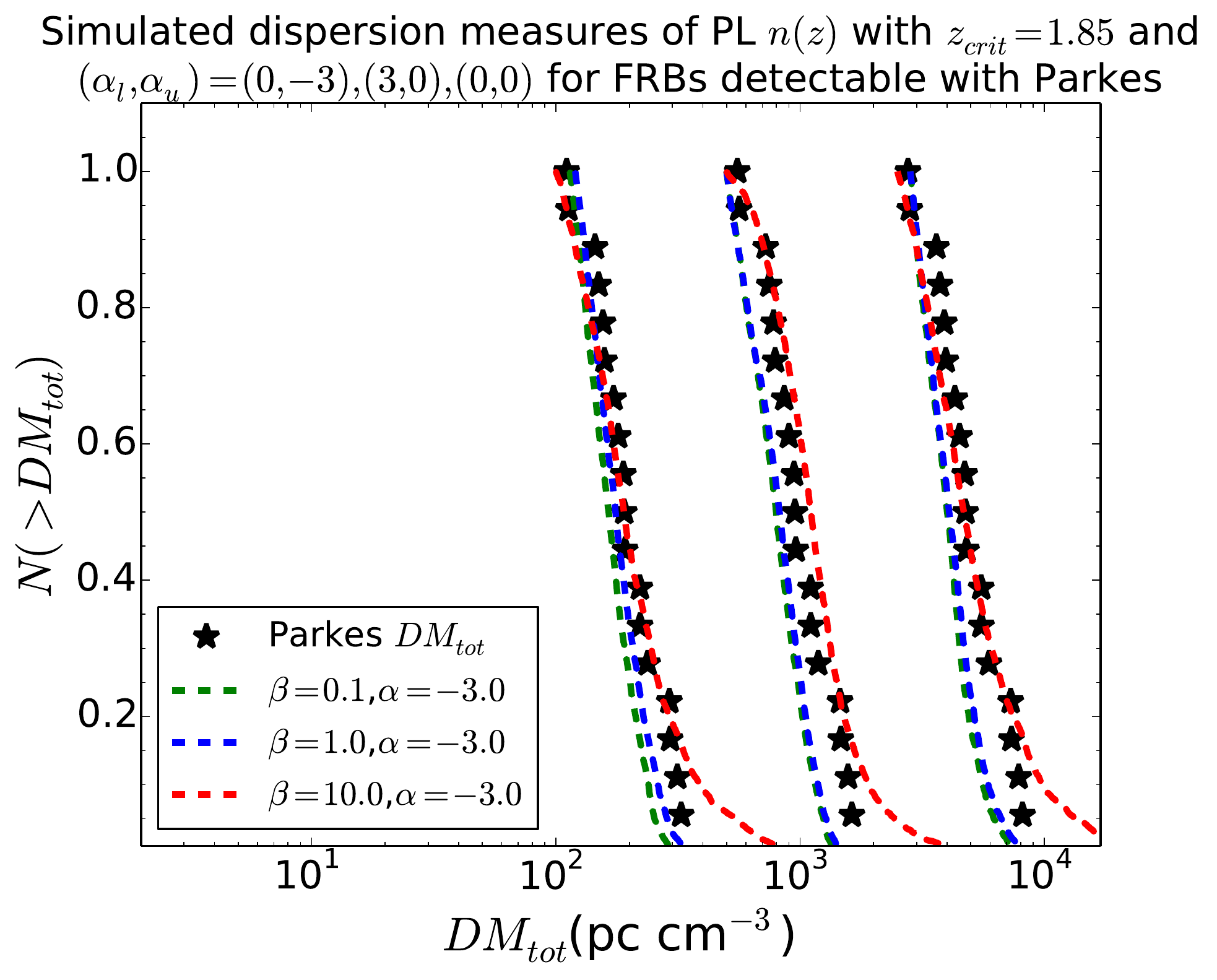}{0.43\textwidth}{}
          }  \vspace{-2.8em}        
  \caption{\emph{Comparison of non-repeating FRBs detected by Parkes and simulated $w_{\rm obs}$, $S_{\rm peak,obs}$ and $DM_{\rm tot}$ for PL FRB population with varying $z_{\rm crit}$ (left-half panels) or varying ($\alpha_{\rm l}$,$\alpha_{\rm u}$) (right-half panels):}	
	The cumulative parameter distributions obtained from the MC simulations for the three cases listed in Section 4.2 are shown for scattering model 2 and intrinsic model w1L1.
%with the index $i=1,2$ in the plot-label 'sc$i$' denoting the specific scattering model. 
The corresponding p-values are presented in the bottom-half of Table \ref{Table5}.
We only show the simulation results for intrinsic model w1L1 as the dependence of FRB observables on the specific $w_{\rm int}$ and $L_{\rm int}$ distribution is found to be relatively weak. 
The FRB parameter values are scaled by a factor 0.2/5.0 for $z_{\rm crit}=1.0/3.0$ in the left-half panels, while the corresponding values are scaled by 0.2/5.0 for ($\alpha_l$,$\alpha_u$)=(0,-3)/(0,0) in the right-half panels to avoid overlap.
	{\it Top-left panel:} Simulated $w_{\rm obs}$ for fixed ($\alpha_{\rm l}$,$\alpha_{\rm u}$)=(2.7,-2.9) and varying $z_{\rm crit}$=1.0,2.0,3.0,
	{\it Top-right panel:} Simulated $w_{\rm obs}$ for fixed $z_{\rm crit}=1.85$ and varying ($\alpha_{\rm l}$,$\alpha_{\rm u}$)=(0,-3),(3,0),(0,0),
	{\it Center-left panel:} Simulated $S_{\rm peak,obs}$ for fixed ($\alpha_{\rm l}$,$\alpha_{\rm u}$) and varying $z_{\rm crit}$,
	{\it Center-right panel:} Simulated $S_{\rm peak,obs}$ for fixed $z_{\rm crit}$ and varying ($\alpha_{\rm l}$,$\alpha_{\rm u}$),
	{\it Bottom-left panel:} Simulated $DM_{\rm tot}$ for fixed ($\alpha_{\rm l}$,$\alpha_{\rm u}$) and varying $z_{\rm crit}$,
	{\it Bottom-right panel:} Simulated $DM_{\rm tot}$ for fixed $z_{\rm crit}$ and varying ($\alpha_{\rm l}$,$\alpha_{\rm u}$). 
}%
\label{fig6} 
\end{figure*}

\subsection{Non-evolving and cosmic star formation history FRB spatial densities}
\label{NE_SFH_nz}
We show the results for the simulated $w_{\rm obs}$, $S_{\rm peak,obs}$ and $DM_{\rm tot}$ distributions of the non-repeating FRBs and compare them with the data from Parkes at $\nu_{\rm obs}=1.4\ {\rm GHz}$ in Figure \ref{fig5}. 
These simulations are performed for NE and SFH $n(z)$ distributions with the intrinsic models w1L1 and w1L2, 
$DM_{\rm host}$ in the range $\beta \sim 0.1-10.0$ and the energy spectral index within range $\alpha \sim -3.0$ to 2.0. 
We do not show the results for intrinsic width model W2 in Figure \ref{fig5} as the corresponding simulated distributions are found to be similar to those for model W1 for a given set of FRB parameters. 
%We {\bf only consider scattering model 2} for the simulations {\bf as the relative difference between .... towards the inferred true properties} is not found to be significant for the parametric space considered (see Section \ref{sec2}). 
We consider $w_{\rm obs}$ and $S_{\rm peak,obs}$ to be the independent parameters among ($w_{\rm obs}$, $S_{\rm peak,obs}$, $\mathcal{F}_{\rm obs}$) in addition to $DM_{\rm tot}$ to evaluate the equivalent p-value $\gamma_{\rm eq} = \sqrt{\gamma_{w_{\rm obs}}^{2} + \gamma_{S_{\rm peak,obs}}^{2} + \gamma_{DM_{\rm tot}}^{2}}$ in each case. The p-values obtained by comparing the simulated parameters for both the NE and SFH $n(z)$ cases with the Parkes observations are listed in the top-half of Table \ref{Table5} in Appendix \ref{AppendixC}. 

We find that the four model distributions (w1L1, w1L2, w2L1 and w2L2) considered for the intrinsic widths and luminosities of non-repeating FRBs are practically indistinguishable for both NE and SFH $n(z)$ populations.
This further indicates the weak dependence of the simulated FRB observable properties on the intrinsic source properties for the range of model parameters that we consider here.  
We find that for both NE and SFH burst spatial densities, the observed $w_{\rm obs}$ from Parkes are in better agreement with the simulated $w_{\rm obs}$ for a host galaxy DM contribution $DM_{\rm host}$ that is comparable to the Galactic contribution with $\beta \sim 0.1-1.0$ as opposed to $\beta \gtrsim 10.0$. 
%relatively small $DM_{\rm host}$ contribution corresponding to $\beta \sim 0.1-1.0$ as opposed to $\beta \gtrsim 10.0$. 
Moreover, the observed $w_{\rm obs}$ distribution suggests a large negative spectral index $\alpha$ for the FRB energies as $\gamma_{w_{\rm obs}}$ gradually decreases with increase in $\alpha$ for both $n(z)$ distributions. 
In case of $S_{\rm peak,obs}$, we also find that a larger $DM_{\rm host}$ contribution $\beta \gtrsim 10.0$ cannot physically explain the observed flux density distribution for either of the FRB population density distributions. 
%the $DM_{\rm host}$ contribution does not significantly affect the observed distribution for either of the FRB population densities. 
Similar to the observed $w_{\rm obs}$ distribution, the observed $S_{\rm peak,obs}$ from Parkes can be better explained for a power-law FRB energy spectrum with a negative spectral index as $\gamma_{S_{\rm peak,obs}}(\alpha = -3.0) \gtrsim \gamma_{S_{\rm peak,obs}}(\alpha = -1.5) > \gamma_{S_{\rm peak,obs}}(\alpha = 2.0)$ for both $n(z)$ considered. 
%However, the observed $S_{\rm peak,obs}$ from Parkes can be better explained for a shallower FRB energy spectrum with 
%$\gamma_{S_{\rm peak,obs}}(\alpha = 1.5) \gg \gamma_{S_{\rm peak,obs}}(\alpha = -1.5) > \gamma_{S_{\rm peak,obs}}(\alpha = \pm 3.0)$ for both $n(z)$ considered. 

For both NE and SFH spatial density distributions used here, $DM_{\rm host} \sim DM_{\rm MW}$ is clearly favored from the $DM_{\rm tot}$ detected at Parkes with $\gamma_{DM_{\rm tot}}(\beta \sim 0.1-1.0) \gg \gamma_{DM_{\rm tot}}(\beta \sim 10.0)$. 
The observed $DM_{\rm tot}$ values at Parkes are expected to arise from a relatively steep FRB energy density distribution with $\gamma_{DM_{\rm tot}}(\alpha = -3.0) > \gamma_{DM_{\rm tot}}(\alpha = -1.5) > \gamma_{DM_{\rm tot}}(\alpha = 2.0)$ for both FRB $n(z)$. 
%While the scattering models for IGM and ISM turbulence cannot be significantly differentiated using the observed non-repeating FRB population at present, {\bf model 1 is found to be in slightly better agreement with the Parkes data}. 
From the comparison of simulated FRB parameters with the Parkes data for NE/SFH FRB spatial density, we find that: 
\begin{itemize}
\item There is a weak dependence of the FRB observable properties such as $w_{\rm obs}$, $S_{\rm peak,obs}$ and $DM_{\rm tot}$ on the intrinsic source properties, particularly for the parametric space that we consider here.
\item For both NE and SFH burst spatial densities, the host galaxy DM contribution is likely to be smaller or comparable to the MW contribution with the likelihood order $\beta = 1.0 \approx \beta = 0.1 > \beta = 10.0$.  
\item The distribution of observable parameters for Parkes FRBs indicates that a larger negative value of $\alpha$ is more likely with the order $\alpha = -3.0 > \alpha = -1.5 \gg \alpha = 2.0$. 
\item The observed sample of Parkes FRBs is statistically insufficient to clearly differentiate between the NE and SFH spatial densities, and a larger observed population of $\sim 50-100$ Parkes FRBs is necessary to distinguish the FRB redshift distribution with confidence \citep{Caleb16}.
\end{itemize} %\\ \\ \\
%While the non-repeating FRB population observed with Parkes that is considered here does not clearly differentiate between the FRB $n(z)$ distributions, {\bf we find the NE FRB spatial density to be more likely.} 
%It should be noted that NE and SFH $n(z)$ have similar redshift distribution for FRBs {\bf up to} $z \approx 2.0$, where most of the FRBs are generated in the MC simulations and also observed. 

\subsection{Power-law FRB spatial density}
\label{PL_nz}
In order to constrain the parameters of the PL $n(z)$ distribution, we consider specific ($\beta$,$\alpha$) combinations for which the KS test likelihood value is found to be significantly large for either NE or SFH $n(z)$ distributions (see the top-half of Table \ref{Table5}). %with $\gamma_{\rm eq} \gtrsim 0.5$
%From the top-half of Table \ref{Table5}, the five cases with either NE or SFH $n(z)$ for which $\gamma_{\rm eq}$ is maximised and $\gamma_{\rm eq} \geq 0.2$ for a given choice of $\beta$, $\alpha$
As the specific choice of the intrinsic width and luminosity model (w1L1/w1L2/w2L1/w2L2) does not influence the likelihood value significantly, we assume the w1L1 model for all our simulations along with:\\ 
%As the choice of the scattering model does not influence the likelihood, we assume scattering model 2 for all our simulations along with:
%\begin{enumerate}[label=(\arabic*)]
%{\setlength\itemindent{15pt} \item 
1. $\beta = 0.1$ and $\alpha=-3.0$, \\
2. $\beta = 1.0$ and $\alpha=-3.0$, \\
3. $\beta = 10.0$ and $\alpha=-3.0$, \\
%\item $\beta = 1.0$ and $\alpha=-1.5$,
%\item $\beta = 10.0$ and $\alpha=1.5$ for scattering model 1
%\item $\beta = 0.1$ and $\alpha=-1.5$ for scattering model 1
%}
%\end{enumerate}
as the three input cases. 
We show the simulated $w_{\rm obs}$, $S_{\rm peak,obs}$ and $DM_{\rm tot}$ distributions corresponding to these cases for comparison with the observed non-repeating FRB population detected by Parkes at $\nu_{\rm obs} = 1.4\ {\rm GHz}$ in Figure \ref{fig6}. 
In order to constrain the parameters ($\alpha_l$,$\alpha_u$,$z_{\rm crit}$) of the PL $n(z)$ distribution, we perform the MC simulations with either varying $z_{\rm crit}$ or varying ($\alpha_l$,$\alpha_u$) for each of these cases listed above. 

We consider $z_{\rm crit} = 1.0,\: 2.0,\: 3.0$ for fixed ($\alpha_l$,$\alpha_u$) = (2.7,-2.9), where the values for the PL indices are motivated from the SFH distribution at asymptotically low and high redshifts. For the fixed $z_{\rm crit}$ case, we consider $(\alpha_l,\alpha_u) = (0,-3),\: (3,0),\: (0,0)$ and the value of $z_{\rm crit} = 1.85$ is chosen to resemble the redshift at which the cosmic SFH distribution peaks. 
The bottom-half of Table \ref{Table5} in Appendix \ref{AppendixC} lists the p-values $\gamma_{\rm eq}$ obtained by comparing the simulated parameters for both the fixed $z_{\rm crit}$ and fixed ($\alpha_l$,$\alpha_u$) cases with the observed FRBs from Parkes. 
%Along with $w_{\rm obs}$, we consider $L_{\rm obs}$ and $E_{\rm obs}$ as the independent parameters for the KS analysis in case of PL $n(z)$ as they are directly correlated with $S_{\rm peak,obs}$ and $\mathcal{F}_{\rm obs}$ for a given $z$, which in turn depends on the specific $n(z)$ distribution. 
%{\bf As earlier, the KS test equivalent p-value is obtained as} $\gamma_{\rm eq} = \sqrt{\gamma_{w_{\rm obs}}^{2} + \gamma_{L_{\rm obs}}^{2} + \gamma_{E_{\rm obs}}^{2}}$. 
From the fixed ($\alpha_l$,$\alpha_u$) and varying $z_{\rm crit}$ case, we find that the agreement of each of the simulated $w_{\rm obs}$, $S_{\rm peak,obs}$ and $DM_{\rm tot}$ distributions with the corresponding observed quantities from Parkes is considerably better for an intermediate value of peak redshift $z_{\rm crit} \approx 2.0$ with $\gamma_{i}(z_{\rm crit} = 2.0) > \gamma_{i}(z_{\rm crit} = 1.0) > \gamma_{i}(z_{\rm crit} = 3.0)$ for $i = w_{\rm obs},\ S_{\rm peak,obs},\ DM_{\rm tot}$ and most of the cases considered. Therefore, the FRB spatial density is expected to peak at a redshift similar to that of the cosmic SFH distribution.  

%\EntryGap
%\begin{savenotes}
\begin{table*}
%\label{Table5}
\begin{center}
\caption{\small The follow-up observation information for repeating FRB 121102 and non-repeating Parkes FRBs listed in Table \ref{Table1}. The redshift $z$ and burst energy $E_{\rm obs}$ are inferred by assuming a fixed host galaxy DM contribution $DM_{\rm host} = 100\ {\rm pc\ cm^{-3}}$. None of the 13 listed FRBs were observed to repeat in spite of dedicated follow-up efforts ranging from few hours to $\sim 100$ hours with Parkes.  
}
\label{Table7}
\bgroup
\def\arraystretch{1.0}
\begin{tabular}{ccccccccccccccccccccc}
\hline
\hline
\centering
FRB & Telescope & $S_{\rm peak,obs}$ & $\mathcal{F}_{\rm obs}$ & $z$ & $E_{\rm obs}$ & $t_{\rm obs}$ & Reference \\ 
        &		    & (Jy) 			& (Jy ms)			   &        & ($10^{42}$ erg) & (hr)  & \\ \hline \hline         
010621 & Parkes & 0.53 & 4.24 & 0.20 & 0.03 & 15.5 & \citet{Keane12} \\ \hline
%010724 & Parkes & 1.57 & 31.48 & 0.35 & 0.78 & 40 & \citet{Lorimer07} \\ \hline
090625 & Parkes & 1.14 & 2.19 & 1.06 & 0.53 & 33.65 & \citet{Petroff15} \\ \hline %\citet{Champion16}
110220 & Parkes & 1.11 & 7.31 & 1.12 & 1.94 & 1.75 & \citet{Petroff15} \\ \hline %\citet{Thornton13}
110626 & Parkes & 0.63 & 0.89 & 0.81 & 0.12 & 11.25 & \citet{Petroff15} \\ \hline %\citet{Thornton13}
110703 & Parkes & 0.45 & 1.75 & 1.33 & 0.65 & 10.1 & \citet{Petroff15} \\ \hline %\citet{Thornton13}
120127 & Parkes & 0.62 & 0.75 & 0.61 & 0.06 & 5.5 & \citet{Petroff15} \\ \hline %\citet{Thornton13}
121002 & Parkes & 0.43 & 2.34 & 2.00 & 1.86 & 10.25 & \citet{Petroff15} \\ \hline %\citet{Champion16}
121102 & Arecibo, GBT, Effelsberg &  &  & 0.19 &  & 235.7 & $\dagger$ \\ \hline
130626 & Parkes & 0.74 & 1.47 & 1.09 & 0.37 & 9.5 & \citet{Petroff15} \\ \hline %\citet{Champion16}
%130628 & Parkes & 1.91 & 1.22 & 0.47 & 0.06 & 9 & \citet{Champion16} \\ \hline
130729 & Parkes & 0.22 & 3.43 & 1.01 & 0.75 & 10 & \citet{Champion16} \\ \hline
131104 & Parkes & 1.16 & 2.75 & 0.85 & 0.43 & 78 & \citet{Ravi15} \\ \hline
140514 & Parkes & 0.47 & 1.32 & 0.62 & 0.11 & 19.2 & \citet{Petroff15} \\ \hline %\citet{Petroff15b}
%141113 & Arecibo & 0.04 & 0.08 & 0.18 & 0.0005 & 24 & \citet{Patel18} \\ \hline
150215 & Parkes & 0.70 & 2.02 & 0.82 & 0.28 & 17.25 & \citet{Petroff17} \\ \hline
%%%151206 & Parkes & 0.30 & 0.90 & 2.28 & 0.90 & 22.3 & \citet{Bhandari18} \\ \hline
151230 & Parkes & 0.42 & 1.90 & 1.13 & 0.52 & 54.9 & \citet{Bhandari18} \\ \hline
%%160317 & UTMOST & 3.0 & 63.00 & 1.03 & 14.32 & 105 & \citet{Caleb17} \\ \hline
%160410 & UTMOST & 7.0 & 28.00 & 0.19 & 0.20 & 43 & \citet{Caleb17} \\ \hline
%%160608 & UTMOST & 4.3 & 38.70 & 0.51 & 2.05 & 35 & \citet{Caleb17} \\ \hline
%%170107 & ASKAP & 22.3 & 57.98 & 0.68 & 5.64 & 669.6 & \citet{Shannon18} \\ \hline %\citet{Bannister17}, 
%%170416 & ASKAP & 19.4 & 97.00 & 0.56 & 6.30 & 388.8 & \citet{Shannon18} \\ \hline
%%170428 & ASKAP & 7.7 & 34.00 & 1.17 & 9.92 & 758.4 & \citet{Shannon18} \\ \hline
%%171116 & ASKAP & 19.6 & 63.00 & 0.69 & 6.32 & 1096.8 & \citet{Shannon18} \\ \hline
%%180110 & ASKAP & 128.1 & 420.00 & 0.82 & 59.18 & 904.8 & \citet{Shannon18} \\ \hline
%%180131 & ASKAP & 22.2 & 100.00 & 0.74 & 11.46 & 758.4 & \citet{Shannon18} \\ \hline
\end{tabular}
\egroup
\end{center}
$\dagger$ The follow-up data for the repeating FRB 121102 is obtained from \citet{Spitler16}, \citet{Scholz16}, \citet{Scholz17}, \citet{Law17}, \citet{Hardy17}, \citet{Michilli18}, \citet{Gajjar18} and \citet{Spitler18}.
\end{table*}
%\end{savenotes}

Moreover, for the fixed $z_{\rm crit}$ and varying ($\alpha_l$,$\alpha_u$) case, we find that the observed $w_{\rm obs}$ from Parkes can be better explained with a FRB spatial density that gradually decreases with the source distance as $\gamma[(\alpha_l,\alpha_u)=(0,-3)] > \gamma[(\alpha_l,\alpha_u)=(0,0)] > \gamma[(\alpha_l,\alpha_u)=(3,0)]$. 
While $S_{\rm peak,obs}$ does not show a clear dependence on the PL $n(z)$ indices ($\alpha_l$,$\alpha_u$) for a fixed $z_{\rm crit}$, a FRB spatial density falling off sharply at large distances is preferred by the current $DM_{\rm tot}$ distribution with the likelihood order for PL indices ($\alpha_l$,$\alpha_u$) being $(0,-3) \gtrsim (0,0) > (3,0)$. 
%In case of $S_{\rm peak,obs}$ and $DM_{\rm tot}$, a FRB spatial density falling off sharply at large distances is preferred by the current data 
Combining the results from the fixed ($\alpha_l$,$\alpha_u$) and fixed $z_{\rm crit}$ cases for PL $n(z)$ obtained here with those for SFH $n(z)$ in Section \ref{NE_SFH_nz}, we find that: 
\begin{itemize}
\item The comparison of simulated distributions with the observed Parkes FRBs for varying $z_{\rm crit}$ suggests that the FRB spatial density is likely to be a PL distribution that peaks at an intermediate value of redshift $z_{\rm crit} = 2.0$ similar to the cosmic SFH. 
\item The observed parametric distributions for Parkes FRBs for varying ($\alpha_l$,$\alpha_u$) can be better explained with a FRB spatial density that decreases with increasing source distance. The reasonable agreement of the burst parameters with the Parkes FRB observables for SFH $n(z)$ also suggests PL indices $\alpha_l \approx 3.0$ and $\alpha_u \approx -3.0$ at asymptotically low and high redshifts, respectively. 
\item The decreasing FRB spatial densities at large distances are constrained to obtain an upper PL index $\alpha_u \sim -3.0$. Therefore, it is likely that the PL indices for the FRB spatial density are $\alpha_l \sim 0-3$ and $\alpha_u \approx -3$ with the distribution peaking at slightly larger redshifts compared to the cosmic SFH. 
\end{itemize} %\\ \\ \\
Similar to the non-repeating FRBs, we also compare the simulated $S_{\rm peak,obs}$ and $\mathcal{F}_{\rm obs}$ distributions of the repeating FRB 121102 sub-bursts with the Arecibo observations. The corresponding results are shown in Figure \ref{fig7} and the p-values listed in Table \ref{Table6} of Appendix \ref{AppendixB}. 

\begin{figure*}
%\vspace{-10em}
\gridline{\fig{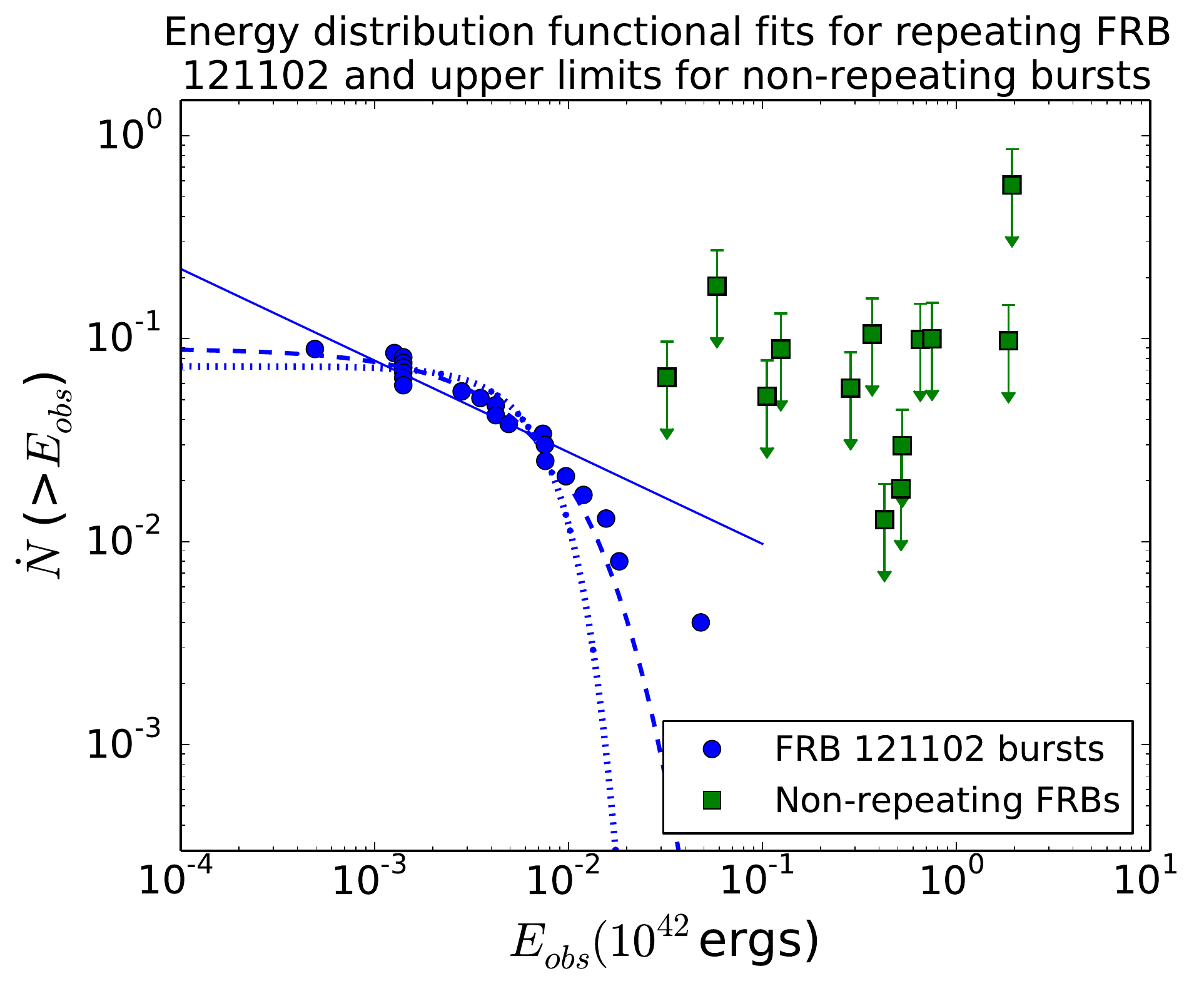}{0.46\textwidth}{}
          \fig{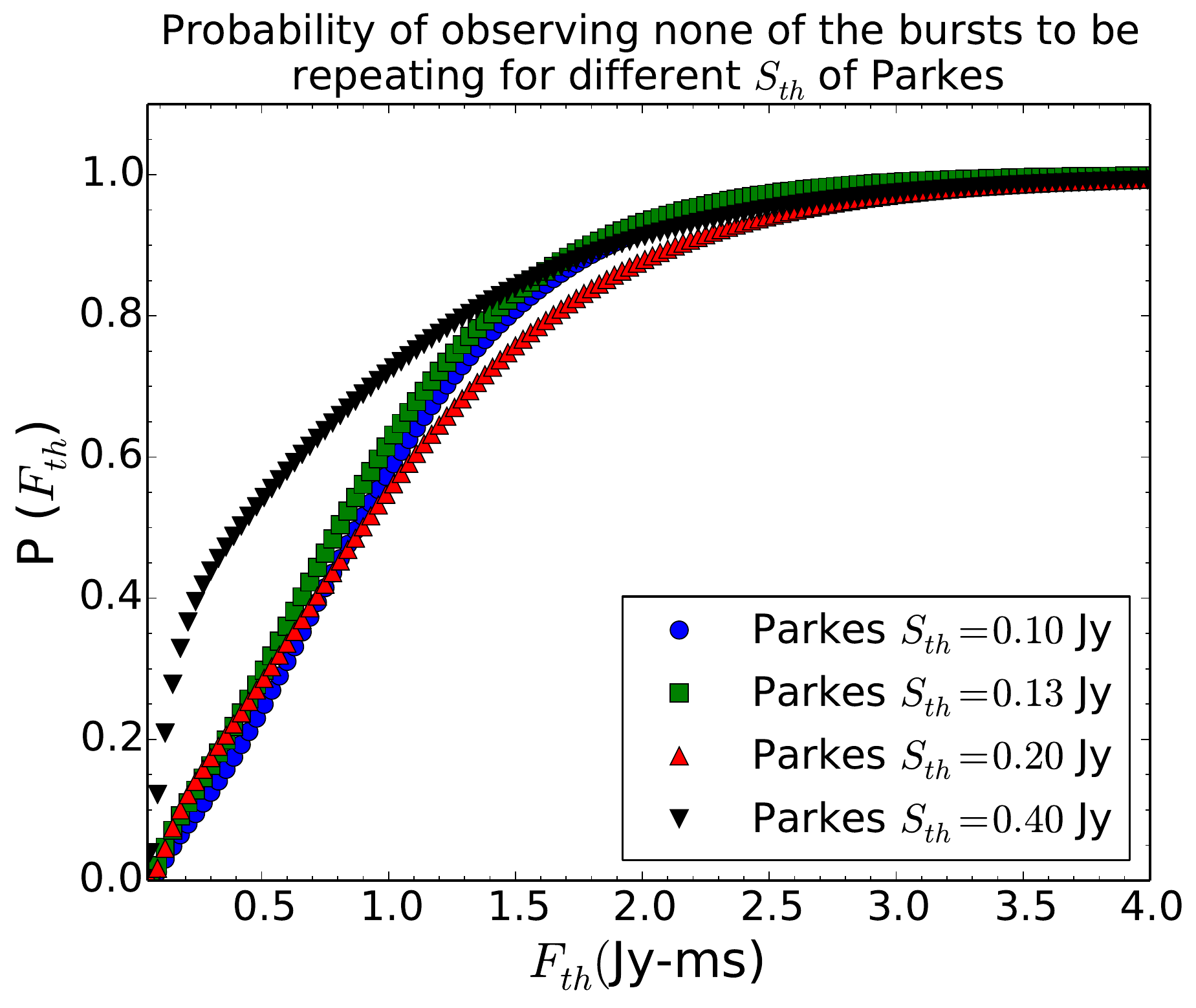}{0.46\textwidth}{}
          }  \vspace{-2.5em}              
  \caption{\emph{Follow-up observations for the repeating FRB 121102 and the non-repeating FRBs:} 
	{\it Left panel:} Power-law, exponential and gaussian fits for the CED of the repeating FRB 121102. The distribution for the repeater is normalized using its follow-up observing time $t_{\rm obs} = 235.7\ {\rm hr}$. The upper limits for $\dot{N}(>E_{\rm obs})$ of the 13 non-repeating Parkes FRBs are also shown and are normalized using their respective $t_{\rm obs}$ values from Table \ref{Table7},
	{\it Right panel:} The probability of observing none of the 13 non-repeating Parkes FRBs to be repeating as a function of the fluence threshold $\mathcal{F}_{\rm th}$ (see equation \ref{P_Fth}). The probability is computed for different sensitivities of Parkes $S_{\rm th} = 0.10,0.13,0.20,0.40\ {\rm Jy}$.	
}%
	\label{fig8}
\end{figure*}

%\SmallEntryGap
%\EntryGap 
\section{FRB population and repeatability}
\label{sec5}
In Section \ref{sec4}, we constrained the FRB host galaxy DM, spectral index of the energy density, spatial distribution of the bursts and also the scattering due to turbulence in the intervening IGM and ISM using the observations for non-repeating FRBs/FRB 121102 bursts from Parkes/Arecibo. Now we use the follow-up data for FRBs in order to investigate whether FRB 121102 is representative of all FRBs repeating with a universal energy distribution function (EDF). Even though most FRBs have been extensively followed up with dedicated surveys ranging from few hours to $\sim 1000\ {\rm hrs}$, only few of them including FRB 180814.J0422+73 were observed to be repeating. Table \ref{Table7} lists the published follow-up observation data for the repeating FRB 121102 and 13 non-repeating Parkes FRBs from Table \ref{Table1}. There can be two possible reasons for the repeating bursts from other FRBs to not get detected in spite of a universal repetitive behaviour for FRBs: (a) the current observing times $t_{\rm obs}$ are smaller compared to the repeating timescale $t_{\rm rep}$ for the FRBs, or (b) the repeating bursts from the other FRBs are very dim and cannot be detected with the typical sensitivities of the current telescopes \citep{CP18}.  

The Parkes sensitivity threshold at $\nu_{\rm obs} = 1.4\ {\rm GHz}$ for $S/N = 9$ and an arbitrary $w_{\rm obs}$ is $S_{\rm th} = 0.36\ {\rm Jy}\ (w_{\rm obs}/1\ {\rm ms})^{-1/2}$ \citep{Caleb16}. For pulse widths in the range $w_{\rm obs} \sim 1.00-8.00\ {\rm ms}$, $S_{\rm th}$ for Parkes varies within the range $\sim 0.1-0.4\ {\rm Jy}$. It should be noted that all the inferred energies in Table \ref{Table1} are based on the lower limits for the fluence due to the uncertainty in the position of the source within a single beam and the assumption of on-axis detection for all bursts. Since Parkes is less sensitive as compared to Arecibo, we only include the repeating FRB 121102 sub-bursts for which $S_{\rm peak,obs}$ exceeds the Parkes $S_{\rm th}$ for evaluating the repeater cumulative energy distribution (CED). Out of the 88 repeating bursts from FRB 121102 that we consider here, $65/55/46/21$ bursts would have been above the Parkes $S_{\rm th} \sim 0.10/0.13/0.20/0.40\ {\rm Jy}$ at $\nu_{\rm obs} = 1.4\ {\rm GHz}$. 
While computing the repeater CED for varying Parkes $S_{\rm th}$ values, we have ignored the effect of different observing frequencies for Parkes as compared to the FRB 121102 observations from Arecibo, GBT and Effelsberg.

The left-hand panel of Figure \ref{fig8} shows the chi-squared PL, exponential and gaussian (with zero mean) fits for the CED of FRB 121102 along with the $\dot{N}(>E_{\rm obs})$ upper limits for the non-repeating FRBs. 
We use the follow-up observations that were published in \citet{Spitler16}, \citet{Scholz16}, \citet{Scholz17}, \citet{Law17}, \citet{Hardy17}, \citet{Michilli18}, \citet{Gajjar18} and \citet{Spitler18} to normalise the repeating FRB 121102 CED with its total observing time $t_{\rm obs} = 235.7\ {\rm hr}$. 
%While the repeating FRB CED is normalized with its total observing time $t_{\rm obs} = 235.7\ {\rm hr}$,
It should be noted that here $t_{\rm obs}$ is essentially a lower limit as not all non-detection results were published in a timely manner. The $\dot{N}(>E_{\rm obs,i})$ values for the individual non-repeating FRBs are normalized as $\dot{N}(>E_{\rm obs,i}) = 1/t_{\rm obs,i}$, where $t_{\rm obs,i}$ is the observing time corresponding to the $i$-th burst (see Table \ref{Table7}). The best fit distribution for the CED of the repeating FRB is found to be: $\dot{N}_0\, {\rm exp}(-E_{\rm obs}/E_0)$ with $(\dot{N}_0,E_0) = (0.283\ {\rm hr^{-1}},4.0\times10^{39}\ {\rm erg})/(0.247\ {\rm hr^{-1}},4.0\times10^{39}\ {\rm erg})/(0.210\ {\rm hr^{-1}},5.0\times10^{39}\ {\rm erg})/(0.090\ {\rm hr^{-1}},6.0\times10^{39}\ {\rm erg})$ for Parkes $S_{\rm th} = 0.10/0.13/0.20/0.40\ {\rm Jy}$. 

For each FRB, the threshold energy corresponding to a fluence completeness threshold $\mathcal{F}_{\rm th}$ is $E_{\rm th}(z_i) = 4\pi D^{2}(z_i) \mathcal{F}_{\rm th} (1+z_i) (\nu_{\rm max}^{\prime} - \nu_{\rm min}^{\prime})$. The average number of repeating events with energy $E \geq E_{\rm th}(z_i)$ within time $t_{\rm obs,i}$ for a burst at redshift $z_i$ is
\begin{align}
\overline{N}_{\rm rep,i} &= \frac{t_{\rm obs,i}}{(1+z_i)} \int_{E_{\rm th}(z_i)}^{\infty} \frac{\dot{N}_0}{E_0}\ {\rm exp}\left(-\frac{E_{\rm obs}}{E_0}\right)dE_{\rm obs}\nonumber \\
&= \frac{\dot{N}_0\ t_{\rm obs,i}}{(1+z_i)}\ {\rm exp}\left[-\frac{E_{\rm th}(z_i)}{E_0}\right].
\label{Nrep}
\end{align}
If all other FRBs repeat at the same rate $\dot{N}_0$ as FRB 121102, the probability of observing none of the detected bursts to be repeating for a given value of $\mathcal{F}_{\rm th}$ is
\begin{align}
{\rm P}(\mathcal{F}_{th}) &= {\displaystyle \prod_{i=1}^{13} {\rm exp}(-\overline{N}_{rep,i})} \nonumber \\
&= {\rm exp}\left[- \sum_{i=1}^{13} \frac{\dot{N}_0\ t_{obs,i}}{(1+z_i)}\ {\rm exp}\left(-\frac{E_{th}(z_i)}{E_0}\right)\right]
%{\rm P}(\mathcal{F}_{\rm th}) &= {\displaystyle \sum_{j=1}^{13} {\rm exp}(-\overline{N}_{\rm rep,j})\overline{N}_{\rm rep,j} \prod_{i \neq j, i=1}^{13} {\rm exp}(-\overline{N}_{\rm rep,i})} \nonumber \\
%&= {\displaystyle \sum_{j=1}^{13} \overline{N}_{\rm rep,j} \prod_{i=1}^{13} {\rm exp}(-\overline{N}_{\rm rep,i})}.
%&= \sum_{j=1}^{13} {\rm exp}\left[-\sum_{i \neq j, i =1}^{13} \frac{\dot{N}_0\ t_{\rm obs,i}}{(1+z_i)}\ {\rm exp}\left(-\frac{E_{\rm th}(z_i)}{E_0}\right)\right].
\label{P_Fth}
\end{align}
Here we assume a simplistic case of Poissonian distribution for the arrival times of the detected FRB 121102 sub-bursts. However, \citet{Opp18} have shown earlier that the observed clustering of arrival times for the initial 17 sub-bursts from FRB 121102 \citep{Spitler16,Scholz16} are better modelled with a Weibull distribution. Although the mean repetition rates obtained from both Poisson and Weibull distributions are found to be similar, the probability of non-detection of bursts is much larger during a continuous observation for Weibull distribution of arrival times due to the clustering behaviour of the bursts. The estimates for the repetition rate are therefore expected to be less certain for the Weibull distribution. As a result, the upper limits on the non-repetition probability ${\rm P}(\mathcal{F}_{\rm th})$ are expected to be weaker than what we obtain for a Poissonian distribution here.

The second repeating radio source, FRB 180916.J0158$+$65, that was detected with CHIME shows a periodicity of $\sim$16.35 days in its repetition behaviour \citep{CHIME20}. Recently, \citet{Rajwade20} have also reported possible periodic activity arising from the initially detected repeating FRB 121102 with an estimated $\sim$160 day periodicity timescale. \citet{James19} have studied clustered distributions of the burst arrival times to show that the repetition behaviour of FRB 121102 is possibly more active as compared to that of the apparently non-repeating sources which are relatively infrequent.

In the right-hand panel of Figure \ref{fig8}, we show the variation of ${\rm P}(\mathcal{F}_{\rm th})$ within a range of $\mathcal{F}_{\rm th}$ for the 13 non-repeating FRBs with $t_{\rm obs,i}$ listed in Table \ref{Table7}. We evaluate the ${\rm P}(\mathcal{F}_{\rm th})$ curves from equation (\ref{P_Fth}) for the $(\dot{N}_0,E_0)$ values corresponding to Parkes $S_{\rm th} = 0.10/0.13/0.20/0.40\ {\rm Jy}$. The fluence completeness threshold for the Parkes FRBs was derived to be $\mathcal{F}_{\rm th} \approx 2\ {\rm Jy\ ms}$ by \citet{KP15}. 
In order to obtain an approximate upper limit on ${\rm P}(\mathcal{F}_{\rm th})$, here we assume this fairly optimistic choice of $\mathcal{F}_{\rm th}$ for the non-localized FRBs in Table \ref{Table7}.
We find that the probability of observing none of the 13 non-repeating FRBs that have been followed up with Parkes so far to be repeating lies within the range ${\rm P}(\mathcal{F}_{\rm th}) \approx 0.8-0.9$ for $\mathcal{F}_{\rm th} \approx 2\ {\rm Jy\ ms}$ and Parkes $S_{\rm th} \sim 0.1-0.4\ {\rm Jy}$. 
The sensitivity threshold range corresponds to the range of observed FRBs at Parkes with $w_{\rm obs} \sim 1.0-8.0$. While the actual experimental thresholds might be somewhat different, the qualitative analysis will not change.

The probability that we obtain for observing none of the Parkes FRBs to be repeating is significantly larger than the probability $\sim 0.05-0.3$ obtained by \citet{LK16} for a smaller sample size and further supports repeating FRB 121102 as being representative of the entire FRB population. Although the follow-up observing times $t_{\rm obs,i}$ have increased for the non-repeating FRBs now, the exponential EDF for the repeating FRB 121102 coupled to the difference in the mean $E_{\rm obs}$ for the non-repeating and FRB 121102 bursts by almost three orders of magnitude implies that the other FRBs need to be followed up for significantly longer before concluding in favour of distinct FRB populations. It should however be noted that $t_{\rm obs,i}$ for the FRBs are spread over different telescopes with a range of observing frequencies and sensitivities, and a more rigorous analysis regarding the repeatability of the FRBs would involve using a uniform sample which is possible once more bursts are followed up in the future.

%\SmallEntryGap
%\EntryGap
\section{Summary and Conclusions}
\label{sec6}
In this paper, we presented a method to study the source properties of the observed non-repeating and repeating FRBs and used MC simulations to constrain the properties of the FRB source, its host galaxy and the intervening turbulent medium from the current observational data. Although the physical origin of these events is still a matter of open debate with no concrete information about the progenitor model and/or radiation mechanism known at present, the population modelling of the FRB parameters helps in extracting useful information regarding the physical properties of these radio bursts from their observations. In this study, we have restricted our sample to non-repeating Parkes FRBs with $DM_{\rm tot}$ exceeding $500\ {\rm pc\ cm^{-3}}$ and derived their true properties self-consistently from the observations without assuming any initial distributions for the FRB parameters. 
%Although more than 20 FRBs have been reported since then, including data from the newly reported bursts will not affect the conclusions of this work significantly. 
%and make observational predictions as the FRB population is expected to grow rapidly over the next few years with many radio transient surveys collecting more data. 

We estimated the individual burst distances and intrinsic pulse widths by assuming a fixed host galaxy DM contribution and two scattering models for the pulse broadening due to multipath propagation through ionized plasma, respectively. While $w_{\rm ISM}$ is suppressed relative to $w_{\rm IGM}$ by the geometrical lever-arm factor $\sim 4f(1-f)$ for both scattering models, $w_{\rm IGM}$ for model 2 is based on a theoretical model for IGM turbulence as opposed to an observationally established empirical fit for model 1. After computing $S_{\rm peak,int}$ from $w_{\rm int}$ and $\mathcal{F}_{\rm obs}$, we obtained the bolometric luminosity and energy for the bursts for a flat FRB energy spectrum with coherent emission within frequency range $\nu_{\rm min}^{\prime} = 300\ {\rm MHz}$ to $\nu_{\rm max}^{\prime} = 8\ {\rm GHz}$. 
%We obtained chi-squared fits for the cumulative distributions of $w_{\rm int}$ and $L_{\rm int}$ of the FRBs, and used them to constrain the physical properties of FRBs with our MC code.
In our analysis, we have obtained the best fit cumulative distributions for $w_{\rm int}$ and $L_{\rm int}$ by including the biases due to the relatively uncertain $DM_{\rm host}$ contribution to $DM_{\rm Ex}$, assumptions about the $DM_{\rm IGM/ISM}$ dependence of $w_{\rm IGM/ISM}$ from the scattering models considered and the peak flux density reduction due to the finite beam size of the telescope.
%included the biases in the $w_{\rm int}$ and $L_{\rm int}$ best fit distributions 

It should be noted that a larger host galaxy DM contribution would result in a smaller inferred $z$ from equation (\ref{DMtot}) leading to a correspondingly small DM contribution from the IGM. The IGM scatter broadening $w_{\rm IGM}$ will then decrease while $w_{\rm ISM,host}$ from the host galaxy ISM increases. As $w_{\rm ISM,host}$ is suppressed significantly relative to $w_{\rm IGM}$ by the geometrical lever-arm factor, there is a net increase in $w_{\rm int}$ with increase in $DM_{\rm host}$. However, as $w_{\rm IGM}$ is almost two orders of magnitude smaller compared to $w_{\rm int}$ for both populations of FRBs (see Figure \ref{wcompNRR}), the resultant increase/decrease in $w_{\rm int}/S_{\rm peak,int}$ is negligible and the reduction in the inferred $L$ and $E$ values from equation (\ref{EdistplLE}) can be ignored. Therefore, the $w_{\rm int}$ and $L_{\rm int}$ distributions derived for a typical MW-like host galaxy with $DM_{\rm host} \approx 100\ {\rm pc\ cm^{-3}}$ in Section \ref{sec2} will not change appreciably. Similarly, the assumption of a flat FRB energy spectrum with $\alpha \approx 0$ within the frequency range $\nu_{\rm min}^{\prime} = 300\ {\rm MHz}$ to $\nu_{\rm max}^{\prime} = 8\ {\rm GHz}$ does not affect the inferred luminosity values significantly relative to the $\alpha \approx -1.4$ case for the Kolmogorov turbulence spectrum. 

In this study, we have used lognormal $w_{\rm int}$ (W1/2: mean $\mu_{1/2} = 0$ and standard deviation $\sigma_{1/2} = 0.25/0.50$) and power-law $L_{\rm int}$ (L1/2: PL index $\alpha_{L,1/2} = -1/-2$) as model input distributions to our MC code in order to constrain the physical properties of the observed FRBs. 
The distances to the simulated bursts are initially determined from the FRB spatial density (NE/SFH/PL) and the IGM contributions to the DM and width of the pulse are computed using $z$. The host galaxy DM contribution is obtained by assuming it to be a MW-like galaxy with the FRB source location similar to the Solar system and scaling $DM_{\rm NE2001}$ with the parameter $\beta \sim 0.1-10$. The telescope beam center flux density is then obtained for a PL FRB energy density $E_{\nu^{\prime}} = k\nu^{\prime \alpha}$, observing frequency bandwidth $(\nu_1,\nu_2) = (\nu_0 - 0.5 \nu_{\rm bw},\nu_0 + 0.5 \nu_{\rm bw})$ and FRB coherent emission frequency range $(\nu_{\rm min}^{\prime},\nu_{\rm max}^{\prime}) = (300\ {\rm MHz},8\ {\rm GHz})$ from equation (\ref{Speak_bc}). We modelled the flux degradation due to finite telescope beam size using a Gaussian beam profile to obtain the $S_{\rm peak,obs}$ from equation (\ref{Speak_obs}). The $S_{\rm peak,obs}$ dependent $S/N$ is computed for every simulated burst and the FRB is detected if its flux density exceeds the telescope sensitivity threshold. We only consider simulated FRBs with $DM_{\rm tot} \geq 500\ {\rm pc\ cm^{-3}}$ for comparison with the observed sample to minimize the error in the estimates of the inferred parameters that are obtained by assuming a specific FRB source location and host galaxy structure.

We compare the properties of the simulated non-repeating FRBs/FRB 121102 bursts with those observed at Parkes/Arecibo in order to constrain the host galaxy DM relative to MW $\beta$, PL energy density spectral index $\alpha$, scattering in the intervening turbulent plasma and the spatial density $n(z)$ of the FRB sources. Lastly, we discuss whether repeating FRB 121102 is representative of the entire FRB population based on its repetition rate $\dot{N}_0$ and a universal EDF. In the following, we summarise the main results of this work: 

\begin{enumerate}[leftmargin=*]

{\setlength\itemindent{0pt} \item 

The $w_{\rm int}$ for non-repeating (repeating) FRBs varies within a broad range $\sim 0.3-10\ {\rm ms}$ ($\sim 0.1-8\ {\rm ms}$) and the cumulative width distribution is an exponential function with a cutoff $w_{\rm int,c} \sim 2.0\ {\rm ms}$ ($\sim 1.6\ {\rm ms}$). 
However, capturing voltage data and sampling with higher temporal resolution has recently resulted in the detection of much smaller pulse intrinsic widths $w_{\rm int} \lesssim 30\ {\rm \mu s}$ \citep{Michilli18,Farah18}, which indicates that both non-repeating and repeating FRBs likely generate significantly narrower bursts than previously expected thereby affecting the $w_{\rm int}$ distribution.
%Although the observed width distribution already spans several orders of magnitude, the true minimum and maximum possible widths for FRBs are not yet known. 
%The narrowest FRB single pulse yet measured is from FRB 121102 observed by Michilli et al. (2018a) to have a width of <=30 µs, although a sub-pulse of FRB 170827 revealed through voltage capture was measured to be 7.5 µs in duration (Farah et al., 2018). 
%FRBs discovered with much smaller inferred (or measured) intrinsic widths due to the new ability to capture voltage data and sample at higher time resolution. 
%---- can impact the conclusions of this work and their relevance to being an accurate representation of the observed population 
The $L_{\rm int}$ for non-repeating (repeating) FRBs varies within $\sim 10^{43} - 10^{46}\ {\rm erg/s}$ ($\sim 10^{41}-10^{43}\ {\rm erg/s}$) with an exponential cumulative distribution and cutoff $L_{\rm int,c} \sim 4.0\times10^{44}\ {\rm erg/s}$ ($\sim 2.7\times10^{42}\ {\rm erg/s}$). 

The ISM contribution to the width broadening is significantly suppressed in comparison to $w_{\rm IGM}$ due to the geometry of the scattering medium along the line of sight to the FRB source with $w_{\rm ISM,MW} \lesssim 10^{-3}\ {\rm ms}$ and $w_{\rm ISM,host} \lesssim 10^{-6}\ {\rm ms}$ for non-repeating FRBs and $w_{\rm ISM,host/MW} \lesssim 10^{-4}\ {\rm ms}$ for repeating FRBs. 
As a result, the pulse width broadening due to scattering $w_{\rm sc} \approx w_{\rm IGM}$ for both classes of bursts. The scatter broadening of the pulse is found to be the smallest contribution to the $w_{\rm obs}$ with $w_{\rm sc} \lesssim 1\ {\rm ms}$ ($w_{\rm sc} \lesssim 2\times10^{-2}\ {\rm ms}$) for the non-repeating (repeating) bursts.

We find that $w_{\rm int}$ is largely scattering model independent for both classes of FRBs, and the average relative temporal broadening $\Delta w_{\rm int}/w_{\rm int} \sim 150\%$ and $\sim 20\%$ for non-repeating and repeating FRB 121102 bursts, respectively. While $w_{\rm DM}$ is the dominant contribution to the temporal broadening for non-repeating FRBs with $w_{\rm obs} \sim w_{\rm int} \sim w_{\rm DM} \gg w_{\rm sc}$, the dispersive smearing in case of FRB 121102 bursts is significantly smaller with $w_{\rm obs} \approx w_{\rm int} \gg w_{\rm DM} \gg w_{\rm sc}$. Due to the small $z \approx 0.19273$ for FRB 121102, $w_{\rm sc}$ and $w_{\rm DM}$ contributions are found to be almost negligible and a considerable fraction of $w_{\rm obs}$ is expected to come from $w_{\rm int}$.} %\\    

{\setlength\itemindent{0pt} \item To constrain the physical parameters of the underlying FRB population, we consider lognormal $w_{\rm int}$ and power-law $L_{\rm int}$ distributions in this work. For the parametric space that we study here, we find that the simulated observable FRB properties exhibit a relatively weak dependence on the specific choice of the intrinsic model distributions. The scattering models 1 and 2 for the pulse temporal broadening %due to multipath propagation 
cannot be distinguished statistically from our Parkes sample. In this analysis, we consider NE, SFH and PL spatial density models for the detected FRB sources. In agreement with the previous studies (see \citealt{Caleb16} and \citealt{Rane17}), we find that a sample of $\lesssim$ 50 Parkes FRBs is insufficient in order to clearly distinguish between the NE and SFH FRB spatial density models.  

For NE/SFH FRB spatial density, the DM contribution from the host galaxy of the FRB source is expected to be roughly comparable to the Galactic contribution with $\beta \sim 0.1-1$. Furthermore, the Parkes observations for the non-repeating FRBs favour a large negative value of the FRB energy density spectral index $\alpha$ within the range -3.0 to -1.5. 
We also compared the simulated FRB parameters with the Parkes data to constrain the peak redshift and low/high-$z$ indices of the power-law FRB spatial density model. The spatial density of FRBs is likely to be a PL distribution peaking at slightly larger redshifts $z_{\rm crit} \sim 2.0$ compared to the cosmic SFH. The FRB density is expected to increase up to $z \approx z_{\rm crit}$ with a PL index $\alpha_l \sim 0-3$ and then drop considerably at larger distances with $\alpha_u \approx -3$.}  %\\     %\\ \\ \\
%For the lognormal $w_{\rm int}$ and the power-law $L_{\rm int}$ distributions that we considered in this work, 
%we find that the simulated observable FRB properties exhibit a relatively weak dependence on the intrinsic model distributions, particularly for the parametric space that we study here.  
%Furthermore, we find that a larger observed sample of Parkes FRBs at $\nu_{\rm obs} = 1.4\ {\rm GHz}$ is required in order to clearly distinguish between the FRB spatial density models (NE and SFH) 
%as well as the pulse scatter broadening models due to multipath propagation (model 1 and model 2).

%We find that the {\bf NE spatial density is preferred over the SFH spatial density} based on the current Parkes observations reported at $\nu_{\rm obs} = 1.4\ {\rm GHz}$. 
%We find that the NE and SFH FRB spatial densities cannot be differentiated with a significant degree of confidence from the current Parkes observations at $\nu_{obs} = 1.4\ {\rm GHz}$. 
%However, the current FRB observations from Parkes do not provide a sufficiently large data sample in order to distinguish substantially between the two IGM scattering broadening models considered here. 

{\setlength\itemindent{0pt} \item We used the published FRB follow-up observing data in order to investigate whether FRB 121102 is representative of the entire population. The CED for the repeating FRB was computed by only including the FRB 121102 bursts for which $S_{\rm peak,obs}$ exceeds the Parkes $S_{\rm th}$. We obtained an exponential CED $\dot{N}_0 {\rm exp}(-E_{\rm obs}/E_0)$, with repetition rate $\dot{N}_0 \sim 0.090-0.283\ {\rm hr^{-1}}$ and cutoff energy $E_{0} \sim (4-6)\times10^{39}\ {\rm erg}$ for Parkes $S_{\rm th} \sim 0.1-0.4\ {\rm Jy}$. We find that if all FRBs repeat at the same rate $\dot{N}_0 \sim 0.090-0.283\ {\rm hr}^{-1}$ and with a universal EDF, the probability of observing none of them to be repeating is $\sim 0.8-0.9$ for $\mathcal{F}_{\rm th} \approx 2\ {\rm Jy-ms}$ and Parkes $S_{\rm th} \sim 0.1-0.4\ {\rm Jy}$. As the universal EDF is an exponential distribution with a cutoff energy that is much smaller compared to the typical non-repeating FRB energies, significantly longer FRB follow-up observations are needed to distinguish between the FRB populations.}

\end{enumerate}

The population synthesis methods that are presented in this work provide useful statistical insights into the physical properties of the underlying source population. 
As the observed FRB sample grows significantly from a few initial detections to hundreds of new bursts reported every year, this analysis will serve as a powerful tool to investigate the source population which is possibly much larger and more diverse in comparison to the observed population. 
The distribution of the observed properties can be directly influenced by the intrinsic source properties (spectral behaviour, redshift and energy distribution), propagation effects (dispersive smearing, scattering and scintillation) as well as selection effects (detection frequency, telescope sensitivity, spectral and temporal resolution). 
%As the sample of reported FRBs will continue to expand rapidly with several broad-band radio surveys becoming operational, the methods presented in this work can be applied in the near future

Currently, performing a robust comparison of FRBs detected with different surveys is difficult essentially due to the instrumental selection effects. 
In a future work, we will extend our analysis to a more diverse and inhomogeneous sample of non-repeating and repeating bursts that are detected with CHIME, ASKAP and UTMOST to investigate the interplay between underlying source properties, instrumental selection biases and propagation effects. 
A detailed maximum likelihood analysis will help to place tighter constraints on the spectral behaviour, redshift and energy distribution for FRB sources as well as the properties of the host galaxy and burst local environment. In particular, mapping the evolution of propagation effects such as scattering and dispersive delay that are significantly more prominent at lower radio frequencies will allow us to disassociate the intrinsic pulse properties from the extrinsic propagation and selection effects. 

Recent localisations of repeating FRBs in spiral/dwarf galaxies with higher star formation rates as opposed to the massive host galaxies of the apparently non-repeating sources naturally raises the question as to whether these two classes of bursts fundamentally originate from different types of source environments and/or host galaxies. 
With the coherent de-dispersion of raw voltage data and high-resolution pulse fits now available from various FRB surveys \citep{Michilli18,Farah18,Ravi19b}, we can probe the sub-millisecond temporal structures in the dynamic spectra \citep{Hessels19} in addition to the clustered repetition rates of the repeating events \citep{Gourdji19} and their relation to the observed temporal pulse widths \citep{CHIME19b}. 
Furthermore, the localised bursts and their association to host galaxies has potential cosmological applications that include studying the baryonic density distribution, IGM turbulence properties and tracking the reionisation history of the Universe. 
As scattering widths directly correlate with the IGM DM contribution, one can further probe the host galaxy DM distribution and establish independent cosmic distance measures based on the localised sources.

\section*{Acknowledgments}
%\begin{comment}
We thank the anonymous reviewer for providing detailed comments and suggestions which improved the analysis presented in this work. We thank Duncan Lorimer, Siddhartha Bhattacharyya, Akshaya Rane, Wenbin Lu, Somnath Bharadwaj and Apurba Bera for useful discussions.

%\end{comment}

\begin{appendix}

\section{Pulse temporal broadening due to IGM turbulence}
\label{AppendixA}
Here we derive the expression (equation \ref{wIGM2}) for the pulse temporal smearing due to IGM turbulence for the theoretical model proposed by \citet{MK13} [henceforth MK13]. Substituting the scattering broadened angular image size $\theta_{\rm sc} = f D_{LS}/D_S k\, r_{\rm diff}$ from equation (13) of MK13, the IGM temporal smearing $w_{\rm IGM}$ from equation (15) of MK13 can be written as
\begin{eqnarray}
w_{\rm IGM} = \frac{D_L D_S \theta_{\rm sc}^{2}}{c D_{LS} (1+z_L)} = \frac{f^2 \lambda_0^2}{c (1+z_L)} \frac{D_{\rm eff}}{4\pi^2 r_{\rm diff}^2}
\label{w_IGMapp}
\end{eqnarray}
where $D_L$/$D_S$/$D_{LS}$ is the angular diameter distance from the observer to the scattering region/observer to the source/scattering region to the source, $z_{L}/z_{S}/z_{LS}$ is the corresponding redshift, $D_{\rm eff} = D_L D_{LS}/D_S$ and $\lambda = 2\pi/k$ is the wavelength in the observer frame.
We consider the case when the diffractive length scale $r_{\rm diff}$ is smaller compared to the inner scale of the scattering region $l$ with the constant $f_K =1.18$ to obtain $r_{\rm diff} = (8.0\times10^9\ {\rm m})\ \lambda_0^{-1}\ {\rm SM_{\rm eff,0}^{-1/2}}\ l_0^{1/6}$ from equation (10a) of MK13, where $\lambda_0 = \lambda/(1\ {\rm m})$, ${\rm SM_{\rm eff,0}} = {\rm SM_{\rm eff}}/(10^{12}\ {\rm m}^{-17/3})$ and $l_0 = l/(1\ {\rm AU})$. The constant $f_K$ is directly associated with the power-law index $\beta_{K}$ for Kolmogorov turbulence spectrum with $f_K =1.18$ corresponding to $\beta_K = 4$ and $r_{\rm diff} < l$. 

Substituting equation (20) of MK13 for the density fluctuation amplitude from turbulence into equation (23) of MK13, the $z$-corrected effective scattering measure can be written as
\begin{eqnarray}
{\rm SM_{\rm eff}}(z) = K_1 \int_0^z (1+z^{\prime})^{3} d_{H}(z^{\prime})dz^{\prime}
\label{SMeff}
\end{eqnarray}
where $d_H(z^{\prime}) = (c/H_0)[\Omega_m (1+z^{\prime})^3 + \Omega_{\Lambda}]^{-1/2}$ and $K_1 \approx (9.42\times10^{-14}\ {\rm m^{-20/3}})$ is constant for the scattering region outer scale $L \sim 1\ {\rm pc}$ and Kolmogorov turbulence spectrum. Further substituting $r_{\rm diff}$ in terms of ${\rm SM_{\rm eff}}$ into equation (\ref{w_IGMapp}) and rewriting $\lambda_0 = c/\nu_0$ gives the temporal smearing timescale (in sec)
\begin{eqnarray}
w_{\rm IGM}(z) = K_2 \frac{D_{\rm eff}}{(1+z_L)} \frac{{\rm SM_{\rm eff}}(z)}{\nu_0^4}
\label{w_IGMapp2}
\end{eqnarray}
where $K_2 \approx (1.56\times10^{-32}\ {\rm s}) (f_K^2 c^3/4\pi^2) (l_0/1\ {\rm AU})^{-1/3}$ is constant and $\nu_0$ is the wave frequency in Hz in the observer frame.

Next in order to simplify $w_{\rm IGM}(z)$ from equation (\ref{w_IGMapp2}) in terms of just the FRB source redshift $z$, we need to obtain $D_{\rm eff}$ directly from cosmology. It is expected that the pulse temporal smearing will be maximized when the scattering screen is placed symmetrically midway along the source to the observer line of sight \citep{Vandenberg76,Lorimer13}. Therefore, in order to further simplify the calculations and obtain the maximum expected value of $w_{\rm IGM}$ for a given FRB line of sight, we use $z_L \approx (1/2)z$ with the condition on comoving distances
\begin{eqnarray}
D_L (1 + z_L) = D_{LS} (1 + z_{LS}) = 0.5 D_S (1 + z). \nonumber 
\end{eqnarray}
Lastly, substituting $D_{\rm eff} = D_L D_{LS}/D_S$ and $SM_{\rm eff}(z)$ from equation (\ref{SMeff}) in equation (\ref{w_IGMapp2}) gives
\begin{eqnarray}
w_{\rm IGM}(z) = \frac{k_{\rm IGM}}{\nu_{\rm 0,GHz}^{4}Z_{L}} \int^{z}_{0}\frac{dz^{\prime}}{\left[\Omega_{m}(1+z^{\prime})^3 + \Omega_{\Lambda}\right]^{0.5}} \nonumber \\
\times \int^{z}_{0}\frac{(1+z^{\prime})^{3}}{\left[\Omega_{m}(1+z^{\prime})^3 + \Omega_{\Lambda}\right]^{0.5}}dz^{\prime}
\end{eqnarray}
where $k_{\rm IGM}$ is the redshift independent normalisation factor including $K_1$ and $K_2$, $\nu_{\rm 0,GHz} = \nu_0/10^{9}$ and $Z_{L} = 1 + (1/2)z$.

\section{Parameter correlation for FRBs}
\label{AppendixD}

\begin{figure*}
%\vspace{-10em}
\gridline{\fig{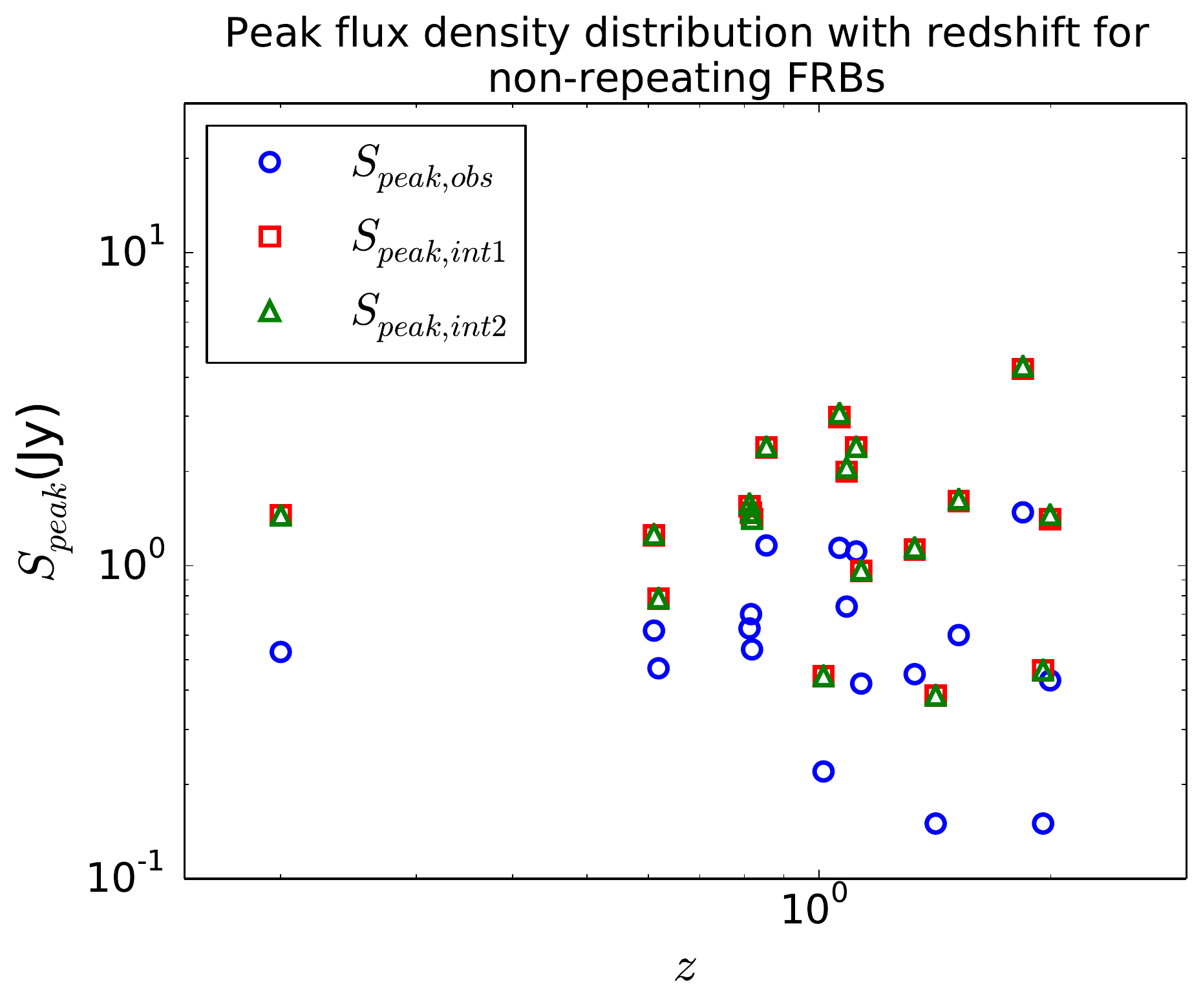}{0.46\textwidth}{}
          \fig{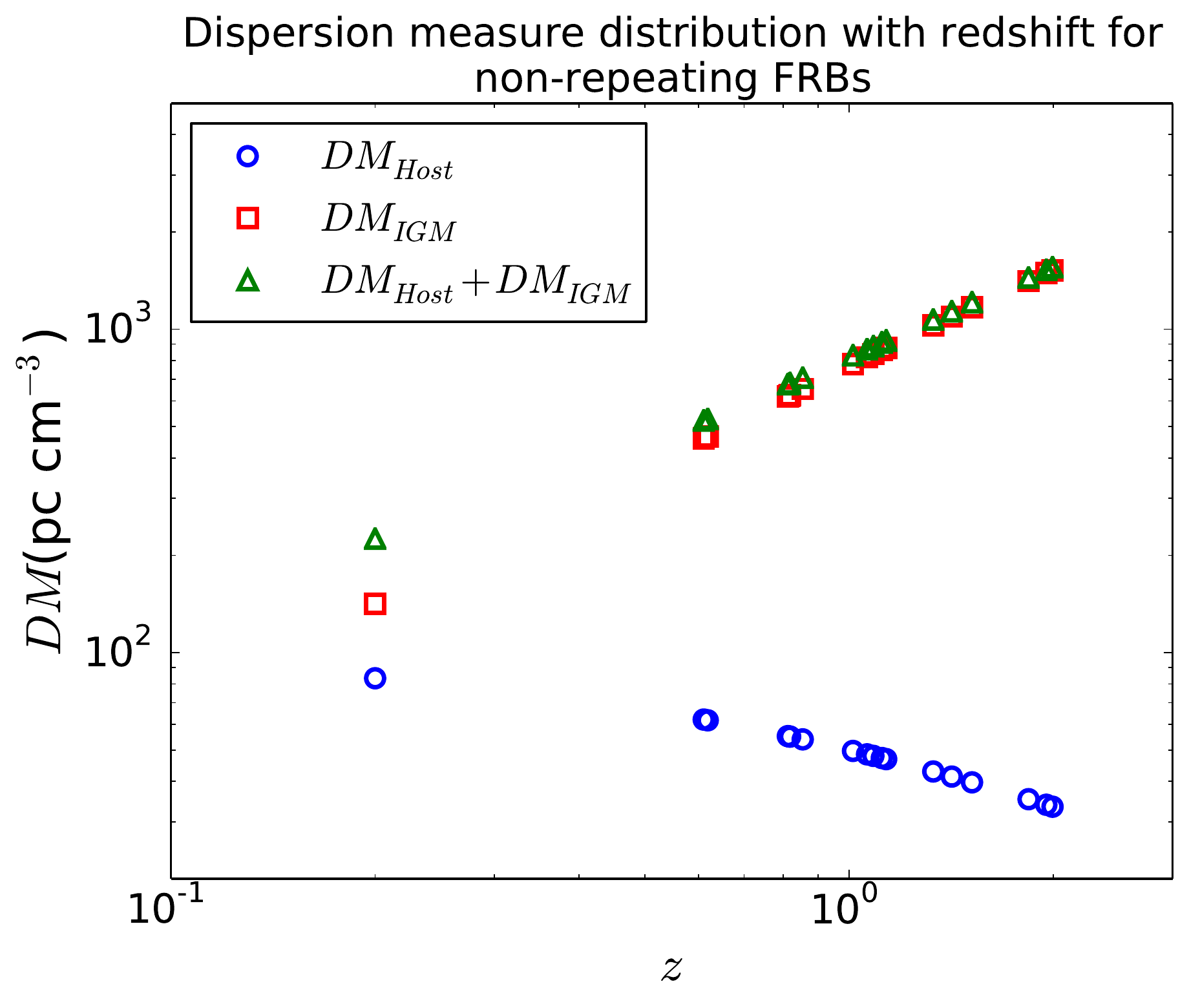}{0.46\textwidth}{}
          }  \vspace{-2.5em}   
\gridline{\fig{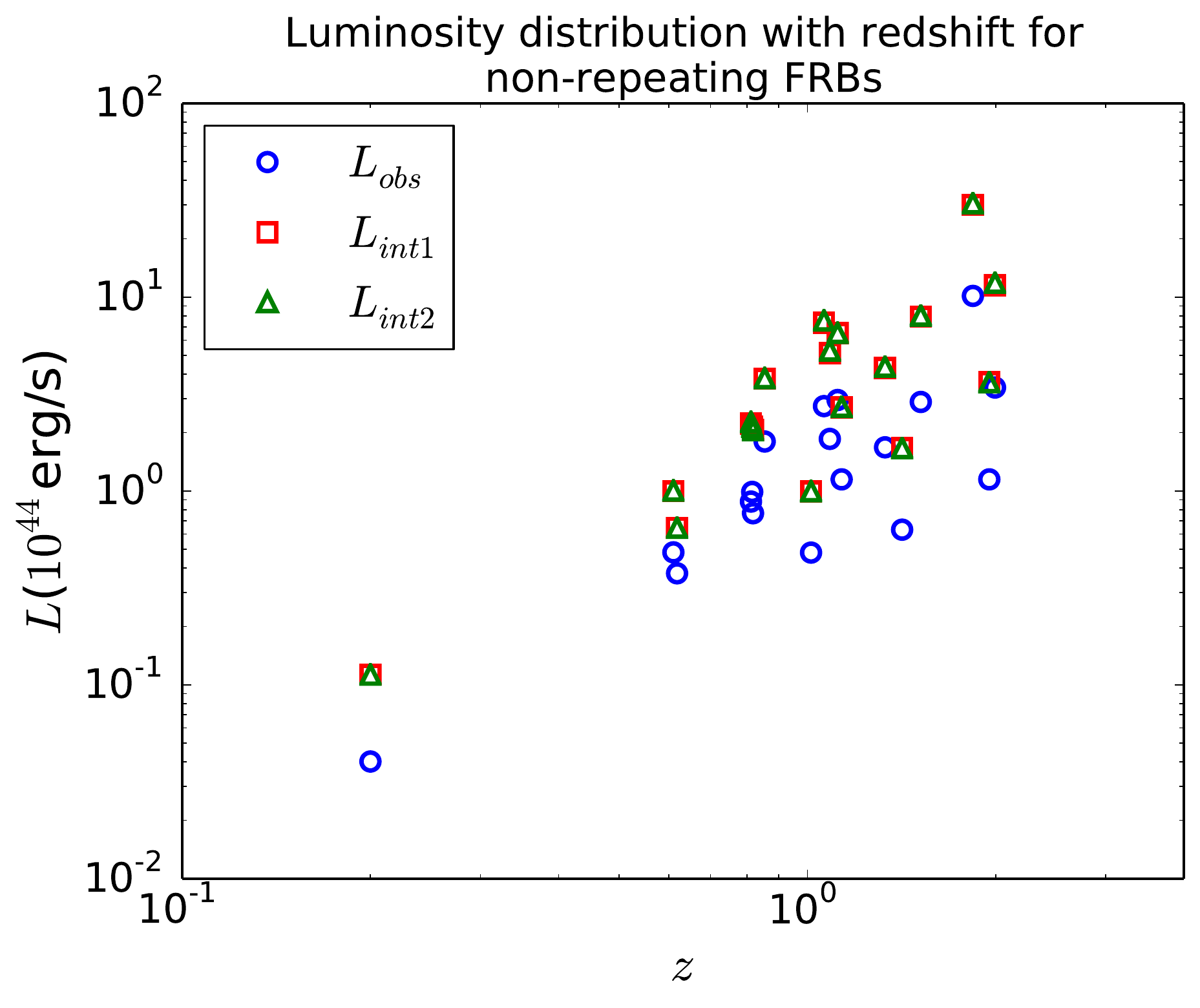}{0.46\textwidth}{}
          \fig{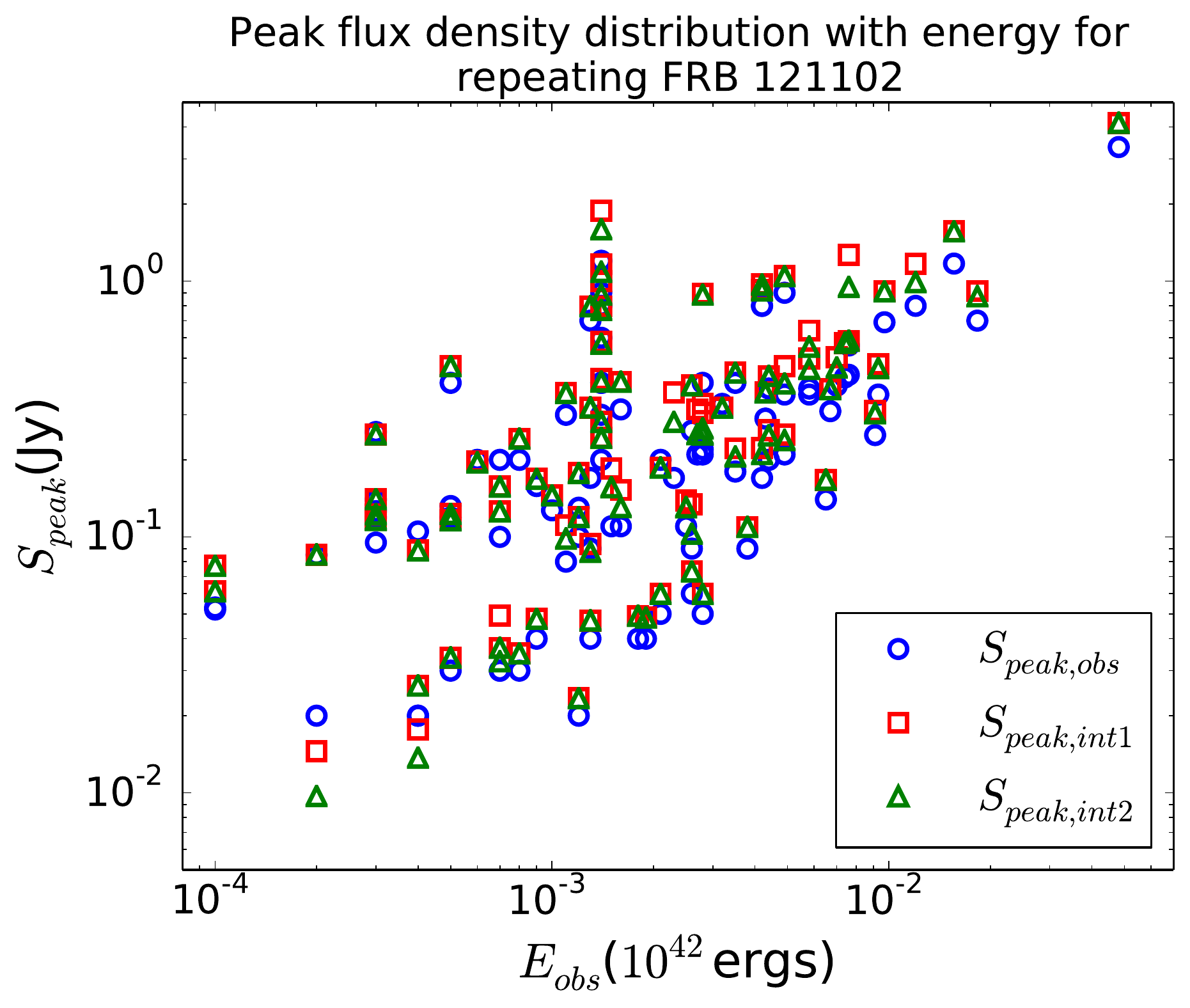}{0.46\textwidth}{}
          }  \vspace{-2.5em}            
  \caption{\emph{Variation of burst parameters with distance for non-repeating bursts and with energy for FRB 121102:} 
	{\it Top-left panel:} Dependence of peak flux densities $S_{\rm peak}$ on distance for Parkes FRBs, 
	{\it Top-right panel:} Dependence of dispersion measure $DM_{\rm tot}$ on distance for Parkes FRBs,
	{\it Bottom-left panel:} Dependence of luminosities $L$ on distance for Parkes FRBs, 
	{\it Bottom-right panel:} Dependence of peak flux densities $S_{\rm peak}$ on burst energy for FRB 121102.	
}%
	\label{parcorNRR}
\end{figure*}

Once the burst parameters are estimated from the FRB observables, we can study the correlation between different parameters and the burst distance/energy for the non-repeating/repeating FRBs. We show the dependence of $S_{\rm peak}$, $DM$ and $L$ on $z$ for the non-repeating Parkes FRBs in the top-left, top-right and bottom-left panels of Figure \ref{parcorNRR}. We find that both the observed and intrinsic $S_{\rm peak}$ of these events have no apparent correlation with the inferred distances. However, the relative scatter in the flux values for a given distance is somewhat larger for intermediate $z \sim 1$. 
This can be due to the fixed $DM_{host}$ contribution that we assume in order to infer FRB distances which possibly leads to a more complete sample around $z \sim 1$ thereby exhibiting larger scatter in the observed data points. 
While the IGM contribution to the $DM$ is found to be comparable to the host galaxy contribution for small $z \approx 0.2$, $DM_{\rm IGM}/DM_{\rm host} \gtrsim 10$ for larger $z \gtrsim 0.6$.

As most of the currently detected bursts have $z \gtrsim 0.6$, small variations in $DM_{\rm host}$ are not expected to significantly affect $DM_{\rm Ex} = DM_{\rm host} + DM_{\rm IGM}$, provided that the typical host galaxy properties are not very different from that of the MW. The burst luminosities increase on an average with an increase in the burst distance, which is expected as $S_{\rm peak}$ for FRBs is almost independent of $z$. 
Moreover, the bursts with higher $L_{\rm int}$ are easier to detect from larger distances compared to dimmer FRBs, for a given telescope sensitivity. 
%Due to the apparent positive correlation of the burst luminosities with their inferred distances, FRBs are not expected to be standard candles as previously expected. 
In the bottom-right panel of Figure \ref{parcorNRR}, the dependence of $S_{\rm peak}$ on $E_{\rm obs}$ is shown for the reported sub-bursts of repeating FRB 121102. We find that the more energetic sub-bursts have a larger value of $S_{\rm peak}$ on average, which is reasonable as brighter bursts detected from a given distance are expected to emit more energy.

\section{FRB intrinsic distributions and KS analysis}
\label{AppendixC}
Here we list the best fit distribution parameters for both non-repeating Parkes FRBs and FRB 121102 bursts as well as the KS test p-values obtained from the comparison between simulated and observed FRB population at Parkes. 
Table \ref{Table2} lists the functional fit parameters with the corresponding chi-squared values for the pulse width and luminosity of the non-repeating FRBs. 
The repeating FRB 121102 width and luminosity distribution fit parameters with the chi-squared values are listed in Table \ref{Table3}. %\\ \\ \\
Table \ref{Table5} lists the KS test p-values $\gamma$ obtained from the comparison of the simulated parameters for NE, SFH and PL spatial densities with those from the observed population at Parkes.
The simulations are done for host galaxy DM contribution $\beta = 0.1,1.0,10.0$, energy density spectral index $\alpha=-3.0,-1.5,2.0$ and scattering model 2. 
While we consider four different intrinsic width and luminosity models (w1L1, w1L2, w2L1 and w2L2) for NE and SFH $n(z)$, we only consider w1L1 for PL $n(z)$ as the relative difference between the intrinsic models is found to be negligible.

%\EntryGap
\begin{table*}
%\label{Table2}
\begin{center}
\caption{\small Power-law, exponential and gaussian fit parameters for the width and luminosity distributions of Parkes FRBs}
\label{Table2}
\bgroup
\def\arraystretch{1.0}
\begin{tabular}{ccccccccccccccccccccc}
\hline
\hline
%\hline
\centering
Width & Distribution & Functional fit & Reduced $\chi^2$ \\ \hline \hline
%\hline
Observed & Power-law & $(18.812\pm1.275)\ w_{\rm obs}^{-0.615\pm0.051}$ & 0.439 \\ \hline
		& Exponential & $(19.837\pm1.064)\ e^{-w_{\rm obs}/(5.877\pm0.486)}$ & 0.174 \\ \hline
		& Gaussian & $(14.185\pm1.056)\ e^{-w_{\rm obs}^2/2(5.707\pm0.602)^2}$ & 0.576 \\ \hline
%\hline
Intrinsic (1/2) & Power-law & $(12.145\pm0.302)\ w_{\rm int1}^{-0.771\pm0.035}/(12.037\pm0.293)\ w_{\rm int2}^{-0.768\pm0.034}$ & 0.303/0.299 \\ \hline
		& Exponential & $(21.339\pm1.242)\ e^{-w_{\rm int1}/(2.161\pm0.181)}/(21.188\pm1.236)\ e^{-w_{\rm int2}/(2.166\pm0.183)}$ & 0.171/0.174 \\ \hline		
		& Gaussian & $(14.982\pm1.065)\ e^{-w_{\rm int1}^2/2(2.089\pm0.191)^2}/(14.908\pm1.062)\ e^{-w_{\rm int2}^2/2(2.093\pm0.193)^2}$ & 0.493/0.500 \\ \hline
%\hline \\
%\hline
%\hline
\hline
Luminosity & Distribution & Functional fit & Reduced $\chi^2$ \\ \hline \hline
%\hline
Observed & Power-law & $(7.859\pm0.880)\ L_{\rm obs}^{-0.392\pm0.061}$ & 1.650 \\ \hline
		& Exponential & $(21.673\pm0.870)\ e^{-L_{\rm obs}/(1.607\pm0.084)}$ & 0.105 \\ \hline	
		& Gaussian & $(15.735\pm0.786)\ e^{-L_{\rm obs}^2/2(1.561\pm0.081)^2}$ & 0.257 \\ \hline		
%\hline
Intrinsic (1/2) & Power-law & $(12.855\pm0.816)\ L_{\rm int1}^{-0.293\pm0.045}/(12.895\pm0.810)\ L_{\rm int2}^{-0.291\pm0.045}$ & 0.531/0.520 \\ \hline
		& Exponential & $(20.375\pm0.442)\ e^{-L_{\rm int1}/(4.250\pm0.166)}/(20.322\pm0.432)\ e^{-L_{\rm int2}/(4.306\pm0.166)}$ & 0.026/0.025 \\ \hline		
		& Gaussian & $(16.073\pm0.705)\ e^{-L_{\rm int1}^2/2(3.521\pm0.225)^2}/(16.040\pm0.706)\ e^{-L_{\rm int2}^2/2(3.564\pm0.232)^2}$ & 0.165/0.167 \\ \hline
\hline
\label{wLdistparNR}
\end{tabular}
\egroup
\end{center}
\end{table*}
%\vspace{-0.2cm}

%\EntryGap
\begin{table*}
%\label{Table3}
\begin{center}
\caption{\small Power-law, exponential and gaussian fit parameters for the width and luminosity distributions of FRB 121102}
\label{Table3}
\bgroup
\def\arraystretch{1.0}
\begin{tabular}{ccccccccccccccccccccc}
\hline
\hline
%\hline
\centering
Width & Distribution & Functional fit & Reduced $\chi^2$ \\ \hline \hline
%\hline
Observed & Power-law & $(40.114\pm1.632)\ w_{\rm obs}^{-0.657\pm0.033}$ & 5.255 \\ \hline
		& Exponential & $(93.887\pm0.757)\ e^{-w_{\rm obs}/(2.135\pm0.023)}$ & 0.113 \\ \hline
		& Gaussian & $(71.173\pm1.438)\ e^{-w_{\rm obs}^2/2(1.992\pm0.043)^2}$ & 0.973 \\ \hline
%\hline
Intrinsic (1/2) & Power-law & $(35.788\pm1.389)\ w_{\rm int1}^{-0.629\pm0.030}/(36.731\pm1.413)\ w_{\rm int2}^{-0.633\pm0.031}$ & 4.355/4.359 \\ \hline
		& Exponential & $(94.470\pm0.836)\ e^{-w_{\rm int1}/(1.626\pm0.019)}/(93.264\pm0.743)\ e^{-w_{\rm int2}/(1.773\pm0.019)}$ & 0.126/0.108 \\ \hline		
		& Gaussian & $(71.814\pm1.489)\ e^{-w_{\rm int1}^2/2(1.483\pm0.033)^2}/(70.624\pm1.464)\ e^{-w_{\rm int2}^2/2(1.651\pm0.037)^2}$ & 1.012/1.026 \\ \hline
%\hline \\
%\hline
%\hline
\hline
Luminosity & Distribution & Functional fit & Reduced $\chi^2$ \\ \hline \hline
%\hline
Observed & Power-law & $(2.436\pm0.351)\ L_{\rm obs}^{-0.593\pm0.028}$ & 4.228 \\ \hline
		& Exponential & $(89.749\pm1.013)\ e^{-L_{\rm obs}/(0.020\pm0.001)}$ & 0.215 \\ \hline
		& Gaussian & $(70.157\pm1.817)\ e^{-L_{\rm obs}^2/2(0.017\pm0.001)^2}$ & 1.579 \\ \hline
%\hline
Intrinsic (1/2) & Power-law & $(3.363\pm0.440)\ L_{\rm int1}^{-0.547\pm0.026}/(3.534\pm0.479)\ L_{\rm int2}^{-0.527\pm0.026}$ & 4.443/5.138 \\ \hline
		& Exponential & $(86.171\pm0.987)\ e^{-L_{\rm int1}/(0.027\pm0.001)}/(86.552\pm1.017)\ e^{-L_{\rm int2}/(0.025\pm0.001)}$ & 0.232/0.246 \\ \hline	
		& Gaussian & $(68.485\pm1.939)\ e^{-L_{\rm int1}^2/2(0.022\pm0.001)^2}/(68.203\pm1.967)\ e^{-L_{\rm int1}^2/2(0.021\pm0.001)^2}$ & 1.905/1.971 \\ \hline
\hline
\label{wLdistparR}
\end{tabular}
\egroup
\end{center}
\end{table*}
%\vspace{-0.2cm}

%\begin{comment}
%\EntryGap
%\begin{multicols}{2}
\begin{table*}
%\label{Table5}
%\onecolumn
\begin{center}
\caption{\small \emph{KS test p-values $\gamma$ from the comparison of simulated FRB parameters with the observed FRB population at Parkes.} The top-half of the table lists the values for $w_{\rm obs}$, $S_{\rm peak,obs}$ and $DM_{\rm tot}$ of the NE and SFH non-repeating FRB population. 
The p-values are obtained for different $\beta$ and $\alpha$ combinations with $\gamma_{\rm eq} = \sqrt{\gamma_{w_{\rm obs}}^{2} + \gamma_{S_{\rm peak,obs}}^{2} + \gamma_{DM_{\rm tot}}^{2}}$. 
The p-values for intrinsic width and luminosity models w1L1/w1L2/w2L1/w2L2 are listed for each entry. 
%scattering model 1/2  
The bottom-half of the table lists the p-values for $w_{\rm obs}$, $S_{\rm peak,obs}$ and $DM_{\rm tot}$ for PL population with varying $z_{\rm crit}$ or varying ($\alpha_l$,$\alpha_u$). 
The p-values for the PL population are obtained for cases 1-3 listed in Section \ref{PL_nz}. Only w1L1 intrinsic model is considered for the PL spatial density case. We assume scattering model 2 for all MC simulations. 
%with $\gamma_{\rm eq} = \sqrt{\gamma_{w_{\rm obs}}^{2} + \gamma_{L_{\rm obs}}^{2} + \gamma_{E_{\rm obs}}^{2}}$. 
}
\label{Table5}
\bgroup
\def\arraystretch{0.8}
\begin{tabular}{ccccccccccccccccccccc}
\hline
\hline
\centering
$n(z)$ & $\beta$ & $\alpha$ & $\gamma_{w_{\rm obs}}$ & $\gamma_{S_{\rm peak,obs}}$ & $\gamma_{DM_{\rm tot}}$ & $\gamma_{\rm eq}$ \\ \hline \hline 

NE & 0.1 & -3.0 & 0.653/0.604/ & 0.138/0.132/ & 0.148/0.151/ & 0.395/0.367/ \\
      &       &        & 0.585/0.612 & 0.121/0.145 & 0.158/0.152 & 0.357/0.374 \vspace{0.1cm} \\    
      &  (1.0)  & & (0.754/0.617/ & (0.197/0.169/ & (0.435/0.407/ & (0.515/0.438/ \\
      &           & & 0.709/0.631) & 0.165/0.202) & 0.361/0.392) & 0.469/0.444) \vspace{0.1cm} \\  
      & [10.0] & & [0.531/0.544/ & [0.005/0.007/ & [0.009/0.009/ & [0.307/0.314/ \\
      &           & & 0.553/0.578] & 0.006/0.008] & 0.007/0.012] & 0.319/0.334] \\ \hline     
      &           & -1.5 & 0.506/0.497/ & 0.113/0.089/ & 0.037/0.081/ & 0.300/0.295/ \\  
      &           &        & 0.512/0.525 & 0.061/0.058 & 0.074/0.065 & 0.301/0.307 \vspace{0.1cm} \\  
      &           & & (0.511/0.647/ & (0.164/0.142/ & (0.061/0.079/ & (0.312/0.385/ \\  
      &           & & 0.642/0.589) & 0.148/0.137) & 0.073/0.083) & 0.383/0.352) \vspace{0.1cm} \\    
      &           & & [0.348/0.356/ & [0.031/0.024/ & [$1.341\times10^{-5}$/$3.462\times10^{-5}$/ & [0.202/0.206/ \\     
      &           & & 0.354/0.405] & 0.019/0.021] & $2.374\times10^{-5}$/$2.658\times10^{-5}$] & 0.205/0.234] \\ \hline                           
      &           &  2.0 & 0.019/0.015/ & 0.029/0.021/ & $2.538\times10^{-4}$/$1.264\times10^{-4}$/ & 0.020/0.015/ \\
      &           &        & 0.021/0.019 & 0.017/0.026 & $1.427\times10^{-4}$/$1.053\times10^{-4}$ & 0.016/0.019 \vspace{0.1cm} \\        
      &           & & (0.046/0.031/ & (0.044/0.035/ & ($1.368\times10^{-4}$/$1.794\times10^{-4}$/ & (0.037/0.027/ \\     
      &           & & 0.036/0.029) & 0.037/0.025) & $8.749\times10^{-5}$/$8.042\times10^{-5}$) & 0.030/0.022) \vspace{0.1cm} \\                      
      &           & & [0.011/0.016/ & [0.015/0.011/ & [$5.164\times10^{-7}$/$6.381\times10^{-7}$/ & [0.011/0.011/ \\
      &           & & 0.009/0.017] & 0.009/0.007] & $1.850\times10^{-7}$/$1.294\times10^{-7}$] & 0.007/0.011] \vspace{0.2cm} \\ \hline \hline                                              
SFH & 0.1 & -3.0 & 0.693/0.717/ & 0.351/0.297/ & 0.340/0.373/ & 0.490/0.497/ \\
        &       &        & 0.645/0.667 & 0.274/0.288 & 0.354/0.339 & 0.453/0.463 \vspace{0.1cm} \\        
      &  (1.0)  & & (0.815/0.784/ & (0.179/0.161/ & (0.456/0.512/ & (0.549/0.549/ \\
      &            & & 0.692/0.741) & 0.138/0.140) & 0.503/0.471) & 0.500/0.513) \vspace{0.1cm} \\  
      & [10.0] & & [0.574/0.598/ & [0.002/$9.602\times10^{-4}$/ & [0.031/0.045/ & [0.332/0.346/ \\      
      &           & & 0.558/0.581] & 0.003/0.002] & 0.028/0.025] & 0.323/0.336] \\ \hline                      
      &           & -1.5 & 0.389/0.394/ & 0.079/0.134/ & 0.076/0.085/ & 0.233/0.245/ \\   
      &           &        & 0.457/0.426 & 0.091/0.116 & 0.056/0.061 & 0.271/0.257 \vspace{0.1cm} \\                          
      &           & & (0.382/0.437/ & (0.116/0.134/ & (0.068/0.071/ & (0.234/0.267/ \\ 
      &           & & 0.465/0.431) & 0.172/0.165) & 0.084/0.067) & 0.290/0.269) \vspace{0.1cm} \\                           
      &           & & [0.342/0.355/ & [0.009/0.011/ & [$3.146\times10^{-5}$/$2.962\times10^{-5}$/ & [0.198/0.205/ \\      
      &           & & 0.336/0.407] & 0.008/0.009] & $2.916\times10^{-5}$/$5.043\times10^{-5}$] & 0.194/0.235] \\ \hline                           
      &           &  2.0 & 0.021/0.017/ & 0.032/0.023/  & $3.159\times10^{-4}$/$2.847\times10^{-4}$/ & 0.022/0.017/ \\       
      &           &        & 0.011/0.025 & 0.027/0.016 & $1.727\times10^{-4}$/$4.769\times10^{-4}$ & 0.017/0.017 \vspace{0.1cm} \\                 
      &           & & (0.048/0.042/ & (0.093/0.062/ & ($9.461\times10^{-5}$/$2.293\times10^{-4}$/ & (0.060/0.043/ \\        
      &           & & 0.059/0.063) & 0.085/0.064) & $1.950\times10^{-4}$/$1.183\times10^{-4}$) & 0.060/0.052) \vspace{0.1cm} \\                  
      &           & & [0.008/0.012/ & [0.025/0.021/ & [$9.529\times10^{-7}$/$6.529\times10^{-7}$/ & [0.015/0.014/ \\      
      &           & & 0.011/0.013] & 0.032/0.019] & $6.147\times10^{-7}$/$8.483\times10^{-7}$] & 0.020/0.013] \vspace{0.2cm} \\       
      \hline \hline                  
PL $n(z)$ & ($\alpha_{l}$,$\alpha_{u}$) & $z_{\rm crit}$ & $\gamma_{w_{\rm obs}}$ & $\gamma_{S_{\rm peak,obs}}$ & $\gamma_{DM_{\rm tot}}$ & $\gamma_{\rm eq}$ \\ \hline \hline
Case 1 & (2.7,-2.9)   & 1.00 (2.00) & 0.507 (0.749) & 0.141 (0.198) & 0.552 (0.715) & 0.440 (0.609)\\ 
            & 		        & [3.00]          & [0.526]           & [0.088]	      & [0.063]	      & [0.310] \\ \hline                                                                                                                          
Case 2 & 		        & 		    & 0.693 (0.847) & 0.312 (0.274) & 0.426 (0.869) & 0.503 (0.718)\\ 
            & 		        &                   & [0.670]           & [0.099]	      & [0.147]	      & [0.400] \\ \hline                                                                                              
Case 3 & 		        & 		    & 0.559 (0.618) & 0.009 (0.008)  & 0.003 (0.003) & 0.323 (0.357)\\ 
            & 		        &                   & [0.537]           & [0.006]	      & [0.013]	      & [0.310] \\ \hline \hline                                                  
%Case 5 & 		        & 		    & 0.672 (0.051) & 0.620 (0.094) 			      & 0.822 ($3.834\times10^{-4}$) & 0.710 (0.062)\\ 
%            & 		        &                   & [0.005]           & [$1.798\times10^{-4}$]	      & [$3.138\times10^{-7}$]	      & [0.003] \\ \hline \hline                     
%Case 6 & 		        & 		    & 0.993 (0.256) & 0.003 (0.001) 			      & 0.006 ($3.217\times10^{-5}$) & 0.573 (0.148)\\ 
%            & 		        &                   & [0.028]           & [$1.808\times10^{-4}$]	      & [$1.935\times10^{-7}$]	      & [0.016] \\ \hline \hline     
Case 1 & (0,-3)\{3,0\}   & 1.85 & 0.613 \{0.297\} & 0.113 \{0.132\} & 0.684 \{0.347\} & 0.534 \{0.274\} \\ 
            & [0,0]              &         & [0.346]              & [0.093]	           & [0.406]	      & [0.313] \\ \hline                                      
Case 2 & 		           & 	      & 0.668 \{0.347\} & 0.161 \{0.203\} & 0.766 \{0.546\} & 0.594 \{0.391\} \\ 
            & 		           &         & [0.570]           & [0.179]	      & [0.532]	      & [0.462] \\ \hline                                                
Case 3 & 		           & 	      & 0.322 \{0.268\} & $6.281\times10^{-5}$ \{0.026\} & 0.159 \{0.061\} & 0.207 \{0.159\} \\ 
            & 		           &         & [0.231]           & [$2.198\times10^{-5}$]	       & [0.094]	      	& [0.144] \\ \hline \hline                     
%Case 5 & 		           & 	      & 0.491 \{0.086\} & 0.942 \{0.218\} & 0.389 \{0.002\} & 0.653 \{0.135\} \\ 
%            & 		           &         & [0.120]           & [0.128]	       & [0.013]	      			        & [0.102] \\ \hline \hline            
%Case 6 & 		           & 	      & 0.960 \{0.208\} & 0.002 \{0.001\} & 0.002 \{$6.286\times10^{-5}$\} & 0.554 \{0.120\} \\ 
%            & 		           &         & [0.508]           & [0.002]	       & [$1.203\times10^{-4}$]	       & [0.293] \\ \hline	       
\end{tabular}
\egroup
\end{center}
\end{table*}
%\end{multicols}
%\twocolumn
%\addtocounter{table}{-1}
%\end{comment}

\section{Simulation results for repeating FRB 121102}
\label{AppendixB}
%\subsection{Repeating FRBs}
%For the repeating FRB 121102, the host galaxy DM contribution and scattering model are fixed in order to constrain the energy spectral index $\alpha$ from the Arecibo $\nu_{obs} = 1.4\ {\rm GHz}$ data. 

\begin{figure*}
%\vspace{-10em}
\gridline{\fig{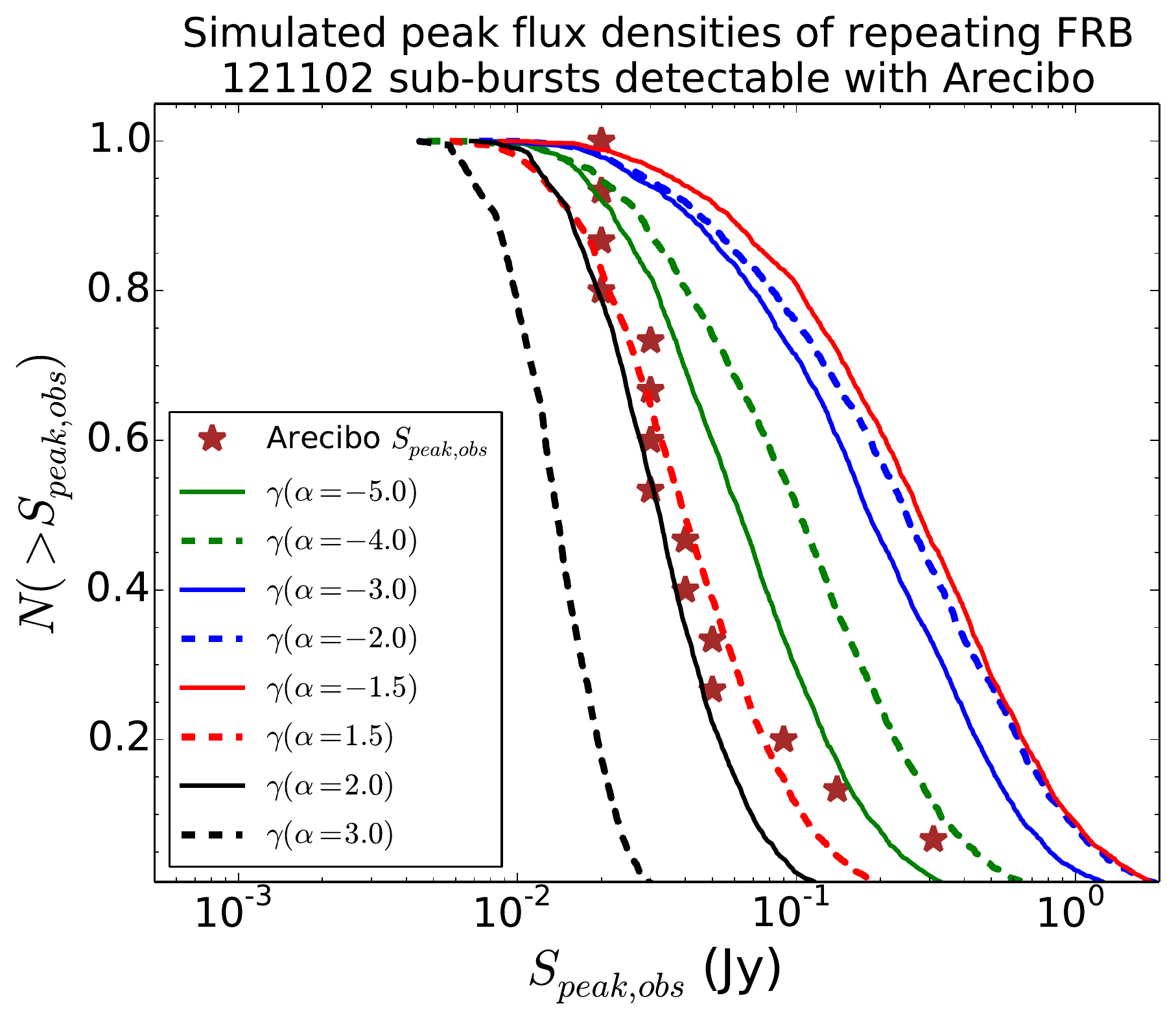}{0.46\textwidth}{}
          \fig{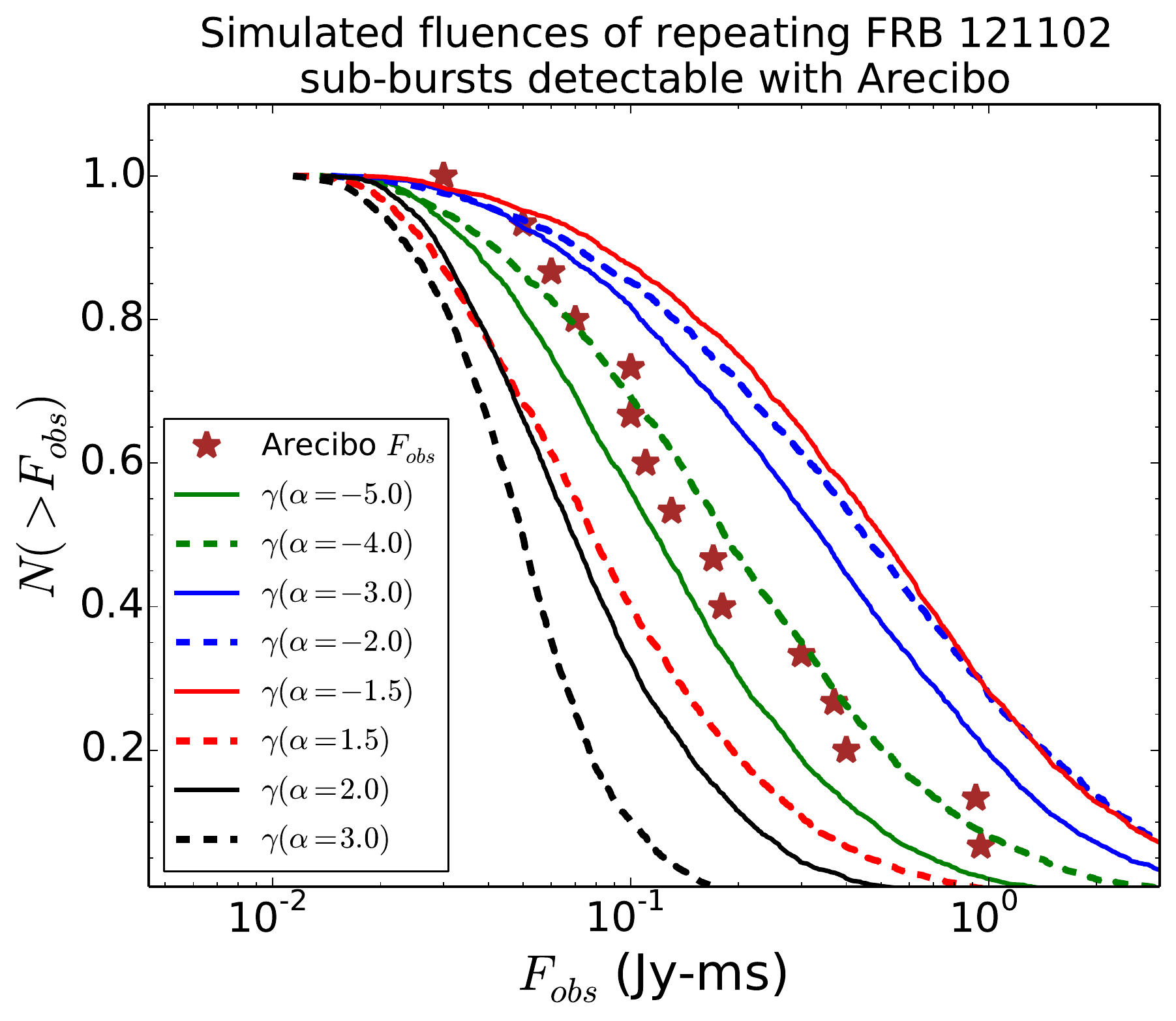}{0.46\textwidth}{}
          }  \vspace{-2.5em}              
  \caption{Comparison of FRB 121102 sub-bursts detected by Arecibo at observing frequency $\nu_{\rm obs} = 1.4\ {\rm GHz}$ and simulated $S_{\rm peak,obs}$ and $\mathcal{F}_{\rm obs}$ for FRB 121102 with different spectral indices. In both the panels, the MC simulation results are shown for $\alpha$ values ranging from -5.0 to 3.0. 
While $\beta$ and $n(z)$ are fixed for the simulations of repeating FRB 121102, we only show the simulation results for intrinsic width and luminosity model w1L1 and scattering model 2. 
%consider model 1 for the IGM and ISM scattering as the difference between the scattering models for the FRB 121102 sub-bursts is found to be negligible (see Section 2.3).
	{\it Left panel:} Simulation results for $S_{\rm peak,obs}$ and different $\alpha$,
	{\it Right panel:} Simulation results for $\mathcal{F}_{\rm obs}$ and different $\alpha$.	
}%
	\label{fig7}
\end{figure*}

%\EntryGap
\begin{table*}
%\label{Table5}
\begin{center}
\caption{\small KS test p-values from the comparison of simulated $S_{\rm peak,obs}$ and $\mathcal{F}_{\rm obs}$ with the observed repeating FRB 121102 sub-burst population detected at Arecibo with $\nu_{\rm obs} = 1.4\ {\rm GHz}$. The p-values are obtained for a fixed $\beta$ and $n(z)$ with $\gamma_{\rm eq} = \sqrt{\gamma_{S_{\rm peak,obs}}^{2} + \gamma_{\mathcal{F}_{\rm obs}}^{2}}$, and are listed for intrinsic width and luminosity models w1L1/w1L2/w2L1/w2L2 for each entry. 
All simulations are performed 
%the values are listed 
for scattering model 2 as the difference between the intrinsic distributions is found to be insignificant between the two scattering models (see Section 2).
%Table \ref{Table3}). 
}
\label{Table6}
\bgroup
\def\arraystretch{1.0}
\begin{tabular}{ccccccccccccccccccccc}
\hline
\hline
\centering
$\alpha$ & $\gamma_{S_{\rm peak,obs}}$ & $\gamma_{F_{\rm obs}}$ & $\gamma_{\rm eq}$ \\ \hline \hline 
-5.0 & 0.012/0.010/0.016/0.011 & 0.721/0.573/0.663/0.628 & 0.510/0.405/0.469/0.444 \\ \hline
-4.0 & $1.697\times10^{-4}$/$1.476\times10^{-4}$/$3.034\times10^{-4}$/$1.297\times10^{-4}$ & 0.683/0.760/0.782/0.699 & 0.483/0.537/0.553/0.494 \\ \hline
-3.0 & $1.278\times10^{-6}$/$1.242\times10^{-6}$/$7.717\times10^{-7}$/$9.317\times10^{-7}$ & 0.044/0.053/0.055/0.047 & 0.031/0.037/0.039/0.033 \\ \hline
-2.0 & $4.974\times10^{-7}$/$4.620\times10^{-7}$/$7.388\times10^{-7}$/$5.595\times10^{-7}$ & 0.011/0.010/0.014/0.011 & 0.008/0.007/0.010/0.008 \\ \hline
-1.5 & $1.281\times10^{-7}$/$1.850\times10^{-7}$/$1.517\times10^{-7}$/$1.878\times10^{-7}$ & 0.004/0.007/0.005/0.005 & 0.003/0.005/0.004/0.004 \\ \hline
1.5 & 0.625/0.676/0.681/0.643 & 0.056/0.077/0.042/0.050 & 0.444/0.481/0.482/0.456 \\ \hline
2.0 & 0.465/0.391/0.483/0.423 & 0.009/0.007/0.006/0.007 & 0.329/0.277/0.342/0.299 \\ \hline
3.0 & $6.796\times10^{-8}$/$1.036\times10^{-7}$/ & $5.075\times10^{-6}$/$5.970\times10^{-6}$/ & $3.589\times10^{-6}$/$4.222\times10^{-6}$/ \\ 
      & $6.112\times10^{-8}$/$8.332\times10^{-8}$ & $4.251\times10^{-6}$/$6.560\times10^{-6}$ & $3.006\times10^{-6}$/$4.639\times10^{-6}$ \\ \hline	       
\end{tabular}
\egroup
\end{center}
\end{table*}

Here we discuss the simulation results for the sub-bursts of the repeating FRB 121102 in order to better constrain the spectral properties of this source. In Figure \ref{fig7}, we show the results for the simulated $S_{\rm peak,obs}$ and $\mathcal{F}_{\rm obs}$ distributions of the FRB 121102 bursts and further compare them with the Arecibo observations at $\nu_{\rm obs} = 1.4\ {\rm GHz}$. We fix the host galaxy DM contribution relative to MW $\beta$ and the spatial density $n(z)$ model for all the simulations as the source redshift and the individual DM components ($DM_{\rm IGM}$, $DM_{\rm MW}$ and $DM_{\rm host}$) along the line of sight are both well known for FRB 121102. 
We only show the simulation results for intrinsic model w1L1 in Figure \ref{fig7} as the relative difference between the intrinsic models is found to be statistically negligible (see Section 4). 
%Also, we only consider model 1 for the IGM and ISM scattering in our simulations as the relative difference between the scattering models is found to be negligible for FRB 121102. 
For the power-law model $L_{\rm int}$ distributions of FRB 121102, we use $L_{\rm min} = 10^{41}\ {\rm erg/s}$ and $L_{\rm min} = 10^{43}\ {\rm erg/s}$.  
We consider the value of the energy spectral index $\alpha$ to be varying within the range of -5.0 to 3.0, as supported by the current observations.

As $z=0.19273$ is fixed for all the FRB 121102 bursts, the distribution for $L_{\rm obs}/E_{\rm obs}$ is essentially the same as that for $S_{\rm peak,obs}/\mathcal{F}_{\rm obs}$ while the $DM_{\rm tot}$ is fixed. This reduces the number of independent parameters among the observed/inferred quantities to only two, and here we consider $S_{\rm peak,obs}$ and $\mathcal{F}_{\rm obs}$ as the independent parameters for our analysis. Table \ref{Table6} lists the p-values from the comparison of the simulated $S_{\rm peak,obs}$ and $\mathcal{F}_{\rm obs}$ with the Arecibo $\nu_{\rm obs} = 1.4\ {\rm GHz}$ population for intrinsic model w1L1/w1L2/w2L1/w2L2. The KS test equivalent p-value is obtained from the two observable parameters as $\gamma_{\rm eq} = \sqrt{\gamma_{S_{\rm peak,obs}}^{2} + \gamma_{\mathcal{F}_{\rm obs}}^{2}}$. %\\ \\ \\
We find that the observed $S_{\rm peak,obs}$ for FRB 121102 bursts detected by Arecibo agree better with the simulated $S_{\rm peak,obs}$ results for moderately positive energy spectral indices, especially $\alpha \sim 1.0 - 2.0$. 
However, the observed $\mathcal{F}_{\rm obs}$ for the Arecibo bursts implies an energy density spectrum that decreases sharply with increasing energy for FRB 121102. 
%either a steep increasing or decreasing energy spectrum for this FRB. 
The $\gamma_{\rm eq}$ values obtained for FRB 121102 suggest either a large negative $\alpha \sim -5.0\ {\rm to}\ -4.0$ or moderately positive $\alpha \sim 1.0-2.0$ for this repeating FRB. As a result, it is very unlikely that the repeating FRB 121102 has a flat energy spectrum across its entire emission range and its spectrum is expected to be better constrained in the future once more bursts are detected by Arecibo at $\nu_{\rm obs} = 1.4\ {\rm GHz}$ and their spectral information are available.
%We find that it is very unlikely for the FRB 121102 energy spectrum to be shallow/flat, with either a large negative $\alpha$ within -5.0 to -4.0 or moderately positive $\alpha \sim 1.5-2.0$ suggested from the current observations.

\end{appendix}

\label{lastpage}


\begin{thebibliography}{10}

\bibitem[Bannister et al. (2017)]{Bannister17} Bannister K. W. et al., 2017,\apj, 841, L12

\bibitem[Bannister et al. (2019)]{Bannister19} Bannister K. W., et al., 2019, Science, 365, 565

\bibitem[Bera et al. (2016)]{Bera16} Bera A., Bhattacharyya S., Bharadwaj S., Bhat N. D. R., Chengalur J. N., 2016,\mnras, 457, 2530

\bibitem[Bhandari et al. (2018)]{Bhandari18} Bhandari S. et al., 2018,\mnras, 475, 1427

\bibitem[Bhandari et al. (2020)]{Bhandari20} Bhandari S., et al., 2020, arXiv e-prints, p. arXiv:2005.13160

\bibitem[Bhat et al. (2004)]{Bhat04} Bhat N. D. R., Cordes J. M., Camilo F., Nice D. J., Lorimer D. R., 2004,\apj, 605, 759

\bibitem[Bhattacharya (2019)]{MB19} Bhattacharya M., 2019, arXiv e-prints, p. arXiv:1907.11992

\bibitem[Caleb et al. (2016)]{Caleb16} Caleb M., Flynn C., Bailes M., Barr E. D., Hunstead R. W., Keane E. F., Ravi V., van Straten W., 2016,\mnras, 458, 708

\bibitem[Caleb et al. (2017)]{Caleb17} Caleb M. et al., 2017,\mnras, 468, 3746

\bibitem[Caleb et al. (2019)]{Caleb19} Caleb M., Stappers B. W., Rajwade K., Flynn C., 2019,\mnras, 484, 5500

\bibitem[Champion et al. (2016)]{Champion16} Champion D. J. et al., 2016,\mnras, 460, L30

\bibitem[Chatterjee et al. (2017)]{Chatterjee17} Chatterjee S. et al., 2017,\nat, 541, 58

\bibitem[CHIME/FRB Collaboration (2019a)]{CHIME19} CHIME/FRB Collaboration et al., 2019a, arXiv:1901.04525

\bibitem[CHIME/FRB Collaboration (2019b)]{CHIME19b} CHIME/FRB Collaboration et al., 2019b, Nature, 566, 230

\bibitem[Chittidi et al. (2020)]{Chittidi20} Chittidi J. S., Simha S., Mannings A., et al., 2020, arXiv e-prints, p. arXiv:2005.13158

\bibitem[Connor et al. (2016)]{Connor16} Connor L., Sievers J., Pen U.-L., 2016,\mnras, 458, L19

\bibitem[Connor \& Petroff (2018)]{CP18} Connor L., Petroff E., 2018,\apj, 861, L1

\bibitem[Connor (2019)]{Connor19} Connor L., 2019,\mnras, 487, 5753

\bibitem[Cordes \& Lazio (2002)]{CL02} Cordes J. M., Lazio T. J. W., 2002, preprint (\href{https://arxiv.org/abs/astro-ph/0207156}{astro-ph/0207156})

\bibitem[Cordes \& Wasserman (2016)]{CW16} Cordes J. M., Wasserman I., 2016,\mnras, 457, 232

\bibitem[Deng \& Zhang (2014)]{DZ14} Deng W., Zhang B., 2014,\apj, 783, L35

\bibitem[Dolag et al. (2015)]{Dolag15} Dolag K., Gaensler B. M., Beck A. M., Beck M. C., 2015,\mnras, 451, 4277

\bibitem[Falcke \& Rezolla (2014)]{FR14} Falcke H., Rezzolla L., 2014,\aap, 562, A137

\bibitem[Farah et al. (2018)]{Farah18} Farah W., Flynn C., Bailes M. et al., 2018,\mnras, 478, 1209

\bibitem[Gajjar et al. (2018)]{Gajjar18} Gajjar V. et al., 2018,\apj, 863, 2

\bibitem[Gao et al. (2014)]{Gao14} Gao H., Li Z., Zhang B., 2014,\apj, 788, 189

\bibitem[Gardenier et al. (2019)]{Gardenier19} Gardenier D. W., van Leeuwen J., Connor L., Petroff E., 2019,\aap, 632, A125

\bibitem[Gourdji et al. (2019)]{Gourdji19} Gourdji K., Michilli D., Spitler L. G., Hessels J. W. T., Seymour A., Cordes J. M., Chatterjee S., 2019,\apj, 877, L19

\bibitem[Hardy et al. (2017)]{Hardy17} Hardy L. K. et al., 2017,\mnras, 472, 2800

\bibitem[Hassall et al. (2013)]{Hassall13} Hassall T. E., Keane E. F., Fender R. P., 2013,\mnras, 436, 371

\bibitem[Hessels et al. (2019)]{Hessels19} Hessels J. W. T., et al., 2019,\apj, 876, L23

\bibitem[Inoue (2004)]{Inoue04} Inoue S., 2004,\mnras, 348, 999

\bibitem[Ioka (2003)]{Ioka03} Ioka K., 2003,\apj, 598, L79

\bibitem[James et al. (2019a)]{James19a} James, C. W., Ekers, R. D., Macquart, J. P., et al. 2019,
MNRAS, 483, 1342

\bibitem[James et al. (2019b)]{James19} James C. W., Oslowski S., Flynn C. et al., 2019, arXiv:1912.07847

\bibitem[Kashiyama et al. (2013)]{Kashiyama13} Kashiyama K., Ioka K., Meszaros P., 2013,\apj, 776, L39

\bibitem[Kashiyama \& Murase (2017)]{Kashiyama17} Kashiyama K., Murase K. 2017,\apj, 839, L3

\bibitem[Katz (2014)]{Katz14} Katz J. I., 2014,\prd, 89, 103009

\bibitem[Katz (2016)]{Katz16} Katz J. I., 2016,\apj, 826, 226

\bibitem[Keane et al. (2012)]{Keane12} Keane E. F. et al., 2012,\mnras, 425, L71

\bibitem[Keane \& Petroff (2015)]{KP15} Keane E. F., Petroff E., 2015,\mnras, 447, 2852

\bibitem[Keane et al. (2016)]{Keane16} Keane E. F. et al., 2016,\nat, 530, 453

\bibitem[Keane et al. (2019)]{Keane19} Keane E. F., Lorimer D. R., Crawford F., 2019, RNAAS, 3, 41

\bibitem[Krishnakumar et al. (2015)]{Krishnakumar15} Krishnakumar M. A., Mitra D., Naidu A., Joshi B. C., Manoharan P. K., 2015,\apj, 804, 23

\bibitem[Kulkarni et al. (2014)]{Kulkarni14} Kulkarni S. R., Ofek E. O., Neill J. D., Zheng Z., Juric M., 2014,\apj, 797, 70

\bibitem[Kumar et al. (2017)]{Kumar17} Kumar P., Lu W., Bhattacharya M., 2017,\mnras, 468, 2726

\bibitem[Law et al. (2017)]{Law17} Law C. J. et al., 2017,\apj, 850, 76

\bibitem[Lawrence et al. (2017)]{Lawrence17} Lawrence E., Vander Wiel S., Law C., Burke Spolaor S., Bower G. C., 2017,\aj, 154, 117

\bibitem[Loeb et al. (2014)]{Loeb14} Loeb A., Shvartzvald Y., Maoz D., 2014,\mnras, 439, L46

\bibitem[Lorimer et al. (2007)]{Lorimer07} Lorimer D. R., Bailes M., McLaughlin M. A., Narkevic D. J., Crawford F., 2007,\sci, 318, 777

\bibitem[Lorimer et al. (2013)]{Lorimer13} Lorimer D. R., Karastergiou A., McLaughlin M. A., Johnston S., 2013,\mnras, 436, L5

\bibitem[Lu \& Kumar (2016)]{LK16} Lu W., Kumar P., 2016,\mnras, 461, L122

\bibitem[Lu \& Kumar (2018)]{LK18} Lu W., Kumar P., 2018,\mnras, 477, 2470

\bibitem[Lu \& Piro (2019)]{LP19} Lu W., Piro A. L., 2019,\apj, 883, 40

\bibitem[Luan \& Goldreich (2014)]{LG14} Luan J., Goldreich P., 2014,\apj, 785, L26

\bibitem[Luo et al. (2018)]{Luo18}  Luo R., Lee K., Lorimer D. R., Zhang B., 2018,\mnras, 481, 2320

\bibitem[Lyubarsky (2014)]{Lyubarsky14} Lyubarsky Y., 2014,\mnras, 442, 9

\bibitem[Lyutikov et al. (2016)]{Lyutikov16} Lyutikov M., Burzawa L., Popov S. B., 2016,\mnras, 462, 941

\bibitem[Macquart \& Ekers (2018)]{ME18} Macquart J. P., Ekers R. D., 2018, MNRAS, 474, 1900

\bibitem[Macquart \& Koay (2013)]{MK13} Macquart J.-P., Koay J. Y., 2013,\apj, 776, 125

\bibitem[Macquart et al. (2019)]{Macquart2019} Macquart J.-P., Shannon R. M., Bannister K. W., James C. W., Ekers R. D., Bunton J. D., 2019,\apj, 872, L19

\bibitem[Macquart et al. (2020)]{Macquart2020} Macquart J. -P., Prochaska J. X., McQuinn M., et al., 2020,\nat, 581, 391

\bibitem[Madau \& Dickinson (2014)]{MD14} Madau P., Dickinson M., 2014,\araa, 52, 415

\bibitem[Marcote et al. (2017)]{Marcote17} Marcote B. et al., 2017,\apj, 834, L8

\bibitem[Marcote et al. (2020)]{Marcote20} Marcote B., Nimmo K., Hessels J. W. T., et al., 2020,\nat, 577,
190

\bibitem[Masui et al. (2015)]{Masui15} Masui K., Lin H.-H., Sievers J. et al. 2015,\nat, 528, 523

\bibitem[McQuinn (2014)]{McQuinn14} McQuinn M., 2014,\apj, 780, L33

\bibitem[Metzger et al. (2017)]{Metzger17} Metzger B. D., Berger E., Margalit B., 2017,\apj, 841, 14

\bibitem[Michilli et al. (2018)]{Michilli18} Michilli D. et al., 2018,\nat, 553, 182

\bibitem[Mottez \& Zarka (2014)]{MZ14} Mottez F., Zarka P., 2014,\aap, 569, A86

\bibitem[Niino (2018)]{Niino18} Niino Y., 2018, ApJ, 858, 4

\bibitem[Oppermann et al. (2018)]{Opp18} Oppermann N., Yu H.-R., Pen U.-L., 2018,\mnras, 475, 5109

\bibitem[Oostrum et al. (2019)]{Oos19} Oostrum L. C., et al., 2019, arXiv e-prints, p. arXiv:1912.12217

\bibitem[Oslowski et al. (2019)]{Oslowski19} Oslowski et al., 2019,\mnras, 488, 868

%\bibitem[Patel et al. (2018)]{Patel18} Patel C., Agarwal D., Bhardwaj M., et al. 2018, arXiv:1808.03710

\bibitem[Petroff et al. (2015a)]{Petroff15} Petroff E. et al., 2015,\mnras, 454, 457

\bibitem[Petroff et al. (2015b)]{Petroff15b} Petroff E. et al., 2015,\mnras, 447, 246

\bibitem[Petroff et al. (2016)]{Petroff16} Petroff E. et al., 2016,\pasa, 33, e045

\bibitem[Petroff et al. (2017)]{Petroff17} Petroff E. et al., 2017,\mnras, 469, 4465

\bibitem[Planck Collaboration et al. (2014)]{Planck14} Planck Collaboration et al., 2014,\aap, 571, 16

\bibitem[Piro (2012)]{Piro12} Piro A. L., 2012,\apj, 755, 80

\bibitem[Popov \& Postnov (2010)]{PP10} Popov S. B., Postnov K. A., 2010, Evolution of Cosmic Objects through their Physical Activity. Publishing House of Nat. Acad. Sci. Rep. Armenia (NAS RA), Gitutyun, p. 129

\bibitem[Prochaska et al. (2019)]{Prochaska19} Prochaska J. X., Macquart J. P., McQuinn M., et al., 2019, Science, 366, 231

\bibitem[Prochaska \& Zheng (2019)]{PZ19} Prochaska J. X., Zheng Y., 2019,\mnras, 485, 648

\bibitem[Rajwade et al. (2020)]{Rajwade20} Rajwade K. M., et al., 2020, arXiv e-prints, p. arXiv:2003.03596

\bibitem[Rane et al. (2016)]{Rane16} Rane A., Lorimer D. R., Bates S. D., McMann N., McLaughlin M. A., Rajwade K., 2016,\mnras, 455, 2207

\bibitem[Rane (2017)]{Rane17} Rane A., 2017, PhD thesis, West Virginia University

\bibitem[Ravi et al. (2015)]{Ravi15} Ravi V., Shannon R. M., Jameson A., 2015,\apj, 799, L5

\bibitem[Ravi (2019a)]{Ravi19} Ravi V., 2019,\apj, 872, 1

\bibitem[Ravi et al. (2019b)]{Ravi19b} Ravi V., et al., 2019,\nat, 572, 352

\bibitem[Scholz et al. (2016)]{Scholz16} Scholz P. et al., 2016,\apj, 833, 177

\bibitem[Scholz et al. (2017)]{Scholz17} Scholz P. et al., 2017,\apj, 846, 80

\bibitem[Scholz \& The CHIME Collaboration (2020)]{ScholzCHIME20} Scholtz P., The CHIME Collaboration 2020, The Astronomer's Telegram, 13681, 1

\bibitem[Shannon et al. (2018)]{Shannon18} Shannon R. M. et al., 2018,\nat, 562, 386

\bibitem[Shull \& Danforth (2018)]{SD18} Shull J. M., Danforth C. W., 2018,\apj, 852, L11

\bibitem[Simha et al. (2020)]{Simha20} Simha S., Burchett J. N., Prochaska J. X., et al., 2020, arXiv e-prints, p. arXiv:2005.13157

\bibitem[Smith et al. (2011)]{Smith11} Smith B. D., Hallman E. J., Shull J. M., O'Shea B. W., 2011,\apj, 731, 6

\bibitem[Spitler et al. (2014)]{Spitler14} Spitler L. G. et al., 2014,\apj, 790, 101

\bibitem[Spitler et al. (2016)]{Spitler16} Spitler L. G. et al., 2016,\nat, 531, 202

\bibitem[Spitler et al. (2018)]{Spitler18} Spitler L. G. et al., 2018,\apj, 863, 150

\bibitem[Staveley-Smith et al. (1996)]{Stav96} Staveley-Smith L. et al., 1996,\pasa, 13, 243

\bibitem[The CHIME/FRB Collaboration et al. (2020)]{CHIME20} The CHIME/FRB Collaboration et al., 2020, arXiv e-prints, p. arXiv:2001.10275

\bibitem[Tendulkar et al. (2017)]{Tendulkar17} Tendulkar S. P. et al., 2017,\apj, 834, L7

\bibitem[Thornton (2013)]{Thorn13} Thornton D., 2013, PhD thesis, The University of Manchester

\bibitem[Thornton et al. (2013)]{Thornton13} Thornton D. et al., 2013,\sci, 341, 53

\bibitem[Totani (2013)]{Totani13} Totani T., 2013,\pasj, 65, L12

\bibitem[Vandenberg (1976)]{Vandenberg76} Vandenberg N. R., 1976,\apj, 209, 578

\bibitem[Vedantham et al. (2016)]{Vedantham16} Vedantham H. K., Ravi V., Hallinan G., Shannon R. M., 2016, ApJ, 830, 75

\bibitem[Williamson (1972)]{Williamson72} Williamson I. P., 1972,\mnras, 157, 55

\bibitem[Zhang (2014)]{Zhang14} Zhang B., 2014,\apj, 780, L21

\bibitem[Zhang (2017)]{Zhang17} Zhang B., 2017,\apj, 836, L32

\bibitem[Zheng et al. (2014)]{Zheng14} Zheng Z., Ofek E. O., Kulkarni S. R., Neill J. D., Juric M., 2014,\apj, 797, 71





\end{thebibliography}
\end{document}